\RequirePackage{fix-cm}
\documentclass[smallextended]{svjour3}       
\smartqed  
\usepackage{graphicx}
\usepackage{txfonts}
\def\arcsec{$^{\prime\prime}$}
\def\arcmin{$^{\prime}$}

\begin{document}

\title{Clusters of galaxies : observational properties of the diffuse
radio emission 
}

\titlerunning{Diffuse radio emission in galaxy clusters}        

\author{Luigina Feretti         \and
        Gabriele Giovannini     \and
        Federica Govoni         \and
        Matteo Murgia
}

\authorrunning{Feretti et al.} 

\institute{L. Feretti \at
          INAF Istituto di Radioastronomia, via Gobetti 101, 40129 Bologna, I
            \\
              \email{lferetti@ira.inaf.it}           
           \and
           G. Giovannini \at
              Dipartimento di Astronomia, via Ranzani 1, 40127 Bologna, I \\
           \and
           F. Govoni \at
        INAF Osservatorio Astronomico di Cagliari, Strada 54,
               Loc. Poggio dei Pini 09012 Capoterra (Ca) I \\
           \and
           M. Murgia \at
        INAF  Osservatorio Astronomico di Cagliari, Strada 54, Loc.
               Poggio dei Pini 09012 Capoterra (Ca) I
}

\date{Received: date / Accepted: date}

\maketitle

\begin{abstract}

Clusters of galaxies, as the largest virialized systems in the
Universe, are ideal laboratories to study the formation and evolution
of cosmic structures.  The luminous matter of clusters consists of
galaxies and of an embedding intracluster medium (ICM), which has been
heated to temperatures of tens of millions degrees, and thus is
detected through its thermal emission in the soft X-ray regime.  Most
of the detailed knowledge of galaxy clusters has been obtained in
recent years from the study of ICM through X-ray Astronomy. At the
same time, radio observations have proved that the ICM is mixed with
non-thermal components, i.e. highly relativistic particles and
large-scale magnetic fields, detected through their synchrotron
emission.

The knowledge of the properties of these non-thermal ICM components
has increased significantly, owing to sensitive radio images and to
the development of theoretical models.  Diffuse synchrotron radio
emission in the central and peripheral cluster regions has been found
in many clusters. Moreover large-scale magnetic fields appear to be
present in all galaxy clusters, as derived from Rotation Measure (RM)
studies.  Non-thermal components are linked to the cluster X-ray
properties, and to the cluster evolutionary stage, and are cucial for
a comprehensive physical description of the intracluster medium. They
play an important role in the cluster formation and evolution.

We review here the observational properties of diffuse non-thermal
sources detected in galaxy clusters: halos, relics and mini-halos. We
discuss their classification and properties.  We report published
results up to date and obtain and discuss statistical properties. We
present the properties of large-scale magnetic fields in clusters and
in even larger structures: filaments connecting galaxy clusters. We
summarize the current models of the origin of these cluster
components, and outline the improvements that are expected in this
area from future developments thanks to the new generation of radio
telescopes.

\end{abstract}

\section{Introduction}
\label{intro}

Clusters of galaxies are the largest gravitationally bound systems in
the Universe, with typical masses of about 10$^{15}$ M$_{sun}$, and
volumes of about 100 Mpc$^3$.  Most of the gravitating matter in any
cluster is in the form of dark matter ($\sim$ 80\%). Some of the
luminous matter is in galaxies ($\sim$ 3\% - 5\%), the rest is in diffuse
hot gas ($\sim$ 15\% - 17\%), detected in X-ray through its thermal
bremsstrahlung emission.  This thermal plasma, consisting of particles
of energies of several keV, is commonly referred to as Intracluster
Medium (ICM). Most of the detailed knowledge of galaxy clusters has been
obtained in recent years  from the study of ICM through X-ray Astronomy.

Clusters are formed by hierarchical structure formation processes.  In
this scenario, smaller units (galaxies, groups and small clusters)
formed first and merged under gravitational pull to larger and larger
units in the course of time.  Cluster mergers are the mechanism by
which clusters are assembled. Denser regions form a filamentary
structure in the Universe, and clusters are formed within filaments, often
at their intersection, by a combination of large and small mergers.
Major cluster mergers are among the most energetic events in the
Universe since the Big Bang \cite{Sarazin2002}.  During mergers,
shocks are driven into the ICM, with the subsequent injection of
turbulence.  The merger activity, which has characterized much of the
history of the Universe, appears to be continuing at the present time
and explains the relative abundance of substructure and temperature
gradients detected in clusters of galaxies by optical and X-ray
observations.

Eventually, clusters reach a relaxed state, characterized by a giant
galaxy at the center, and enhanced X-ray surface brightness peak in
the core.  The hot gas in the centre has a high density, which implies
short radiative cooling times, typically one or two order of
magnitudes smaller than the Hubble time. Therefore energy losses due
to X-ray emission are dramatic and produce a temperature drop towards
the center.  It was formerly suggested that in these conditions, the
ICM plasma in the cluster core should cool and condense, giving rise
to a steady, pressure-driven inward flow caused by the compression of
the hot surrounding gas. Relaxed clusters were then classified as
``cooling flow'' clusters \cite{Fabian1994}, with predicted gas mass
deposition rates in the cluster center up to 1000 M$_{sun}$
year$^{-1}$.  This model was the subject of much debate, when
XMM-Newton spectral results failed to confirm the lines and features
expected as a product of a steady state cooling flow
\cite{Peterson2001,Peterson2006}.  X-ray and optical studies showed
that the gas cooling rates were overestimated by an order of magnitude
or more \cite{Mcnamara2006}.  The classical ``cooling-flow'' model has
finally been replaced by the paradigm of ``cool-core''.
Observationally cool-cores are characterized by a strong peak in the
surface brightness, a significant drop in the temperature and a peak
in the metal distribution (e.g. \cite{Degrandi2001}). The cooling time
is often much shorter than 1 Gyr, therefore some source of heating is
necessary to balance the radiation losses. At present, it is widely
accepted that the source of heating in cool-core clusters is the AGN
activity of the brightest cluster galaxy at the center (see
\cite{Mcnamara2007,Bohringer2010} for recent reviews).

Galaxy clusters are also characterized by emission in the radio
band. Obvious radio sources are the individual galaxies, whose
emission has been observed in recent decades with sensitive radio
telescopes. It often extends well beyond the galaxy optical
boundaries, out to hundreds of kiloparsec, and hence it is expected
that the radio emitting regions interact with the ICM. This
interaction is indeed observed in tailed radio galaxies, and radio
sources filling X-ray cavities at the centre of cool-core clusters
\cite{Mcnamara2007,Feretti2008}.

More puzzling are diffuse extended radio sources, which cannot be
obviously ascribed to individual galaxies, but are instead associated
with the ICM.  This radio emission represents a striking feature of
clusters, since it demonstrates that the thermal ICM plasma is mixed
with non-thermal components. Such components are large-scale magnetic
fields with a population of relativistic electrons in the cluster
volume.  Further demonstration of the existence of magnetic fields in
the ICM is obtained by studies of the Faraday rotation of polarized
radio galaxies, lying in the background or embedded within the
magnetized intracluster medium.

The number density of relativistic particles is of the order of
10$^{-10}$ cm$^{-3}$, while their Lorentz factors are $\gamma >>$
1000.  Magnetic field intensities are around $\sim$ 0.1-1 $\mu$G, and
the energy density of the relativistic plasma is globally $\lesssim$
1\% than that of the thermal gas.  Non-thermal components are
nevertheless important for a comprehensive physical description of the
intracluster medium in galaxy clusters, and play a major role in the
evolution of large-scale structures in the Universe.  They can have
dynamic and thermodynamic effects: magnetic fields affect the heat
conduction in the ICM and the gas dynamics, relativistic particles are
sources of additional pressure and undergo acceleration processes that
can modify the details of the ICM heating process. Realistic
cosmological simulations of galaxy cluster formation also include
non-thermal components, thus a deep knowledge of the properties of
these components and of their interplay with the thermal gas are
needed to properly constrain the large-scale structure formation
scenario.
The discovery of diffuse cluster radio emission has represented an
important step in the understanding of the physical processes in
clusters of galaxies.  Diffuse synchrotron sources are sensitive to
the turbulence and shock structures of large-scale environments and
provide essential complements to studies at other wavebands. Studies
in the radio domain will fill essential gaps in both cluster
astrophysics and in the growth of structure in the Universe,
especially where the signatures of shocks and turbulence, or even the
thermal plasma itself, may be otherwise undetectable. 

The aim of this review is to present the observational results
obtained in recent years in the radio domain related to the diffuse
radio sources, in order to give an overview of the state of the art of
the current knowledge of non-thermal cluster components.  We will show
how the radio properties can be linked to the X-ray properties and to
the cluster evolutionary state, and will discuss the physical
implications.

Cluster non-thermal emission of inverse-Compton (IC) origin is
expected in the hard X-ray domain, due to scattering of the cosmic
microwave background photons by the synchrotron relativistic electrons
\cite{Sarazin1999}. To present days, the detection of a non-thermal
hard tail in the X-ray spectrum of clusters of galaxies has been
reported in the literature 
(e.g. \cite{Fusco1999,Rephaeli2002,Fusco2004,Fusco2005,Fusco2007,Wik2009,Fusco2011,Wik2011}), and it is still debated in some cases.  
The presence of a high frequency non-thermal emission is clearly
confirmed in the Ophiuchus cluster \cite{Murgia2010b,Eckert2011}.  
The discussion of this topic is beyond the scope of this review, which is
focussed on the results in the radio band.

The organization of this paper is as follows: In Sect. \ref{sec:2} the
general properties of diffuse cluster radio sources and their
classification are presented; in Sect. \ref{sec:3} the problems
related to the detection of diffuse emission are outlined;
Sects. \ref{sec:4}, \ref{sec:5}, \ref{sec:6} give the observational
properties of radio halos, relics and mini-halo, respectively.  In
Sect. \ref{sec:7} the current results about magnetic fields are
presented.  Sect. \ref{sec:8} presents the evidence for radio emission
and magnetic fields beyond clusters of galaxies.  Sect. \ref{sec:9}
briefly summarizes the current models, and Sect. \ref{sec:10} depicts
the future prospects.

The intrinsic parameters quoted in this paper are computed with
$\Lambda$CDM cosmology with $H_0$ = 71 km s$^{-1}$Mpc$^{-1}$,
$\Omega_m$ = 0.27, and $\Omega_{\Lambda}$ = 0.73.  Values taken from
the literature have been scaled to this cosmology.

\section{Synchrotron emission and diffuse cluster radio sources}
\label{sec:2}

Synchrotron emission is produced by the
spiraling motion of relativistic electrons in a magnetic field.
The power emitted by a relativistic electron depends on its energy
(Lorentz factor) and on the magnetic field strength.  The higher
the magnetic field strength, the lower the electron energy needed
to produce emission at a given frequency.
The case of astrophysical interest is that of a homogeneous and
isotropic population of electrons with a power-law energy
distribution with index $\delta$. In this case, if the plasma is
optically thin, as in diffuse cluster radio sources, the diagnostics
of synchrotron radiation are the following:

i) the emissivity is related to the number density of the
relativistic electrons and to the intensity of the magnetic field, and
the monochromatic emissivity follows a power-law with spectral index
related to the index of the electron energy distribution
$\alpha=(\delta-1)/2$; typical observed radio spectra are around 0.7 - 0.8,
consistent with typical values of $\delta$ around 2.5.

ii) energy losses of the radio emitting particles produce, with passing
time, a change in the overall particle energy distribution, which in
turn produces a modification of the emitted radio spectrum.  In
particular, the spectrum shows a cutoff at frequencies higher than a
critical frequency $\nu^*$, related to the particle lifetime.
Therefore aged radio sources show curved spectra and/or
spectra steeper that the values 0.7 - 0.8 indicated above;

iii) the synchrotron radiation from a population of
relativistic electrons in a uniform magnetic field is linearly
polarized, with the electric (polarization) vector perpendicular to
the projection of the magnetic field onto the plane of the sky.  The
intrinsic polarization degree depends on the
particle energy distribution, and equals $\sim$ 75\% - 80\% for typical
values of the spectral index.  Complex and disordered magnetic field
structures decrease the observed  degree of polarization;

iv) the total energy content in a synchrotron radio source depends on
the contributions from relativistic particles (electrons plus protons)
and from magnetic fields (considering the fraction of the source
volume occupied by magnetic fields, i.e. the filling factor).  From
radio data, it is possible to derive, the minimum value of the total
energy, which occurs when the contributions of magnetic fields and
relativistic particles are approximately equal (equipartition
condition). The corresponding magnetic field is commonly referred to
as equipartition value B$_{eq}$.

Rigorous treatments of the synchrotron theory and radiation
processes are in \cite{Pacholczyk1970,Tucker1975}. Useful formulas 
of common use can be  found in \cite{Govoni2004b,Feretti2008}.

\begin{figure*}
 \includegraphics[width= 1\textwidth]{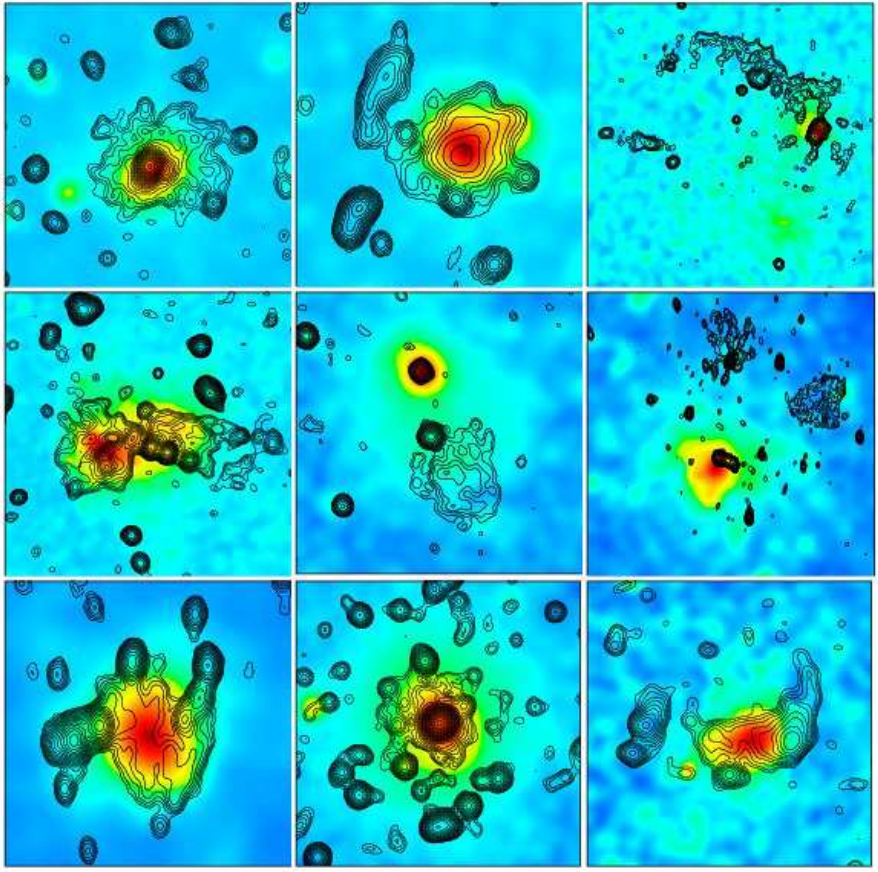}
\caption
{Collection of clusters showing several types of radio emission, shown
  in contours, overlaid onto the X-ray emission, shown in
  colors. Clusters are (from left to right and from top to bottom)
  A\,2219 (halo), A\,2744 (halo + relic), A\,115 (relic),
  A\,754 (complex, halo plus relic), A\,1664 (relic), A\,548b (relic),
  A\,520 (halo), A\,2029 (mini-halo), RXCJ1314.4-2515 (halo
  plus double relics). For references to these objects see
  Tab. \ref{tab-halo}, \ref{table-rel}, and \ref{tab-minih}.
}
\label{fig:clcollection}
\end{figure*}

In recent years, there has been growing evidence for the existence of
cluster large-scale diffuse radio sources, of synchrotron origin,
which have no optical counterpart and no obvious connection to the
cluster galaxies, and are therefore associated with the ICM
\cite{Giovannini2002,Feretti2008,Ferrari2008}.  The first diffuse
radio source detected in a cluster of galaxies is the giant radio halo
at the center of the Coma cluster \cite{Large1959,Willson1970}.
Several years later, extended diffuse emission was also detected at
the periphery of the Coma cluster \cite{Ballarati1981,Giovannini1991}
and at the center of the Perseus cluster
\cite{Miley1975,Noordam1982,Burns1992}.  Nowadays, diffuse radio
emission with surface brightness down to 0.1 $\mu$Jy arcsec$^{-2}$ at
1.4 GHz is known in about 80 clusters, under several cluster
evolutionary conditions (merging and relaxed clusters), at different
cluster locations (center, periphery, intermediate distance), and on
very different size scales: 100 kpc to $>$Mpc, (see
Fig. \ref{fig:clcollection} for several examples).

Diffuse sources are typically grouped in halos, relics and mini-halos
\cite{Feretti1996}, according to their location in the cluster and
to the cluster type (merging or cool-core). The three classes will be
extensively discussed in the next sections. We anticipate that
halos are hosted in clusters showing merging processes
\cite{Feretti2000,Giovannini2002} and are located at the cluster
center. Relics are located in cluster peripheral regions of both
merging and relaxed clusters. Prototypical examples of these two classes
are both in the Coma cluster: the radio halo Coma C and the relic
1253+275.  Mini-halos are hosted in relaxed cool-core
clusters,  are centrally located, and usually surround a
powerful radio galaxy; the protoype of this class is in the Perseus
cluster. 
It should be remarked here that non-thermal diffuse emission can also
be present beyond clusters, indicating the presence of relativistic
particles and magnetic fields in very low density environments.

All diffuse radio sources in clusters have very steep
radio spectra in common.
A clear difference between halos and relics is in polarization
property, with peripheral sources being more strongly polarized than
the central ones.  A possible problem in the classification of halos
and relics could be represented by projection effects. In other words,
halos could be actually located at cluster periphery, and simply
projected onto the cluster center.  This issue has been
analyzed several years ago \cite{Feretti2000}, using data at the time on
about 30 clusters. From the distribution of projected distances of
halos and relics from the cluster center, it was shown that there is a
large excess of sources at the center, thus
diffuse sources are not located at random position in clusters.
This indicates that central halos are truly at the cluster center and not
simply projected onto it. Moreover, detailed radio images show
different morphologies of halos and relics.
There are 42 halos currently known, while 39 clusters show at
least 1 relic source (for a total of 50 relics).
This confirms the findings above. A powerful way
to distinguish between the two classes is through the polarization,
this will become easy with new generation radio telescopes.

Giovannini et al. \cite{Giovannini1999} showed that diffuse cluster
sources are not present in all galaxy clusters, but are not a
phenomenon as rare as formerly believed. They first showed that diffuse
sources with the surface brightness limit of the NRAO VLA Sky Survey
(NVSS \cite{Condon1998}) are present in 6\%-9\% of clusters with
L$_x$ $<$ 5 $\times$ 10$^{44}$ erg/s, but are present in 27\% -- 44\%
of clusters with L$_x$ $>$ 5 $\times$ 10$^{44}$ erg/s.  This trend is
confirmed for the occurrence of halos and relics, as reported in
Sect. \ref{sec:4} and Sect. \ref{sec:5}, respectively.

\section{Detection of diffuse emission}
\label{sec:3}

For the detection of diffuse radio sources, the sensitivity to low
surface brightness emission is the critical parameter.  Single dish
observations are in principle best suited to detect extended emission,
however they are generally limited by the high confusion, because of the
large beam. Moreover, low angular resolution may
prevent the separation of embedded unrelated discrete sources.  On the
other hand, an interferometer is characterized by a smaller observing
beam, and thus by lower confusion, but its brightness sensitivity, set
primarily by the aperture filling factor and the shortest spacings,
may be insufficient.  A proper sampling of the short spacings is vital not
only for the detection of an extended source, but more crucially for
the determination of the full size and the measurement of the source
flux density.

Multi-frequency images with a similar uv-coverage are needed to derive
spectral index distributions, which can provide information on the age
and energy of radio emitting particles.  The total spectral index can
be easily obtained, but in many sources confusion problems do not
allow a proper determination of low frequency fluxes. 

\begin{figure*}
  \includegraphics[width=0.7\textwidth]{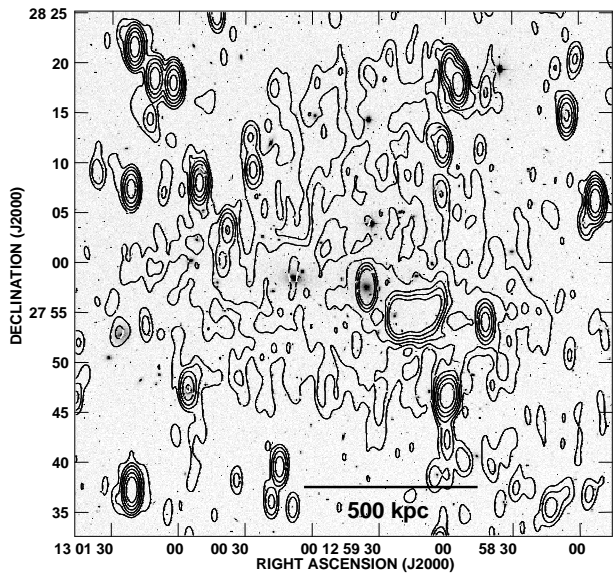}
\caption
{Diffuse radio halo Coma C in the Coma cluster (z = 0.023) at 0.3 GHz
  obtained with the WSRT, superimposed onto the optical image from the
  Digitized Sky Survey DSS1.  The resolution of the radio image is
  55\arcsec$\times$ 125\arcsec (FWHM, RA $\times$ DEC); contour levels
  are: 3, 6, 12, 25, 50, 100 mJy/beam.
}
\label{fig:coma327}
\end{figure*}

To summarize, the instrument best suited to detect cluster diffuse
radio sources should provide a compromise between high sensitivity to
surface brightness, and high angular resolution.  Moreover, since
diffuse emission is generally characterized by steep radio spectra,
the observations should be carried out at low frequencies ($\lesssim$
1.4 GHz). The Very Large Array (VLA) at 1.4 GHz
has been demonstrated to be a very good instrument to detect and study
radio halos, owing to its
superb sensitivity.  However, it should be noted, as a caveat, that the Coma
radio halo (see Fig. \ref{fig:coma327}) is not detected by 
Westerbork Synthesis Radio Telescope (WSRT) and VLA observations at
1.4 GHz because of the missing short spacings.

After the discovery of the Coma radio halo at 408 MHz with the 250-ft
radio telescope at Jodrell Bank \cite{Large1959}, confirmed from
interferometric Cambridge One-Mile telescope radio data
\cite{Willson1970}, observations carried out mostly with single dish
radio telescopes (Arecibo, Green Bank 300-ft, Effelsberg and others)
and the WSRT found about 10 other clusters with a diffuse halo-type
radio emission (see \cite{Hanisch1982}).  Jaffe \& Rudnick
\cite{Jaffe1979}, in their search for radio halos, found an extended
emission region in the peripheral region of the Coma cluster near the
strong radio source Coma A (3C 277.3). This was suggested to be a
feature of the Coma cluster, in the following classified as a relic,
thanks to observations at 408 MHz with the Northern Cross Radio
Telescope at better angular resolution \cite{Ballarati1981}.

\begin{figure*}
  \includegraphics[width=1\textwidth]{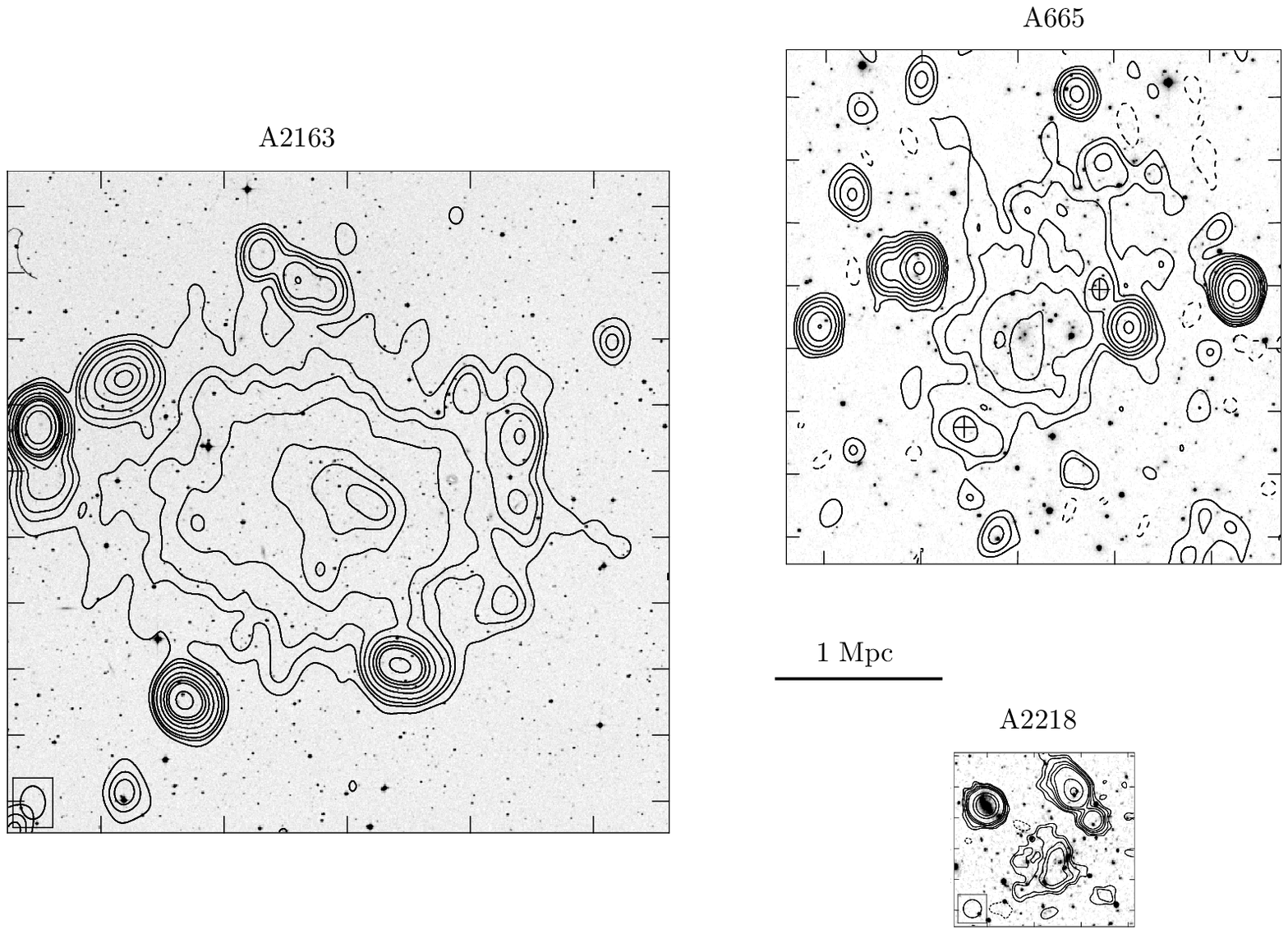}
\caption{Images of the
clusters A\,2163, A\,665 and A\,2218, hosting radio halos:
radio emission is represented by contours, which are overlaid onto
the optical image. The maps are all scaled to the same linear scale.
}
\label{fig:hacollection}
\end{figure*}

A significant breakthrough in the study of radio halos and other
diffuse cluster radio sources was obtained thanks to the all-sky
surveys with the VLA and WSRT.  Giovannini et
al. \cite{Giovannini1999}, combining radio data from the NVSS at 1.4
GHz and X-ray catalogues, detected 18 new halo and relic candidates,
in addition to the 11 already known, owing to the large beam and the
good surface brightness sensitivity of the VLA D configuration. All
new candidates were confirmed by more sensitive targeted observations,
mostly performed with the VLA.  Kempner \& Sarazin \cite{Kempner2001}
presented 7 new candidates from a search in the Westerbork Northern
Sky Survey (WENSS \cite{Rengelink1997}) at 327 MHz; Rudnick \&
Lemmerman \cite{Rudnick2009} searched for Mpc-scale radio emission
reprocessing WENSS radio images.

Recently, other extensive radio observations have been published.  We
note the Giant Metrewave Radio Telescope (GMRT) survey of massive
galaxy clusters at z = 0.2 - 0.4 by Venturi and collaborators
\cite{Venturi2007,Venturi2008}, who
found diffuse sources in 10 clusters (out of which 3 new halos, 2 new
core-halo sources and 3 new halo or relic candidates), the GMRT follow
up at low frequencies of all clusters known to contain radio halos and
relics up to z = 0.4 \cite{Giacintucci2011a}, and the observing
campaign carried out with the WSRT, VLA and GMRT by Van Weeren et al.
\cite{Vanweeren2011a}, who discovery  6 new radio
relics, including a probable double system, and 2 radio halos.  The
above mentioned studies and their follow-ups have produced rich
information in a short time in this field.

\section{Radio halos}
\label{sec:4}

Radio halos are diffuse radio sources of low surface brightness
($\sim$ 1 -- 0.1 $\mu$Jy arcsec$^{-2}$ at 1.4 GHz) permeating the
central volume of a cluster. They are typically extended with sizes of
$\gtrsim$ 1 Mpc, regular in morphology, and are unpolarized down to a
few percent level, probably because of internal or beam
depolarization.  The prototype of this class, Coma C at the center of
the Coma cluster (Fig. \ref{fig:coma327}) has been studied in detail
by many authors (see
e.g. \cite{Willson1970,Giovannini1993,Kronberg2007,Brown2011a}).

\begin{figure*}
  \includegraphics[width=0.8\textwidth]{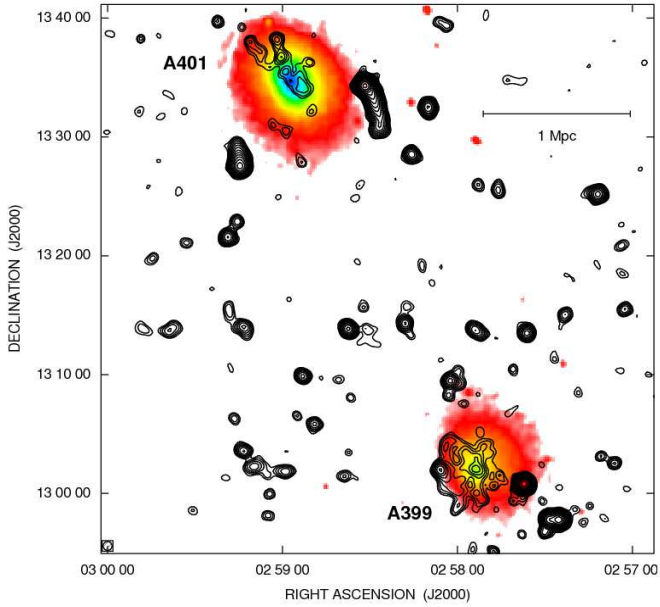}
\caption{
Total intensity radio contours of the complex A401-A399. The radio
image, obtained with the VLA at 1.4 GHz, is shown by the iso-contours
and has a FWHM of 45$''\times 45''$.  The first contour level is drawn
at 120 $\mu$Jy/beam and the rest are spaced by\ a factor of
$\sqrt{2}$.  The color scale represents the XMM X-ray image in the
0.2$-\ $12 keV band. }
\label{fig:hadouble}
\end{figure*}

Thanks to the improvements in observations and data reduction
procedures, radio halos with a smaller size and irregular
morphology have also been detected in rich galaxy clusters. Their
properties in most cases are similar to those of giant radio
halos. Fig. \ref{fig:hacollection} presents images of well known radio
halos of different sizes.
A spectacular case is represented by the double merging system, A399
and A401, which both contain a radio halo and can be considered the
only case so far of a double radio halo system (\cite{Murgia2010a},
Fig. \ref{fig:hadouble}).

\begin{table*}
\caption{September2011-Halo collection (published halos, Sept. 2011)}
\label{tab-halo}
\centering
\begin{tabular}{lclrrrrcl}
\hline\hline
Name    & z & kpc/$''$ & S(1.4)   &Log P(1.4) & LLS & Lx(10$^{44}$)  & Ref
 & Notes \\
        &   &          & mJy      & W/Hz      & Mpc & erg/sec        &    
 &       \\
\hline
A209    & 0.2060 & 3.34 & 16.9  & 24.31 & 1.40 & 6.17 &
\cite{Giovannini2009} &a relic could be present \\
A399    & 0.0718 & 1.35 & 16.0  & 23.30 & 0.57 & 3.80 & \cite{Murgia2010a}
& double with A401\\
A401    & 0.0737 & 1.38 & 17.0  & 23.34 & 0.52 & 6.52 & \cite{Bacchi2003}
& double with A399 \\
A520    & 0.1990 & 3.25 & 34.4  & 24.58 & 1.11 & 8.30 & \cite{Govoni2001c}
& \\
A521    & 0.2533 & 3.91 &  5.9  & 24.05 & 1.17 & 8.47 &
\cite{Giovannini2009} & a relic is also present \\
A523    & 0.1036 & 1.88 & 59.0  & 24.17 & 1.3  & 1.07 &
\cite{Giovannini2011} &  \\
A545    & 0.1540 & 2.64 & 23.0  & 24.16 & 0.89 & 5.55 & \cite{Bacchi2003}
& \\
A665    & 0.1819 & 3.03 & 43.1  & 24.59 & 1.82 & 9.65 &
\cite{Giovannini2000} & \\
A697    & 0.2820 & 4.23 &  5.2  & 24.11 & 0.65 &10.40 &
\cite{Vanweeren2011a} & \\
A746    & 0.232  & 3.67 & 18.0  & 24.58 & 0.85 & 3.68 &
\cite{Vanweeren2011a} & a relic is also present \\
A754    & 0.0542 & 1.04 & 86.0  & 23.77 & 0.99 & 2.21 & \cite{Bacchi2003}
& a relic is also present \\
A773    & 0.2170 & 3.48 & 12.7  & 24.23 & 1.25 & 7.95 & \cite{Govoni2001c}
& \\
A781    & 0.3004 & 4.42 & 20.5  & 24.77 & 1.60 & 4.6  & \cite{Govoni2011}
& a relic is also present \\
A851    & 0.4069 & 5.40 &  3.7  & 24.33 & 1.08 & 5.04 &
\cite{Giovannini2009} &   \\
A1213   & 0.0469 & 0.91 & 72.2  & 23.56 & 0.22 & 0.10 &
\cite{Giovannini2009} &  \\
A1300   & 0.3072 & 4.49 & 20.0  & 24.78 & 1.3  &13.73 & \cite{Reid1999}& a
relic is also present \\
A1351   & 0.3224 & 4.64 & 39.6  & 25.12 & 0.84 & 5.47 &
\cite{Giovannini2009} &         \\
A1656   & 0.0231 & 0.46 &530.0  & 23.80 & 0.83 & 3.99 & \cite{Kim1990} & a
relic is also present \\
A1689   & 0.1832 & 3.05 &  9.9  & 23.96 & 0.73 &12.00 & \cite{Vacca2011} & \\
A1758a  & 0.2790 & 4.20 & 16.7  & 24.60 & 1.51 & 7.09 &
\cite{Giovannini2009} &total diffuse emission \\
A1914   & 0.1712 & 2.88 & 64.0  & 24.71 & 1.28 &10.42 & \cite{Bacchi2003}
& \\
A1995   & 0.3186 & 4.61 &  4.1  & 24.13 & 0.83 & 8.83 &
\cite{Giovannini2009} &   \\
A2034   & 0.1130 & 2.03 & 13.6  & 23.64 & 0.61 & 3.81 &
\cite{Giovannini2009} &  a relic is also present \\
A2163   & 0.2030 & 3.31 &155.0  & 25.26 & 2.28 &22.73 &
\cite{Feretti2001}& a relic is also present \\
A2218   & 0.1756 & 2.94 &  4.7  & 23.60 & 0.38 & 5.77 &
\cite{Giovannini2000} &  \\
A2219   & 0.2256 & 3.59 & 81.0  & 25.08 & 1.72 &12.19 & \cite{Bacchi2003}
&   \\
A2254   & 0.1780 & 2.98 & 33.7  & 24.47 & 0.89 & 4.55 & \cite{Govoni2001c}
& \\
A2255   & 0.0806 & 1.50 & 56.0  & 23.94 & 0.90 & 2.64 & \cite{Govoni2005}
& a relic is also present \\
A2256   & 0.0581 & 1.11 &103.4  & 23.91 & 0.81 & 3.75 & \cite{Clarke2006}
& a relic is also present \\
A2294   & 0.1780 & 2.98 &  5.8  & 23.71 & 0.54 & 3.90 &
\cite{Giovannini2009} &   \\
A2319   & 0.0557 & 1.07 &153.0  & 24.04 & 1.02 & 8.46 & \cite{Feretti1997}
& \\
A2744   & 0.3080 & 4.50 & 57.1  & 25.24 & 1.89 &12.86 & \cite{Govoni2001c}
& a relic is also present \\
A3562   & 0.0490 & 0.95 & 20.0  & 23.04 & 0.28 & 1.57 &
\cite{Venturi2003}&  \\
RX J0107.7+5408&0.1066&1.93&55  & 24.26 & 1.10 & 5.42 &
\cite{Vanweeren2011a} &  \\
CL0016+16 & 0.5456 & 6.37 &  5.5 & 24.81 & 0.96 &19.6  &
\cite{Giovannini2000} & MACSJ0018.5+1626 \\
CL0217+70 & 0.0655 & 1.24 & 58.6  & 23.74 & 0.74 & 0.63 &
\cite{Brown2011b} & double relics also present \\
1E0657-56&0.2960 & 4.38 & 78.0  & 25.33 & 2.1  &22.59 & \cite{Liang2000} &
bullet cluster \\
MACSJ0717.5+3745&0.5458&6.37&118.0 & 26.20 & 1.5&24.46 &
\cite{Bonafede2009b}  \\
RXCJ1314.4-2515&0.2439&3.81&20.3 & 24.55 &1.60  &10.75 &
\cite{Feretti2005} & double relics are present\\
RXCJ1514.9-1523&0.22&3.52& 10.0 & 24.23 &1.5   & 7.2  &
\cite{Giacintucci2011b} & 1.4 GHz flux density from NVSS\\
RXCJ2003.5-2323&0.3171&4.59&35.0 &25.09  &1.40  & 9.12 &
\cite{Giacintucci2009b} &  \\
CIZAJ2242.8+5301&0.1921&3.17&--   &  --   & 3.1  & 6.8  &
\cite{Vanweeren2010} & double relics also present\\
\hline
\multicolumn{9}{l}{\scriptsize Col. 1: Cluster Name; Col. 2: Redshift;
Col. 3:
Angular to linear conversion; Col. 4: Radio flux density at 1.4 GHz;}\\
\multicolumn{9}{l}{\scriptsize Col. 5: Logarithm of radio power at 1.4 GHz;
Col. 6: Radio largest linear size;
Col. 7: X-ray luminosity in the 0.1-2.4 keV band in 10$^{44}$ units; }\\
\multicolumn{9}{l}{\scriptsize Col. 8: Reference to the radio flux
density;  Col. 9: Notes}\\

\end{tabular}
\end{table*}

Tab. \ref{tab-halo} reports all clusters known so far to host a radio
halo as of September, 2011.  This sample will be
henceforth referred to as September2011-Halo collection.  All halo
clusters are undergoing merger processes, as demonstrated by the
dynamical activity detected from X-ray and optical data (see
Sect. \ref{sec:432}).

The occurrence of halos in the NVSS sample \cite{Giovannini1999}
increases with the X-ray luminosity of the host clusters, reaching
about 25\% in clusters with L$_x$ $>$ 5 $\times$ 10$^{44}$ erg/s
\cite{Giovannini2002}.  This trend is confirmed by the recent work of
Cassano et al.  \cite{Cassano2011}, who, with improved statistics,
determined that radio halos are present in $\sim$ 30\% of clusters
with L$_x$ $>$ 5 $\times$ 10$^{44}$ erg/s.

\begin{figure*}
  \includegraphics[width=0.7\textwidth]{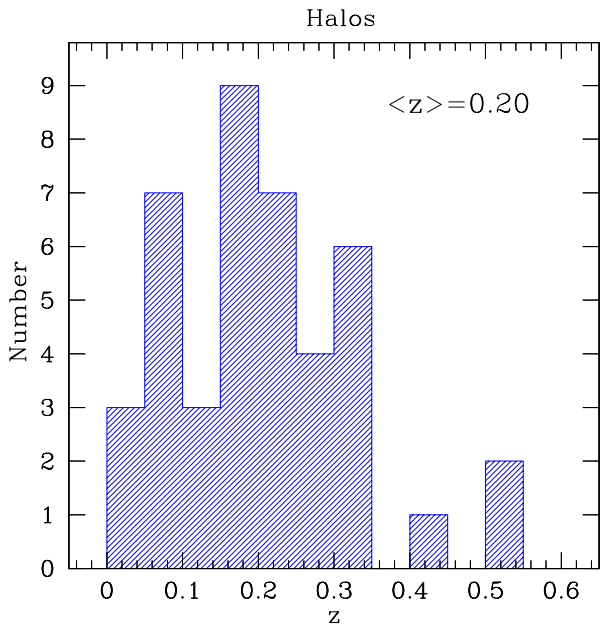}
\caption{Distribution of known clusters with radio halos as a function of
redshift.
}
\label{fig:haz}
\end{figure*}

Fig. \ref{fig:haz} reports the redshift distribution of clusters with
a radio halo.  This distribution is rather homogeneous up to z =
0.35. The scarce number of halos beyond this value is likely due to
sensitivity limitations, and/or selection effects \cite{Giovannini2009}, 
although a real lack of high
redshift radio halos cannot be excluded. This might possibly be
related with the cluster evolution, or with strong Inverse Compton
losses with the cosmic microwave background radiation.  There are
currently three high redshift (z $>$ 0.4) clusters known to host radio
halos: A851 at z = 0.4069 \cite{Giovannini2009}, CL\,0016+16
\cite{Giovannini2000}, and MACS\,J0717.5+3745
\cite{Bonafede2009b,Vanweeren2009b}, both at z = 0.55.  The latter
shows also significant polarization, as also found in A\,2255
\cite{Govoni2005,Pizzo2011}, unlike most radio halos which are
unpolarized.

The physical parameters in radio halos can be estimated assuming
equipartition conditions, and further assuming equal energy in
relativistic protons and electrons, a volume filling factor of 1, a
low frequency cut-off of 10 MHz, and a high frequency cut-off of 10
GHz.  The derived minimum energy densities are of the order of
10$^{-14}$ - 10$^{-13}$ erg cm$^{-3}$, i.e.  much lower than the
energy density in the thermal gas.  The corresponding equipartition
magnetic field strengths range from 0.1 to 1 $\mu$G.

\subsection{Spectra of radio halos}
\label{sec:41}

\subsubsection{Integrated spectrum}
\label{sec:411}

The integrated radio spectra of halo sources are still poorly known.  The
difficulty of spectral studies are that i) only in a few cases the
spectrum is obtained with more than three flux density measurements at
different frequencies, ii) for most sources the highest available
frequency is 1.4 GHz, therefore it is difficult to determine the
presence of a spectral steepening, crucial to discriminate between
different reacceleration models.

Spectral data available on the objects of the September2011-Halo
collection are reported in Tab. \ref{tab-spec}, where we arrange the
information according to the number of available frequency
measurements.  Halos always show a steep spectrum
($\alpha$\footnote{S($\nu$) $\propto$ $\nu^{-\alpha}$ is assumed
throughout the paper} $ \gtrsim$ 1).  
The best studied integrated spectrum is that
of the Coma cluster where clear evidence of a high frequency
steepening is present (see \cite{Thierbach2003} for a detailed
discussion).  Indications of high frequency spectral steepening are
reported in 4 more cases. A very steep spectrum is shown
by the radio halo in A1914 \cite{Bacchi2003} where 9 different points
show a straight spectrum with $\alpha$ = 1.88, although a possible
high frequency curvature has been suggested \cite{Komissarov1994}.

Halo spectra are typical of aged radio sources (see Sect. \ref{sec:2}).
In general, it
is estimated that the radiative lifetime of relativistic electrons
from synchrotron and inverse Compton energy losses is of the
order of $\sim$ 10$^8$ yr \cite{Sarazin1999}. Since the expected
diffusion velocity of the electron population is of the order of the
Alfv\'en speed ($\sim$ 100 km s$^{-1}$), the radiative electron
lifetime is too short to allow the particle diffusion throughout the
cluster volume. Thus, the radiating electrons cannot have been
produced at some localized points of the cluster. They must undergo
{\it in situ} energization, with an efficiency comparable to
the energy loss processes \cite{Petrosian2001}, or be continuously
injected into the cluster volume \cite{Jaffe1977}.
In Sect. \ref{sec:432}, we will highlight that recent 
cluster mergers are likely to supply energy to the halos, through
the reacceleration of radio emitting electrons.

In the framework of the link between halos and clusters mergers, the
existence of a possible correlation between radio spectral index and
cluster temperature has been investigated in the literature
\cite{Feretti2004a,Giovannini2009}, and a marginal evidence
that clusters at higher temperature tend to host halos with a flatter
spectrum has been found.
By using the most recent results (September2011-Halo collection), and
considering the average spectral index value in the frequency range
0.3 - 1.4 GHz, it is obtained that:

\begin{itemize}
\item
radio halos in clusters with an average temperature less than 8 keV show
an average spectral index of 1.7$\pm$0.2;
\item
radio halos in clusters
with a temperature in the range from 8 to 10 keV show an average spectral
index of 1.4$\pm0.4$;
\item
halos in clusters with a temperature greater than 10 keV show an average
spectral index of 1.2$\pm$0.2.
\end{itemize}

This trend indicates that hotter clusters are likely to host halos with
flatter spectra, i.e. halos where energy gains are more significant.  
Unfortunately, the statistics is still poor, because
of the low number of data and highly inhomogeneous spectral index
measurements.  This result, if confirmed, is a very important clue in
favor of reacceleration models: indeed hottest clusters are more
massive and undergoing more violent mergers, thus supplying more
energy to the radio emitting particles
\cite{Feretti2004b,Feretti2004c,Cassano2005,Cassano2008}.

\begin{table*}
\caption{Spectral index of radio halos. The cluster data are grouped
according to how many flux density measurements  at different
frequencies are available for the spectrum determination: the
first group reports clusters with more than 3 frequency measurements, 
the second group reports clusters with 3 frequency measurements, 
the third group reports  clusters with 2 frequency measurements.}
\label{tab-spec}
\centering
\begin{tabular}{lcccc}
\hline\hline
Name     & $\alpha$1     & $\alpha$2 & Temperature        & Ref \\
         &               &           &  keV        &              \\
\hline\hline
A1656    & $\alpha_{0.31}^{1.4}$ = 1.16 &$\alpha_{1.4}^{4.8}$ = 2.28&
8.4&\cite{Thierbach2003} \\
A1914    & $\alpha_{0.26}^{1.4}$ = 1.88 &                   &
7.9&\cite{Bacchi2003}    \\
A2256    & $\alpha_{0.22}^{1.4}$ = 1.61 &                  &
6.6&\cite{Brentjens2008}\\
A3562    & $\alpha_{0.3}^{0.8}$  = 1.3 &$\alpha_{0.8}^{1.4}$ = 2.1  &
5.2&\cite{Giacintucci2005} \\
1E0657-56  & $\alpha_{0.84}^{5.9}$ = 1.3  &                
&10.6&\cite{Liang2000}       \\
\hline
\hline
A521     & $\alpha_{0.24}^{1.4}$ = 1.8 &                    &
5.9&\cite{Giovannini2009} and ref therein \\
A754     & $\alpha_{0.07}^{0.3}$ = 1.1 &$\alpha_{0.3}^{1.4}$ = 1.5  &
9.5&\cite{Bacchi2003} \\
A2163    & $\alpha_{0.3}^{1.3}$  = 1.1 &$\alpha_{1.3}^{1.6}$ = 1.5 
&13.3&\cite{Feretti2001} \\
A2319    & $\alpha_{0.4}^{0.6}$  = 0.9 &$\alpha_{0.6}^{1.4}$ = 2.2  &
8.8&\cite{Feretti1997} \\
RXCJ2003.5-2323& $\alpha_{0.2}^{1.4}$ = 1.3 &               & - 
&\cite{Venturi2009}       \\
\hline
\hline
A545     & $\alpha_{1.3}^{1.6}$ $>$ 1.4 &                           &
5.5&\cite{Bacchi2003}       \\
A665     & $\alpha_{0.3}^{1.4}$ = 1.0 &                             &
9.0&\cite{Giovannini2000}       \\
A697     & $\alpha_{0.3}^{1.4}$ = 1.5  &                          
&10.2&\cite{Giovannini2009,Vanweeren2011a,Macario2011a}\\
A1300    & $\alpha_{0.3}^{1.4}$ = 1.3 &                             &
9.2&\cite{Giacintucci2011a}       \\
A1758    & $\alpha_{0.3}^{1.4}$ = 1.5 &                             &
7.1&\cite{Giacintucci2011a}       \\
A2218    & $\alpha_{1.4}^{5.0}$ = 1.6 &                             &
7.1&\cite{Giovannini2000}      \\
A2219    & $\alpha_{0.3}^{1.4}$ = 0.9  &                           
&12.4&\cite{Orru2007}       \\
A2255    & $\alpha_{0.3}^{1.4}$ = 1.7 &                             &
6.9&\cite{Feretti1997b}      \\
A2744    & $\alpha_{0.3}^{1.4}$ = 1.0 &                            
&10.1&\cite{Orru2007}      \\
MACSJ0717.5+3745 & $\alpha_{1.4}^{4.7}$ = 1.3 &                      &11
&\cite{Bonafede2009b}, and ref therein   \\
CL0217+70  & $\alpha_{0.3}^{1.4}$ = 1.2 &                              &--
&\cite{Brown2011b}  \\
RXCJ1514.9-1523 & $\alpha_{0.3}^{1.4}$ = 1.6 &                       &--
&\cite{Giacintucci2011b}  \\
\hline
\multicolumn{5}{l}{\scriptsize Col. 1: Cluster name; Col.2:
Total spectral index value for straight spectra, or low
frequency spectral index value for curved spectra;}\\
\multicolumn{5}{l}{\scriptsize Col. 3: High frequency spectral index value 
for curved spectra; Col. 4: Average cluster
temperature in keV from literature data; }\\
\multicolumn{5}{l}{\scriptsize Col. 5: References to spectral index
information.}\\
\end{tabular}
\end{table*}

\subsubsection{Spectral index distribution}
\label{sec:412}

A spectral index image is a powerful tool to understand the physical
properties of radio halos, since it reflects the variations of the
shape of the electron energy distribution and of the magnetic field in
which they emit.

The first spectral index map of a radio halo was obtained for the
Coma cluster \cite{Giovannini1993}, where the high sensitivity of 
images allowed the computation of spectral index
up to $\sim$ 30' from the cluster
center.  The spectral distribution is smooth with a steepening
from the center to the peripheral regions, where the spectral index is
$\alpha$ $\sim$ 2. 
The spectral index images of giant radio halos in A665 and A2163, with
an angular resolution $\sim$ 1', show flattening and patches, and
radial steepening in the undisturbed cluster regions \cite{Feretti2004b}.

\begin{figure*}
  \includegraphics[width=0.5\textwidth]{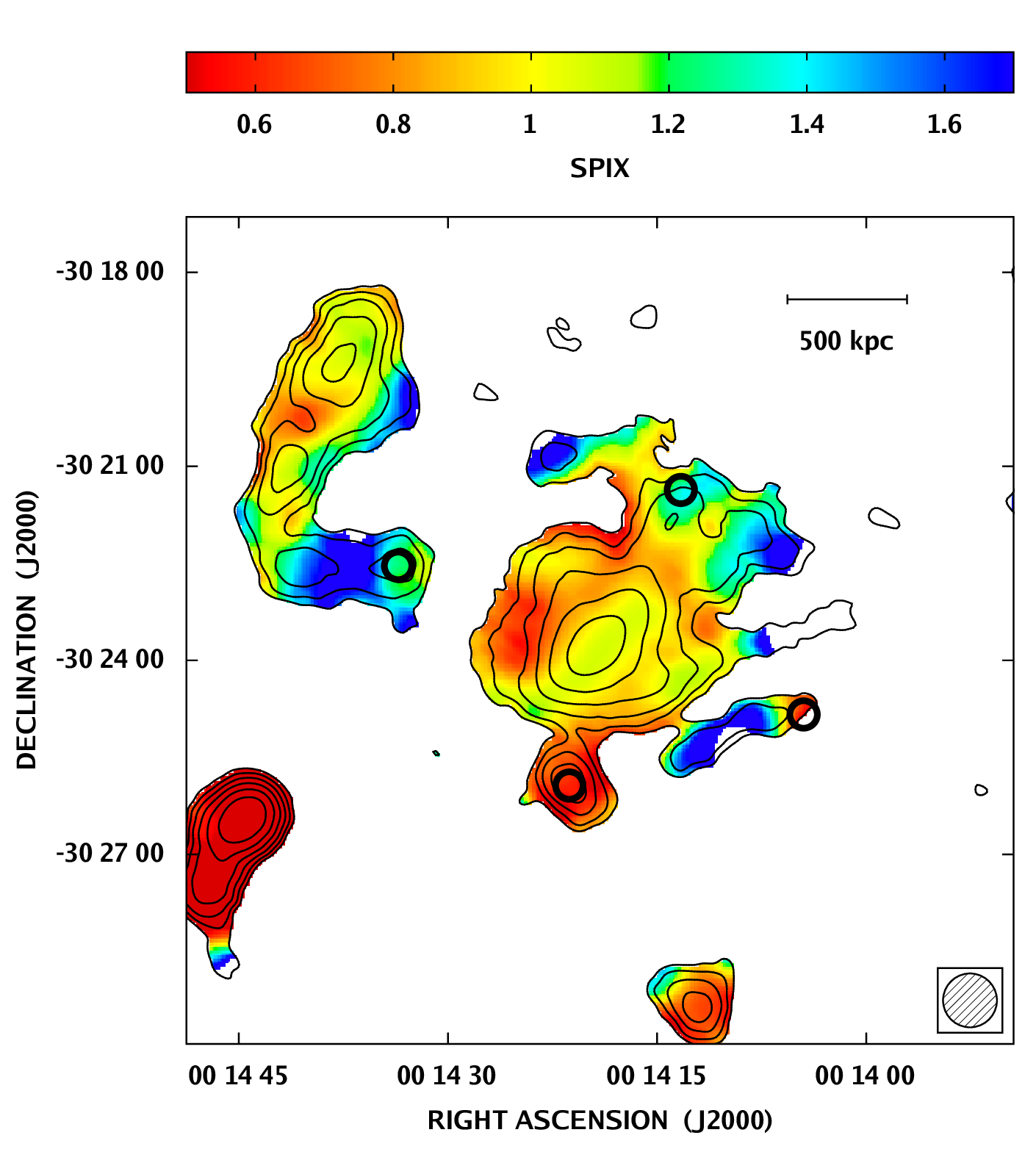}
  \includegraphics[width=0.5\textwidth]{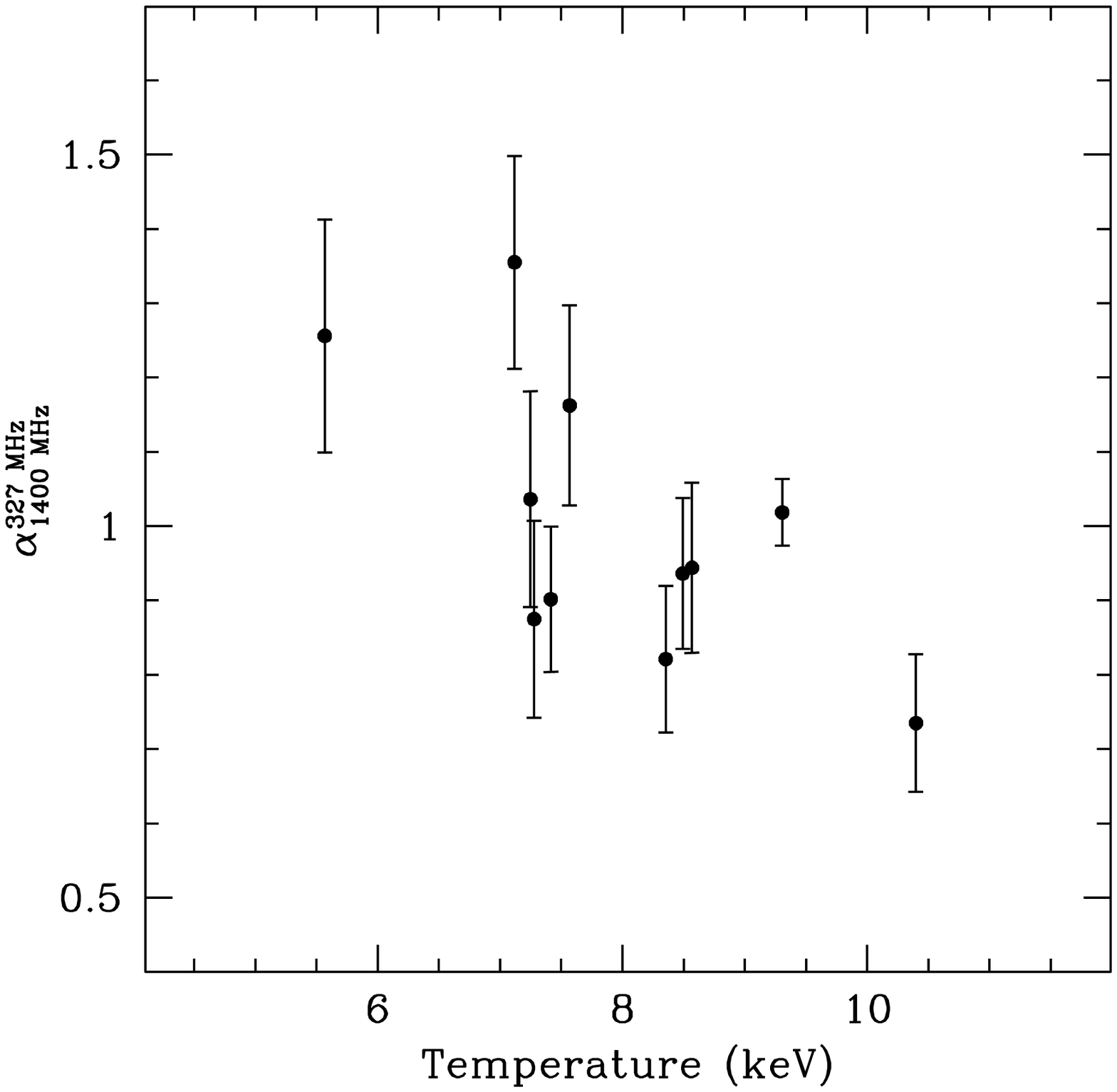}
\caption{{\bf Left panel}:
Spectral index image of A2744 between 325\,MHz and 1.4\,GHz,
with a resolution of 50\arcsec$\times$50\arcsec.
Circles indicate the positions of discrete sources. The radio halo is
at the center, while the elongated feature to the NE is the cluster relic.
{\bf Right panel}:
Plot of the radio  halo spectral index between 325\,MHz and 1.4\,GHz
versus gas temperature, obtained by the {\it Chandra} temperature image
from Kempner \& David \cite{Kempner2004a}.
The values of the spectral index and of
temperature have been derived in a grid of boxes of about 1\arcmin in size.
Both panels are from \cite{Orru2007}.}
\label{fig:alfa2744}
\end{figure*}

In the bullet cluster 1E0657-56, no evidence of spectral steepening 
in the outer regions of the radio halo is found
\cite{Liang2000}.  The spectral index distribution in the A3562 radio
halo shows a very complex structure, with an average value of $\alpha$
$\sim$ 1.5 and a number of knots steepening up to $\alpha$ $\sim$ 2
\cite{Giacintucci2005}.  The spectral index images of A2744 and A2219
\cite{Orru2007} show the presence of localized regions in which the
radio spectrum is significantly different from the average.  From the
comparison of the spectral index map of A2744 (Fig.
\ref{fig:alfa2744}, left panel) and the X-ray data from Chandra, it is
found for the first time that the flat spectrum regions of the radio
halo tend to have higher temperature (Fig.  \ref{fig:alfa2744}, right
panel).

In the radio halo of A2255, the spectrum is steeper at the
center and flatter at the locations of the radio filaments
\cite{Pizzo2009}, indicating either that we are looking at a
superposition of different structures (filaments in the foreground
plus real halo in the background) seen in projection across the
cluster center, or that the halo is intrinsically peculiar.
The complex cluster A2256  shows flat spectral
index in the region NW of the cluster center, and steep spectral index
regions toward the SE of the cluster center \cite{Kale2010}.
The authors interpret this
spectral behavior as resulting from two populations of relativistic
electrons created at two epochs (two mergers).

Spectral variations are likely to reflect the energy losses/gains of
the radiating electrons.  It can be shown \cite{Feretti2004b} that, in
regions of identical volume and identical radio brightness, a
flattening of the spectral index from 1.3 to 0.8 requires 
an amount of energy injected into the electron population larger
by a factor of $\sim$ 2.5. Alternatively, if electrons are no longer
gaining energy, a flatter spectrum would indicate that the last
reacceleration is more recent.  Therefore, it can be argued that the
radio spectrum is flatter in the regions influenced by merger
processes. The data available so far are in  support of electron reacceleration
models, with energy supplied by cluster merger \cite{Cassano2010a}.
In particular the spectral radial steepening is interpreted as
due to the combined effect of a radial decrease of the cluster
magnetic field strength and of the presence of a high energy break in
the energy distribution of the reaccelerated electron population
\cite{Brunetti2001}.

\begin{figure*}
  \includegraphics[width=0.7\textwidth]{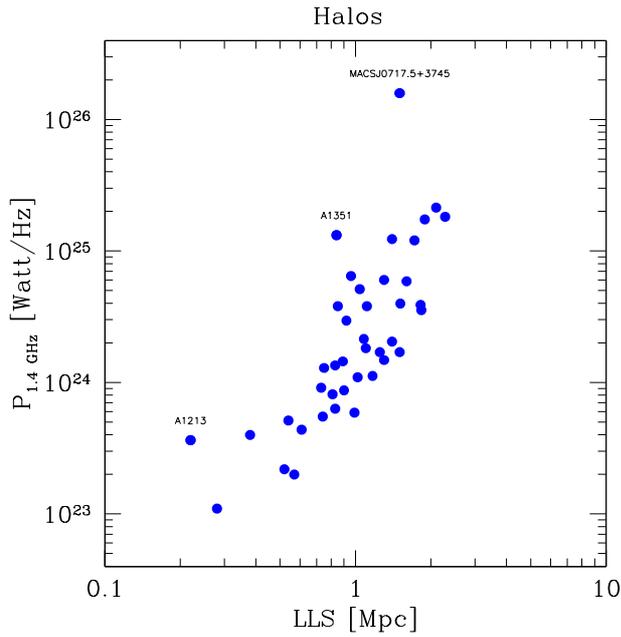}
\caption{Monochromatic radio power of halos at 1.4 GHz versus their
largest linear size (LLS) measured at the same frequency. Names of peculiar
objects are indicated (see text).
}
\label{fig:hasize}
\end{figure*}

\subsubsection{Ultra-steep spectrum halos}
\label{sec:413}

In recent years thanks to the improvement in low frequency radio
observations, a large attention has been devoted to the search for and
discussion of ultra-steep spectrum radio halos (see e.g.
\cite{Venturi2011a}).

From the synchrotron emission theory, it is derived that strong energy
losses in the radiating electrons produce a cutoff at high frequency,
while the radio spectrum at low frequency reflects the original energy
distribution of electrons.  The existence of radio halos showing very
steep spectra also in the low frequency range would indicate objects
characterized by dramatic energy losses.  Cassano \cite{Cassano2010a}
argues that radio halos with spectral index $\alpha$ $\sim$ 1.7, and
steeper, are likely missed by present observations at GHz frequencies.  This
radio halo population can only be revealed with highly sensitive
observations at very low radio frequencies.

Tab. \ref{tab-spec} shows that only a few
ultra-steep spectrum radio halos are currently known: A1914
\cite{Bacchi2003}, A521 \cite{Dallacasa2009,Giovannini2009}, and A2255
\cite{Feretti1997b}.  The cluster A697 was previously classified among
ultra-steep spectrum halos \cite{Macario2010}, however according to
recent flux measurements  \cite{Vanweeren2011a}
the spectral index is around 1.5. These objects may be the tip
of the iceberg of a population which should be detected by future
instruments at low frequencies (LOFAR, LWA, SKA).  We note that
the few ultra-steep radio halos are in relatively low temperature
clusters, in agreement with the correlation between radio halo
spectral index and cluster temperature (see Sect. \ref{sec:411}).

\begin{figure*}
  \includegraphics[width=0.7\textwidth]{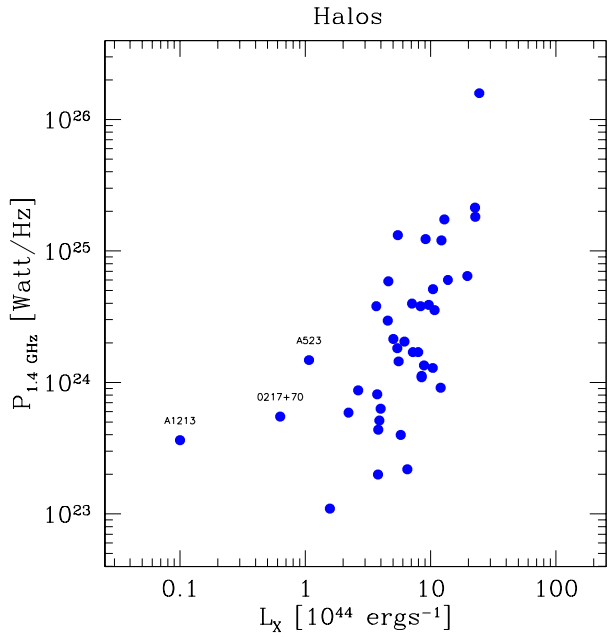}
\caption{Monochromatic radio power of radio halos at 1.4 GHz versus
the cluster X-ray luminosity between 0.1 and 2.4 keV, obtained for the
merging clusters with radio halos. Names of peculiar objects are 
indicated (see text).
}
\label{fig:haradiox}
\end{figure*}

\subsection{Halo size}
\label{sec:42}

Classical halos, i.e. the first which have been detected, are giant
radio sources of about 1 Mpc, or more, in size.  In recent years,
however, several small size halos have been revealed down to few
hundred kpc, owing to the improvement in sensitivity and angular
resolution of instruments.  Giovannini et al. \cite{Giovannini2009}
reported the correlation between the radio halo linear size and the
total radio power, and discussed that this correlation is continuous
from small halos up to giant Mpc-scale radio halos.  The correlation
obtained for the September2011-Halo collection is presented in
Fig. \ref{fig:hasize}.

The slope of the correlation is in agreement with results obtained
using only giant radio halos by Cassano et al. \cite{Cassano2007}.
Murgia et al. \cite{Murgia2009} fitting the azimuthally averaged
brightness profile with an exponential, found that radio halo
emissivity is remarkably similar from one halo to the other, despite
quite different length-scales, in agreement with the correlation
in which larger halos show a higher radio power (see the discussion in
Sect.  \ref{sec:43}). Possible evolutionary effects with redshift
cannot be analyzed with the current data, because of poor statistics.

There are 3 outliers with respect to the general correlation: one is
MACS J0717.5+3745 \cite{Bonafede2009b}, which is the most powerful
radio halo, but not the most extended; the second is the halo in
A1351, which shows a size smaller than expected from its radio power.
It is an irregular radio halo \cite{Giovannini2009,Giacintucci2009a}
and could be related to a recent merger.  The third one is the so far
smallest radio halo, located in the poor Abell cluster A1213 (see
also Sect. \ref{sec:431}). As discussed in \cite{Giovannini2009}, more data
are necessary to better understand its properties.

Except for the few cases outlined above, the correlation of radio size
versus radio power shows a small dispersion, confirming that giant and
smaller radio halos belong to the same class of sources: halos as
small as a few hundred kpc show the same properties as Mpc size giant
halos.  This is in support of a common origin and physical mechanism
for giant and small-size radio halos.

\subsection{Radio - X-ray connection}
\label{sec:43}

The radio properties of halos are linked to the cluster properties.
The probability of detecting radio halos is highest in the clusters with the
highest X-ray luminosity (see Sect. \ref{sec:4}). Moreover, the
monochromatic radio power of a halo at 1.4 GHz correlates with the
cluster X-ray luminosity, and mass and temperature.  In
Fig. \ref{fig:haradiox} we present the correlation between the halo
radio power at 1.4 GHz and the X ray luminosity between 0.1 - 2.4
keV. We note that the correlation is obtained for merging clusters
with radio halos and cannot be generalized to all clusters (see
Sect. \ref{sec:432}).

The most powerful radio halo is in the cluster MACS J0717.5+3745.  It
is remarkable that its radio power is about an order of magnitude
higher than that of the second most powerful known halo.  In the low
X-ray luminosity regime, there are a few clusters well above the
extrapolation of the correlation, which will be discussed in the next
sub-section. If we exclude the three outliers named in
Fig. \ref{fig:haradiox} we find P$_{1.4} \propto L_x^2$.  On the other
hand, ultra-steep spectrum radio halos, discussed in
Sect. \ref{sec:413}, are expected to be less luminous than predicted
by the radio - X-ray correlation, according to Cassano
\cite{Cassano2010a}.

An extrapolation of the above correlation to low radio and X-ray
luminosities indicates that clusters with L$_{X}$ $\lesssim$ 10$^{44}$
erg s$^{-1}$ should host halos of power $<$ 10$^{23}$ W Hz$^{-1}$.
With a typical size of 1 Mpc, they would have a radio surface
brightness lower than current limits obtained in the literature.  It
should be stressed, however, that the less powerful radio halos are
also smaller in size (Fig. \ref{fig:hasize}), implying that actually
it may not be appropriate to assume a typical size $\gtrsim$ 1 Mpc for
low power radio halos.

Brunetti et al. \cite{Brunetti2007,Brunetti2009} report the same
correlation using all clusters searched for radio halos with the GMRT
and derived upper limits to the radio power of undetected halos, by
adopting a halo size of 1 Mpc. They found that upper limits are all in
the region of high X-ray luminosity, and are all about one order of
magnitude below the correlation, indicating a difference
between clusters with and without halos. Their conclusion is that the
distribution is likely related to the dynamical state of the cluster
(see Sect. \ref{sec:432}):
radio halos are associated with dynamically disturbed clusters,
while upper limits refer to  relaxed clusters. This is in
agreement with the result that halos are only present in merging
clusters. In Fig. \ref{fig:haradiox} only the clusters with halos
are reported, thus no upper limits are shown.

Since cluster X-ray luminosity and mass are correlated
\cite{Reiprich2002}, the correlation between radio power ($P_{\rm
  1.4~GHz}$) and X-ray luminosity could reflect a dependence of the
radio power on the cluster mass $M$. A correlation of the type $P_{\rm
  1.4~GHz}$ $\propto$ $M^{2.3}$ has been derived
\cite{Govoni2001c,Feretti2003}, where $M$ is the total gravitational
mass within a radius of 3($H_0$/50)$^{-1}$ Mpc.  Using the cluster
mass within the viral radius, the correlation is steeper
\cite{Cassano2006,Cassano2007}.

Another link between radio and X-ray properties is suggested by the
close similarity of the radio and X-ray structure found in a number of
well resolved clusters.  The similarity was quantitatively confirmed
(see e.g. \cite{Govoni2001b,Feretti2001,Giacintucci2005}), by
comparing the point-to-point surface brightness of the radio and X-ray
emission.  A nearly linear relationship was found in three clusters
(A2255, A2744 and A3562), while a power-law relation with index $<$1
is present in Coma, A2319 and A2163.  As discussed in
\cite{Govoni2001b}, the correlation suggests a direct connection
between the thermal X-ray plasma and the non-thermal radio plasma; its
slope is that expected by electron reacceleration models.

\begin{figure*}
  \includegraphics[width=1\textwidth, bb=50 150 580
340,clip]{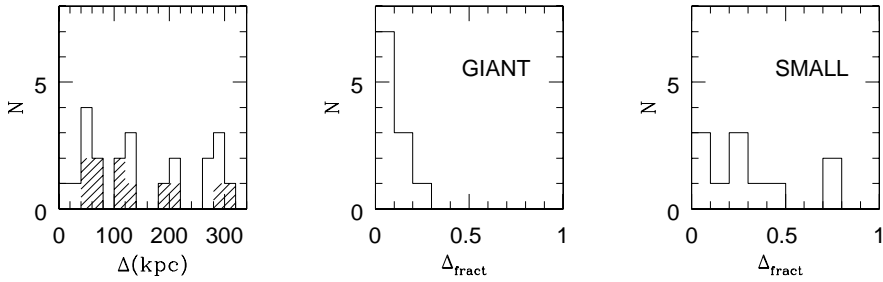}
\caption{{\bf Left panel:} Values of the offset $\Delta$ between the
radio and X-ray centroids in kpc. The dashed area refers to giant halos.
{\bf Middle} and {\bf Right panels:}
Fractional offset ($\Delta$/radio halo size)
for giant halos (size $\ge$ 1 Mpc)
and small halos (size $<$ 1 Mpc), respectively.
}
\label{fig:deltarx}
\end{figure*}

The radio - X-ray brightness relation discussed above is valid for
giant and regular halos. Actually, we have noticed that, owing to
recent deep observations, more irregular and asymmetric halos have
been found. In these halos, the radio emission may show significant
displacement from the X-ray emission. This has been investigated by
comparing the center of the radio halo with that of the X-ray gas
distribution in a sample of clusters with good radio and X-ray data
\cite{Feretti2010}. The left panel of Fig. \ref{fig:deltarx} shows
that both giant and small radio halos can be significantly shifted, up
to hundreds kpc, with respect to the centroid of the host cluster.  To
highlight radio halos with the most pronounced asymmetric
distribution, the ratio between the radio-X-ray offset and the halo
size is plotted in the middle and right panels of
Fig. \ref{fig:deltarx}. It is deduced that halos can be quite
asymmetric with respect to the X-ray gas distribution, and this
becomes more relevant when halos of smaller size are considered.  A
possibility is that the asymmetry in the structure is caused by
magnetic field fluctuations as large as hundreds of kpc, as suggested
by \cite{Vacca2010} on the basis of magnetic field modeling.

\subsubsection{Halos in poor clusters}
\label{sec:431}

Giovannini et al. \cite{Giovannini2009} were the first to discuss
the possible presence of a radio halo in A1213, an under-luminous X-ray
cluster i.e. well outside the correlation between radio power and
X-ray luminosity.  A similar case is found in the cluster CL0217+70 by
Brown et al. \cite{Brown2011b}.  Clear evidence of a radio halo more
luminous than predicted by the radio - X-ray correlation is detected
in A523 \cite{Giovannini2011}.

The number of these cases is still low, and some observational
uncertainties may be present (see \cite{Giovannini2011} for a more
detailed discussion), however there is observational evidence that
in a few cases giant radio halos can be associated with relatively low
luminosity X-ray clusters, i.e. low mass clusters and low density
environments.  These halos are over-luminous in radio by at least an
order of magnitude with respect to what is expected from the
extrapolation of the observed radio power - X-ray luminosity.  These
objects are likely to represent a new class of halos that are difficult
to explain by classical radio halo models.  We note that
ultra-steep spectrum halos, presented in Sect. \ref{sec:413} and
predicted by reacceleration models, show the opposite behaviour,
i.e. they should be less luminous in radio than predicted by the radio
- X-ray correlation.  Giovannini et al \cite{Giovannini2011} suggest
that these powerful radio halos associated with low density
environments could be either young halos or related to clusters at a
special time of the merger event, when particle acceleration processes
have a higher efficiency (see e.g.  \cite{Brunetti2011b}). Another
possibility is that the X-ray luminosity might not be a
good indicator of the previous history of cluster merging activity in
these cases.

\subsubsection{Halos and cluster mergers}
\label{sec:432}

Circumstantial evidence of the link between cluster merging and radio
halos has been outlined in early works on this topic (see e.g.
\cite{Feretti1999a,Feretti2000,Giovannini2002,Schuecker2001,Bohringer2002}),
on the basis of the presence of radio halos only in clusters showing
X-ray and substructure, X-ray temperature gradients and weak (or
absent) cooling flow.  
This has been strongly supported by optical data
(see e.g. \cite{Girardi2002,Ferrari2003,Boschin2004,Boschin2006}).
The first quantitative analysis of the cluster
dynamical state was performed by Buote \cite{Buote2001}, who obtained
the cluster dipole ratio from X-ray images and found that the
strongest radio halos appear only in those clusters experiencing the
largest departures from a virialized state.  Cassano et
al. \cite{Cassano2010b} recently confirmed the connection between
radio halos and cluster mergers, from a quantitative study on a larger
sample, using Chandra X-ray data and adopting three methods to
characterize the cluster state, i.e. the power ratio, the centroid
shift and the X-ray brightness concentration parameter.

The link between radio halos and cluster merger is well supported by
several arguments, and recent findings obtained with improved
statistics support the early suggestion that there is a clear
separation between relaxed and non relaxed clusters, in the radio
behaviour.  The bi-modality between halo and non halo clusters,
discussed by \cite{Brunetti2009}, reflects the dichotomy existing in
the evolutionary state of clusters, between non relaxed and relaxed
clusters, as also discussed by \cite{Rossetti2011}.

What is intriguing is the existence of dynamically disturbed clusters
which do not show any evidence of radio halo.  Cassano et
al. \cite{Cassano2010b} find 4 such clusters: A141, A781, A2631 and
MACS J2228.5+2036.  A 781 has been found to host a radio halo
\cite{Govoni2011}, but there are still 3 outliers (out of them one
cluster is at z $>$ 0.32).  Some other well known merging clusters do
not host a radio halo, e.g. A119 \cite{Giovannini2000} and A2146
\cite{Russel2011}.

The existence of merging clusters with halos, and merging clusters
without halo, indicates that there is a dichotomy in merging
clusters, which is not currently understood.  It may be possible that
giant halos are only present above a threshold of mass or temperature.
Future radio data with next generation instruments (LOFAR, LWA, SKA)
will allow the detection of low brightness/low power large halos, in
order to clarify if halos are present in all merging clusters or only
in the most massive ones.  The dichotomy has to be solved to get a
full and comprehensive understanding of these phenomena.

\section{Relics}
\label{sec:5}

Relic sources are diffuse extended sources similar to radio halos in
their low surface brightness, large size ($\gtrsim$ 1 Mpc) and steep
spectrum ($\alpha$ $\gtrsim$ 1), but, unlike halos, they are located
in cluster peripheral regions and are strongly polarized ($\sim$
20-30\%).

Observations of relics provide the best indications for the presence
of $\mu$G level magnetic fields and relativistic particles in
cluster outskirts. They provide evidence for the
acceleration of relativistic particles at shock fronts at large
distance (Mpc scale) from the cluster centers (see
e.g. \cite{Bruggen2011}).

\begin{figure*}
  \includegraphics[width=0.7\textwidth]{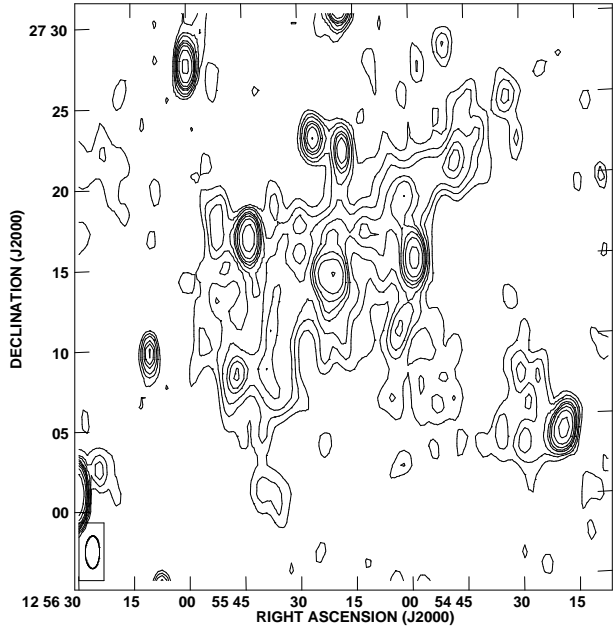}
\caption{
Image at 327 MHz, obtained with the WSRT, of the relic 1253+275
in the Coma Cluster. The beam is 50\arcsec $\times$ 120\arcsec
(RA $\times$ DEC). Contour levels are 3, 5, 7, 10, 12, 15, 20, 30, 50
mJy/beam.
The cluster center is at about 1.2 Mpc distance in the NE direction.
See \cite{Giovannini1991} for more details.
}
\label{fig:comarelic}
\end{figure*}

The prototype source of this class is the relic 1253+275 detected in
the Coma cluster (see \cite{Giovannini1991} and references therein),
shown in Fig. \ref{fig:comarelic}.  The Coma relic and most relics
show an elongated radio structure, with the major axis roughly
perpendicular to the direction of the cluster center.  Recently, new
spectacular relics have been detected and studied, increasing our
knowledge of these diffuse sources.  We quote as one of the most
interesting cases, the giant narrow relic in CIZA J2242.8+5301
(Sect. \ref{sec:511}). In addition, sources with a more roundish
structure, as well as sources of small size, have been detected and
classified as cluster relics.  Different relic morphologies and
related properties will be discussed in the next sub-section.

Relics are found in both merging and cool-core clusters,
suggesting that they may be related to minor or off-axis mergers, as
well as to major mergers.  Theoretical models propose that they are
tracers of shock waves in merger events.  This is consistent with
their elongated structure, almost perpendicular to the merger axis,
and is confirmed by observational results (see
e.g. \cite{Solovyeva2008,Finoguenov2010}, and the review by
Br{\"u}ggen et al. \cite{Bruggen2011}).

\begin{table*}
\caption{September2011-Relic collection (published relics, Sept. 2011)}
\label{table-rel}
\centering
\begin{tabular}{lclrrrrrcl}
\hline\hline
Name    & z  & kpc/$''$ & S(1.4)& Log P(1.4)  & Rcc  & LLS & Lx(10$^{44}$)&
Ref & Notes \\
        &    &          & mJy   & W/Hz       & Mpc  & Mpc & erg/sec      &
     &       \\
\hline
A13     & 0.0943 & 1.67 & 31.0 & 23.80 & 0.20 & 0.25 & 1.18
&\cite{Slee2001}        & 2 \\
A85     & 0.0551 & 1.01 & 43.0 & 23.44 & 0.43 & 0.35 & 4.10
&\cite{Slee2001}        & 2 \\
A115    & 0.1971 & 3.13 &147.0 & 25.18 & 1.32 & 2.44 & 8.91
&\cite{Govoni2001c}     & 1; average Rcc \\
A521    & 0.2533 & 3.81 & 15.0  & 24.41 & 0.93 & 1.00 & 8.47
&\cite{Giacintucci2008} & 1 + halo\\
A548b-NW& 0.0424 & 0.81 & 61.0 & 23.38 & 0.43 & 0.26 & 0.15
&\cite{Feretti2006a}     & 2 \\
A548b-N &        &      & 60.0 & 23.37 & 0.50 & 0.31 & 0.15
&\cite{Feretti2006a}     & 2 \\
A610    & 0.0954 & 1.71 & 18.6 & 23.60 & 0.31 & 0.33 & --  
&\cite{Giovannini2000}  & 2 \\
A746    & 0.232  & 3.67 & 24.5 & 24.83 & 1.61 & 1.10 & 3.68
&\cite{Vanweeren2011a}  & 1 + halo \\
AS753   & 0.014  & 0.29 &460.0 & 23.32 & 0.41 & 0.35 & --  
&\cite{Subrahmanyan2003}& 2 \\
A754    & 0.0542 & 1.03 & 69.0 & 23.67 & 0.50 & 0.80 & 2.21
&\cite{Bacchi2003}      & 1; complex + halo  \\
A781    & 0.3004 & 4.42 & 15.5 & 24.65 & 0.66 & 0.44 & 4.6 
&\cite{Govoni2011}      & 1 + halo \\
A1240-N & 0.1590 & 2.65 &  6.0 & 23.59 & 0.70 & 0.65 & 1.00
&\cite{Bonafede2009a}   & 1  \\
A1240-S &        &      & 10.1 & 23.81 & 1.10 & 1.25 & 1.00
&\cite{Bonafede2009a}   & 1 \\
A1300   & 0.3072 & 4.39 & 20.0 & 24.76 & 0.79 & 0.70 &13.73
&\cite{Reid1999}        & 1 + halo \\
A1367   & 0.0220 & 0.45 & 35.0 & 22.59 & 0.62 & 0.22 & 0.64
&\cite{Gavazzi1983}     & 2 \\
A1612   & 0.179  & 2.99 & 62.8 & 24.90 & 0.89 & 0.78 & 2.41
&\cite{Vanweeren2011a}  & 1 \\
A1656   & 0.0231 & 0.47 &260.0 & 23.49 & 2.20 & 0.85 & 3.99
&\cite{Giovannini1991}  & 1 + halo\\
A1664   & 0.1283 & 2.22 &107.0 & 24.64 & 1.03 & 1.07 & 3.09
&\cite{Govoni2001c}     & 2 \\
A2034   & 0.1130 & 2.03 & 24.0 & 23.95 & 0.73 & 0.22 & 3.81
&\cite{Vanweeren2011a}  & 1 + halo  \\
A2048   & 0.0972 & 1.74 & 19.0 & 23.63 & 0.33 & 0.31 & 1.91
&\cite{Vanweeren2011b}  & 2 \\
A2061   & 0.0784 & 1.46 & 27.6 & 23.65 & 1.56 & 0.68 & 3.95
&\cite{Vanweeren2011a}  & 1 \\
A2063   & 0.0349 & 0.68 & 67.0 & 23.26 & 0.04 & 0.04 & 0.98
&\cite{Komissarov1994}  & 2 (3C318.1) \\
A2163   & 0.2030 & 3.22 & 18.7 & 24.32 & 1.17 & 0.48 &22.73
&\cite{Feretti2001}     & 1 \\
A2255   & 0.0806 & 1.46 & 12.0 & 23.25 & 0.90 & 0.70 & 2.64
&\cite{Feretti1997b}     & 1 + halo\\
A2256   & 0.0581 & 1.08 & 462  & 24.56 & 0.44 & 1.13 & 3.75
&\cite{Clarke2006}      & 2 + halo \\
A2345-W & 0.1765 & 2.86 & 30.0 & 24.38 & 1.00 & 1.15 & 5.90
&\cite{Bonafede2009a}   & 2 \\
A2345-E &        &      & 29.0 & 24.36 & 0.89 & 1.50 & 5.90
&\cite{Bonafede2009a}   & 1 \\
A2443   & 0.1080 & 1.88 &  6.5 & 23.25 & 0.23 & 0.43 & 1.90
&\cite{Cohen2011}       & 2; complex cluster \\
A2744   & 0.3080 & 4.37 & 18.2 & 24.71 & 1.56 & 1.62 &12.86
&\cite{Govoni2001c}     & 1 + halo \\
A3365-E & 0.0926 & 1.66 & 42.6 & 24.04 & 1.08 & 0.56 & 0.86
&\cite{Vanweeren2011a}  & 1 \\
A3365-W &        &      &  5.3 & 23.15 & 0.5  & 0.24 & 0.86
&\cite{Vanweeren2011a}  & 1 \\
A3376-W & 0.0456 & 0.87 &166.0 & 23.88 & 1.43 & 0.80 & 1.08
&\cite{Bagchi2006}      & 1 \\
A3376-E &        &      &136.0 & 23.79 & 0.52 & 0.95 & 1.08
&\cite{Bagchi2006}      & 1 \\
A3667-NW& 0.0556 & 1.03 &2400.0& 25.21 & 2.05 & 1.86 & 4.73
&\cite{Roettgering1997} & 1  \\
A3667-SE&        &      &  --   &  --   & 1.36 & 1.30 & 1.37
&\cite{Roettgering1997} & 1 \\
A4038   & 0.0300 & 0.56 & 49.0 & 22.94 & 0.04 & 0.13 & 1.97
&\cite{Slee2001}  & 2 \\
ZwCl0008.8+5215-W&0.1032&1.82&11.0&23.44&0.85 & 0.29 & 0.50
&\cite{Vanweeren2011c}  & 1  \\
ZwCl0008.8+5215-E&      &    &56.0&24.15&0.85 & 1.40 & 0.50
&\cite{Vanweeren2011c}  & 1  \\
S1081   & 0.2200 & 3.52 &  2.4 & 23.17 & 2.00 & 0.28 &  -- 
&\cite{Mao2010,Middelberg2008}  & 1 \\
CL0217+70-NW&0.0655& 1.26 &--    &  --   & 1.00 & 0.95 & 0.63
&\cite{Brown2011b}      & 1; confused \\
CL0217+70-SE&      &      & --   &  --   & 1.00 & 0.38 & 0.63
&\cite{Brown2011b}      & 1; confused \\
CIZAJ0649.3+1801&0.064&1.22&125&24.08  & 0.80 & 0.8  & 2.38
&\cite{Vanweeren2011a}  & 1 \\
RXCJ1053.7+5452&0.0704&1.33&15.0&23.30 & 0.98 & 0.60 & 3.69
&\cite{Vanweeren2011a}  & 1 \\
RXCJ1314.4-2515-E&0.2474&3.76&10.1&24.27&0.90 & 0.56 &10.75
&\cite{Feretti2005}     & 1 + halo \\
RXCJ1314.4-2515-W&      &    &20.2&24.57&0.55 & 1.01 &     
&\cite{Feretti2005}     & 1 + halo \\
CL1446+26& 0.3700& 5.08 &  9.2 & 24.61 & 0.66 & 0.36 & 3.42
&\cite{Giovannini2009}  & 1 \\
CIZAJ2242.8+5301-N&0.1921&3.07&--&--    & 1.57 & 1.70 & 6.80
&\cite{Vanweeren2011d}  & 1 + halo \\
CIZAJ2242.8+5301-S&      &    &--&--     & 1.06 & 1.45 & 6.80
&\cite{Vanweeren2011d}  & 1 + halo \\
PLCK G287.0+32.9-N&0.39&5.26&33.0&25.24& 1.58 & 1.4  &17.2 
&\cite{Bagchi2011}      & 1 \\
PLCK G287.0+32.9-S&    &    &25.0&25.12& 3.00 & 1.1  &17.2 
&\cite{Bagchi2011}      & 1 \\

\hline
\multicolumn{10}{l}{\scriptsize Col. 1: Cluster Name; Col. 2: Redshift;
Col. 3:
Angular to linear conversion; Col. 4: Radio flux density at 1.4 GHz;
}\\
\multicolumn{10}{l}{\scriptsize  Col. 5: Logarithm of radio power at 1.4 GHz;
Col. 6: Distance from the cluster center;
Col. 7: Radio largest linear size;
 }\\
\multicolumn{10}{l}{\scriptsize Col. 8: Cluster X-ray luminosity in the 
0.1-2.4 keV band in 10$^{44}$ units; 
Col. 9: References to the radio flux density;}\\
\multicolumn{10}{l}{\scriptsize Col. 10: Relic type 
(1 = elongated, 2 = roundish) and notes on the cluster diffuse emission} \\
\end{tabular}
\end{table*}

\begin{figure*}
  \includegraphics[width=0.7\textwidth]{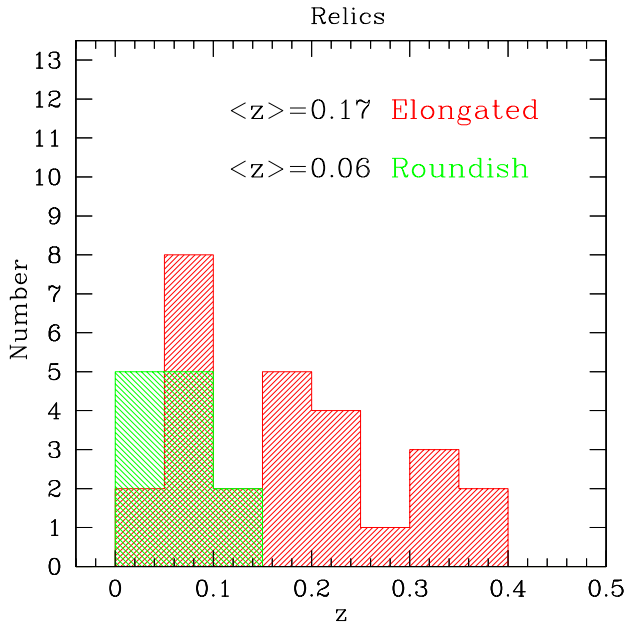}
\caption{
Distribution of known clusters with
radio relics from Tab. \ref{table-rel} as
a function of the cluster redshift.
For the classification of elongated (red) and roundish (green)
sources see Sects. \ref{sec:511} and \ref{sec:512}, respectively.
}
\label{fig:relz}
\end{figure*}

Currently we know 50 relics hosted in 39 clusters of galaxies: they
are reported in Tab. \ref{table-rel}. This collection, which is
updated to September, 2011, will be referred to henceforth as
September2011-Relic collection.  Most clusters are rich Abell
clusters, but relics have also been detected in X-Ray selected
clusters (e.g. RXS J131423.6-251521, \cite{Valtchanov2002}), and in
poor clusters (e.g. S0753, \cite{Subrahmanyan2003}).

The classification of relics may be troublesome in some cases.
This may cause small differences from other lists in
the literature (see e.g. \cite{Nuza2011}).
In particular, A133 is excluded from the September2011-Relic collection
of Tab. \ref{table-rel} after the discussion given in
\cite{Randall2010}, who suggest that the source
formerly classified as a relic is one of the two lobes of a double
radio galaxy, identified with the central cD galaxy.  Moreover, we do
not consider the relics in ZwCl 2341.1+0000, since the diffuse radio
emission in this cluster is likely originating from a large-scale
filament \cite{Giovannini2010} (see Sect. \ref{sec:8} and
Fig. \ref{fig:filament}), rather than from double relics
\cite{Vanweeren2009d}.

We also note that diffuse sources which arise from AGN, in which the
nuclear activity has ceased, may be sometimes classified as cluster
relics.  Actually, these sources are not originating from the ICM, and
should be more properly considered as {\it dying sources} or {\it AGN
  relics} (see \cite{Murgia2011}).  They will not be considered here
because they are not originated by the ICM.

The occurrence of relics in the NVSS sample \cite{Giovannini1999} is
higher in clusters showing high X-ray luminosity, reaching about 25\%
in clusters with L$_x$ $>$ 5 $\times$ 10$^{44}$ erg/s
\cite{Giovannini2002}.  The redshift distribution of clusters of
galaxies, hosting at least one relic, is given in Fig. \ref{fig:relz},
where different colors refer to different structures
(Sect. \ref{sec:51}).  As discussed by \cite{Giovannini2004}, relic
sources at very low redshift could have a large angular extension, and
therefore may be missed in interferometric observations because of the
poor coverage of short spacings.  However, owing to the irregular and
elongated structure of most relics, this observational problem is not
as crucial as in the more regular and diffuse halo sources.  Relics
are detected up to z $\sim$ 0.4, and their distribution with redshift
is similar to that of radio halos.  The low number of high redshift
relics could be due to selection effects, but could also be a real
trend, due e.g. to the evolution of cluster mergers, or to the inverse
Compton losses which become more relevant at increasing redshift.
Deeper radio surveys and cluster catalogues are necessary to
investigate this point.

\subsection{Structure}
\label{sec:51}

After the discovery of the prototypical relic 1253+275 in the Coma
cluster (Fig. \ref{fig:comarelic}), different morphologies have been
found to be associated with relics (see e.g. \cite{Giovannini2004}):
classical sources are elongated and located in the
periphery of galaxy clusters, but  more regular and roundish structures
have also been detected, which are located nearer the cluster
center and/or in peripheral regions.

Kempner et al. \cite{Kempner2004b}, in a classification of cluster radio
sources, suggested that radio relics could be distinguished
in two classes: those associated with a previous AGN activity,
called radio Phoenix, and those related to the ICM, named radio Gischt.
This classification has been used somewhat in the literature.
However, it implies that the origin of a relic source has to be
established, in order to properly classify it.

To be conservative, we prefer to rely on a classification that is
based only on the morphology, which is an observable.  Therefore we
divide relics in {\it elongated} and {\it roundish}, and obtain the
properties of these two classes, to find out whether they may be
indications of different origin and physical parameters.  In all the
following plots, as done in Fig. \ref{fig:relz}, we will use different
symbols for the elongated and roundish relics.

\subsubsection{Elongated Relics}
\label{sec:511}

\begin{figure*}
  \includegraphics[width=0.3\textwidth,angle=90]{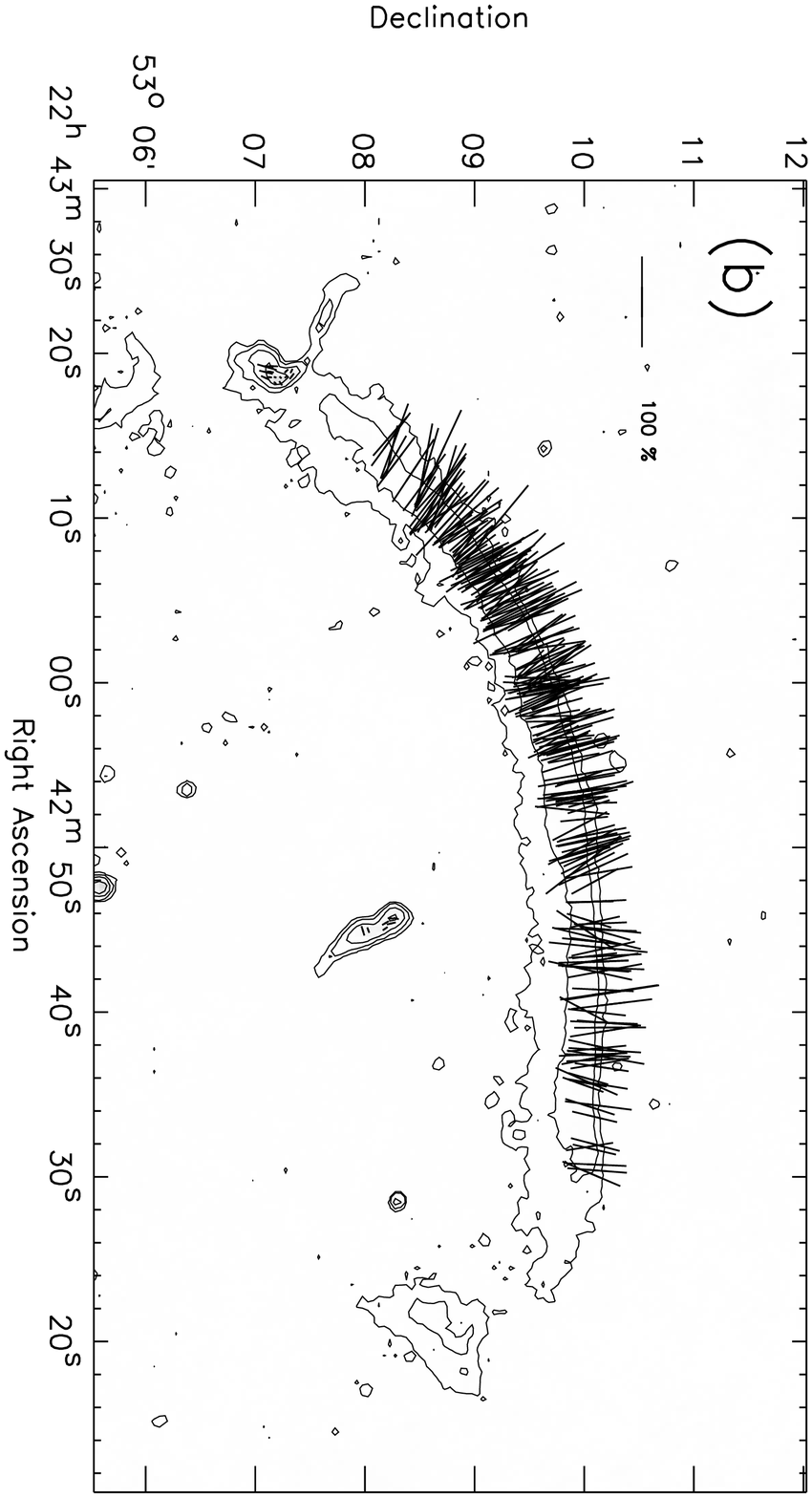}
  \includegraphics[width=0.3\textwidth,angle=90]{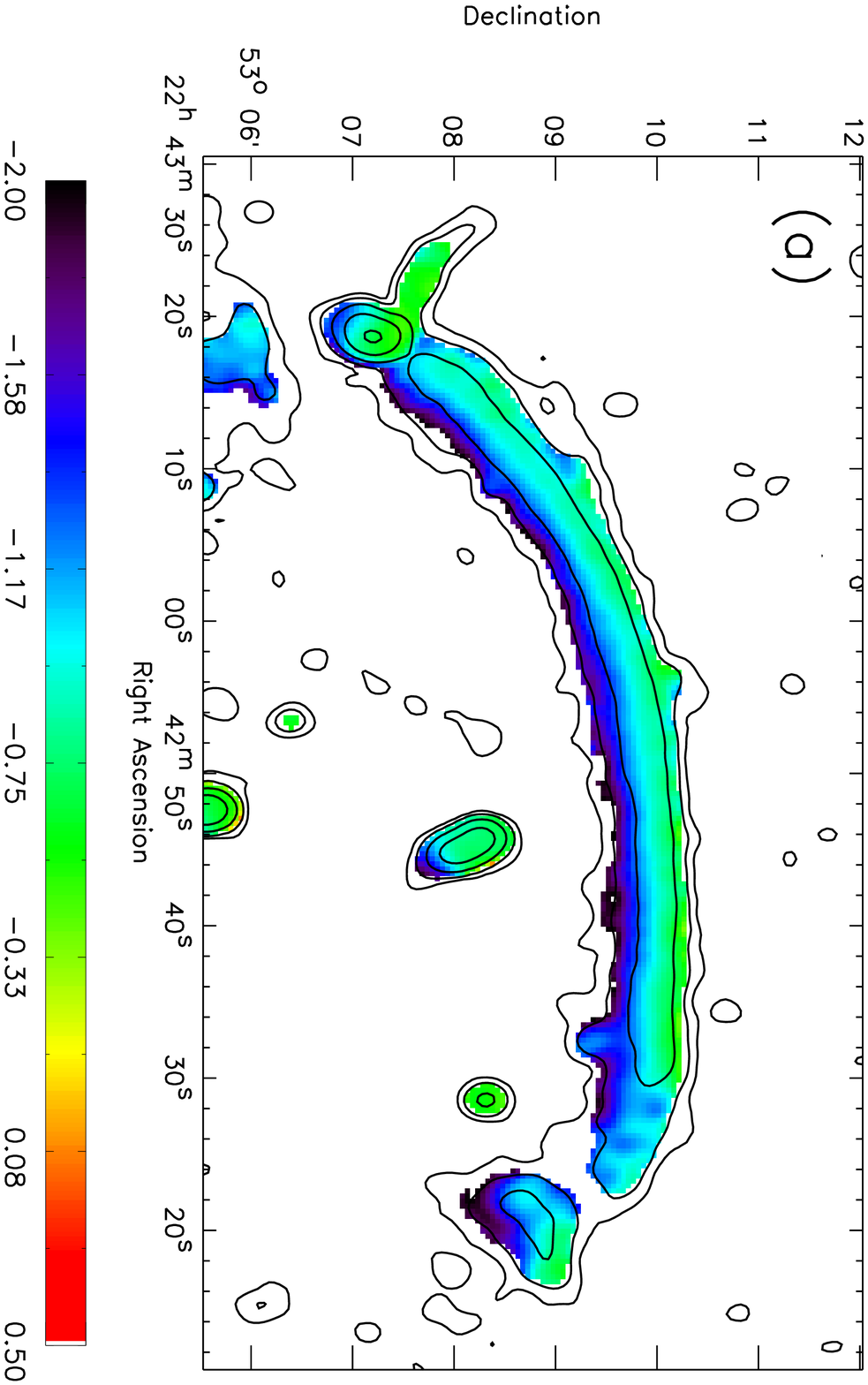}
\caption{
{\bf Left panel:} Contours of the radio emission of the northern relic
in CIZAJ2242.8+5301, obtained with the GMRT at 610~MHz.  Superimposed
lines represent the polarization electric field vectors obtained with
the VLA at 4.9~GHz. The length of the vectors are proportional in length
to the polarization fraction. {\bf Right panel:} Radio spectral index map of
the same relic, obtained with a power-law fit to measurements at 5
frequencies between 2.3 and 0.61 GHz.  Contours of the radio emission are
from WSRT 1.4~GHz. Images are from \cite{Vanweeren2010}, whom we refer for
any detail.
}
\label{fig:sausage}
\end{figure*}

Elongated relics are the classical extended objects, located in
cluster peripheral regions, characterized by an elongated shape
roughly perpendicular to the cluster center direction, as the relic
1253+275 in the Coma cluster (\cite{Giovannini1991,Brown2011a};
Fig. \ref{fig:comarelic}).  They do not show any evident
substructures, and in some case their transverse size is very small.
When observed with high angular resolution, they show an asymmetric
transverse profile, with a sharp edge usually on the side toward the
cluster outer edge and the radio emission is usually highly polarized.
A spectacular example, detected very recently, is represented by the
northern relic in CIZAJ2242.8+5301 \cite{Vanweeren2010}. This giant
relic is very narrow and slightly curved following the cluster
boundary. It  is strongly polarized, with highly regular polarization 
vectors,  and shows steep spectrum (see Fig. \ref{fig:sausage}).  
To the south of the cluster, at about 2.8
Mpc distance, a second (fainter) relic structure is present.

\begin{figure*}
  \includegraphics[width=0.5\textwidth]{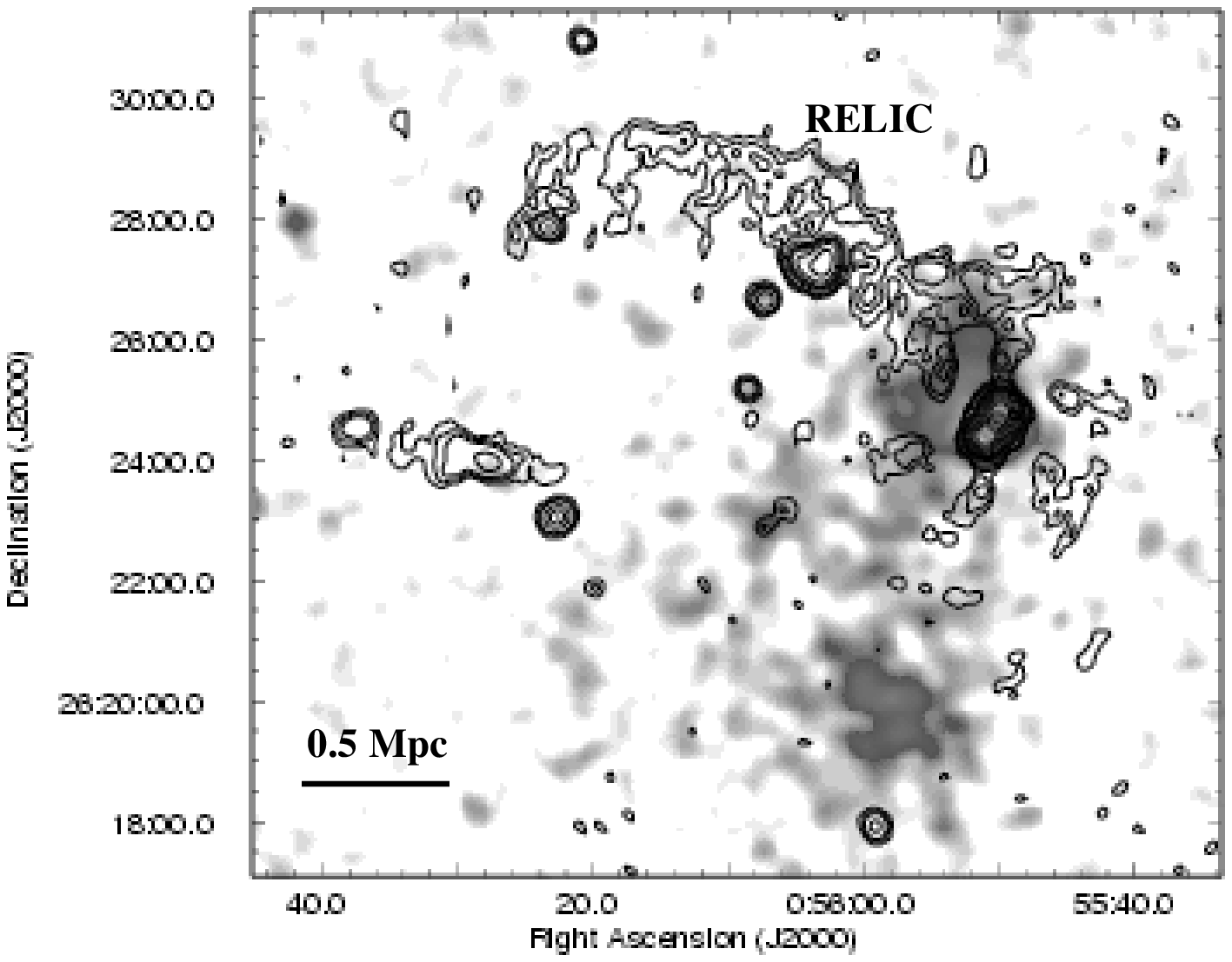}
  \includegraphics[width=0.5\textwidth]{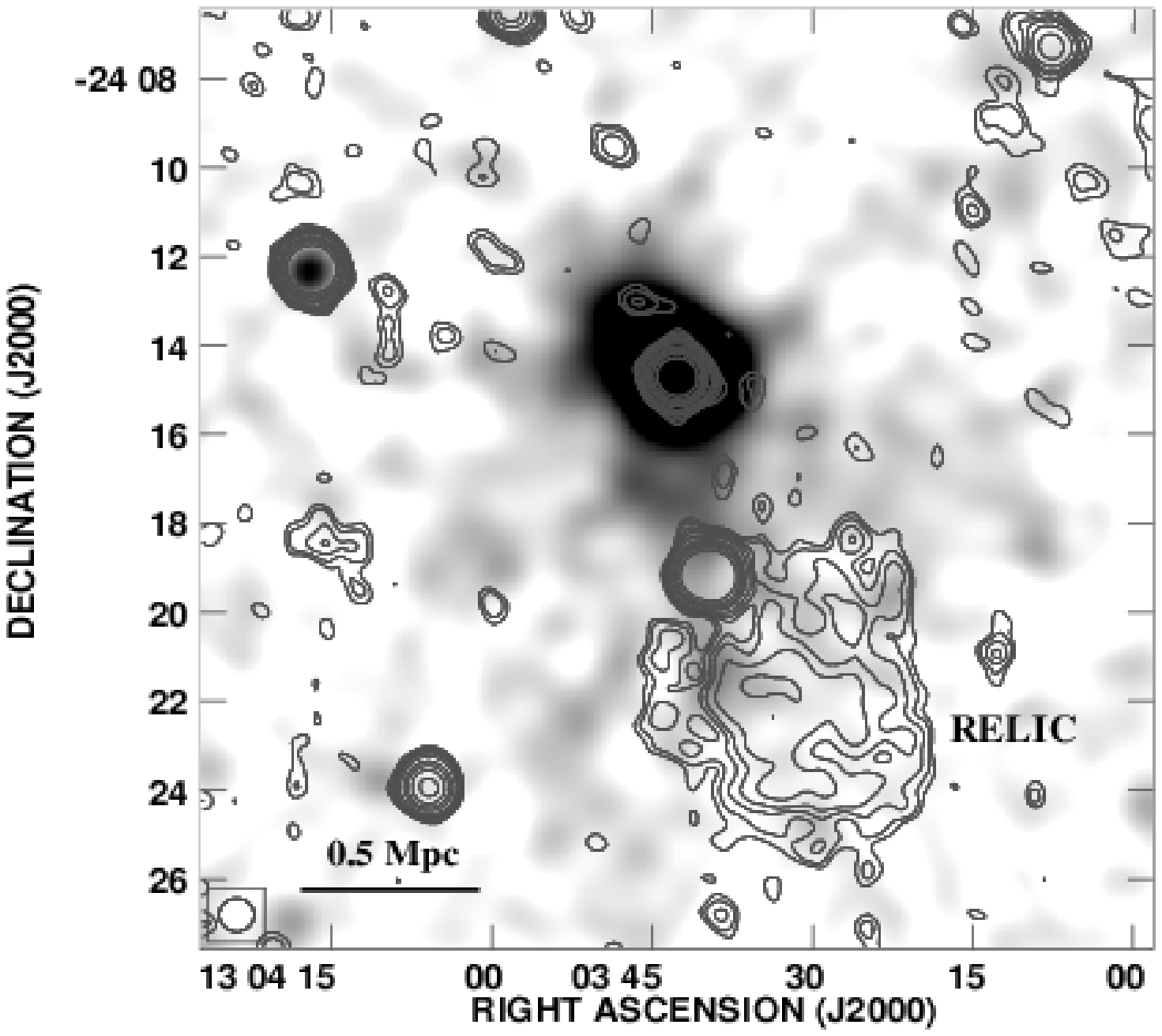}
\caption{Radio relics in the clusters:
{\bf Left panel:} A115 (z = 0.197), extended from the
cluster center toward the cluster periphery, probably because
of projection effects:  {\bf Right panel:} A1664 (z = 0.128),
exhibiting a diffuse roundish structure.
In both panels the radio emission at 20 cm obtained with the VLA
is in contours,
while the superimposed grey-scale refers to the X-ray emission
detected from ROSAT PSPC (see \cite{Govoni2001c}).
}
\label{fig:relstructures}
\end{figure*}

We note that projection effects may be important in these elongated
structures. An extreme, and unique, case in the double cluster A115
\cite{Govoni2001c}, where the relic apparently starts at the center of
the northern cluster and is elongated towards the cluster periphery
(Fig. \ref{fig:relstructures}, left panel).

These morphologies are in very good agreement with models predicting
that these sources are related to large-scale shocks generated during
cluster merger events. Indeed, these shocks expanding with high
velocity (Mach number $\sim$ 1 - 3) can accelerate electrons to high
energies, and compress magnetic fields, giving rise to large regions
emitting synchrotron radiation.  The accelerated particles will have a
power-law energy distribution, and magnetic fields aligned parallel to
the shock front. This is also consistent with the observed spectral
index (see Sect. \ref{sec:53} and Fig. \ref{fig:sausage}, right panel)
and polarization properties, and
explain the shape and location of these sources (see
e.g. \cite{Vanweeren2011a}, and references therein).

\subsubsection{Roundish Relics}
\label{sec:512}

In addition to elongated relics, diffuse extended radio sources with a
more regular and roundish structure have been detected off-center in
clusters: an example is in A1664 \cite{Govoni2001c}, shown in the
right panel of Fig. \ref{fig:relstructures}.  Despite their different
structure, these relics share with the elongated relics the location
at the cluster periphery and the lack of an optical identification.
The main properties of roundish relics are the diffuse morphology
resolved in filamentary sub-structures and a very steep curved
spectrum (see e.g. \cite{Slee2001}).

The detection of roundish relics might lead to the obvious
interpretation that these sources would be elongated relics seen
face-on.  This is however very unlikely, indeed the number of relics
with elongated shape is too high to be consistent with simple
projection effects.  Moreover, we will show in the following and in
the next subsections that the properties of elongated and roundish
relics may be different.

The class of roundish relics defined here includes the sources named
Radio Phoenix by \cite{Kempner2004b}, or relic Sources near the First
Ranked Galaxy (FRG) by \cite{Giovannini2004}. These objects are
located near the central FRG, usually a cD galaxy, but not coincident
with it. Their size is in a large range of values, i.e. from
$\lesssim$ 100 kpc (as in A2063 and A4038) to $\sim$ 350 kpc (as in
A85).  These sources, when observed at higher resolution, show evident
filamentary substructures (e.g. \cite{Govoni2001c}).  Owing to the
proximity of these sources to an AGN, an obvious interpretation is
that these sources consist of old radio lobes originated by previous
AGN activity, which have become no longer visible because of strong
radiation losses, and have been revived by the energy supplied by
shock waves.  A caveat to this interpretation is that these relics are
always located only on one side with respect to the FRG.  We note,
however, the case of A133, which was classified in the past as a relic
(radio Phoenix) but has been recently shown to be an old radio galaxy
\cite{Randall2010}.  Another  possible explanation of the nature of these
sources has been supplied by Mathews \& Brighenti
\cite{Mathews2008}, who, in their study of the connection between
X-ray cavities and radio lobes, suggest that radio bubbles, blown into
the cluster gas by a radio galaxy will rise buoyantly and may
eventually expand to the cluster outskirts, where the cosmic rays
would impact into the surrounding medium, giving rise to small relics
located near the AGN.

Among relics with a roundish shape, there are also some objects
which are located at large distance from the FRG or any bright galaxy,
therefore difficult to reconcile with any model involving a previous
radio activity: we already noted A1664 shown in the right panel of
Fig. \ref{fig:relstructures}; another case is in A548b, where two
roundish relics are present, on the same side of the cluster
\cite{Feretti2006a,Solovyeva2008} (see Fig. \ref{fig:clcollection}).

More data are necessary to understand the origin and nature of these
objects.  Interestingly, we note that roundish relics are peaked at
lower redshift with respect to the total relic redshift distribution
and that no roundish relics at z $>$ 0.2 are known (Fig. \ref{fig:relz}).

\subsubsection{Clusters with double relics}
\label{sec:513}

\begin{figure*}
  \includegraphics[width=0.7\textwidth]{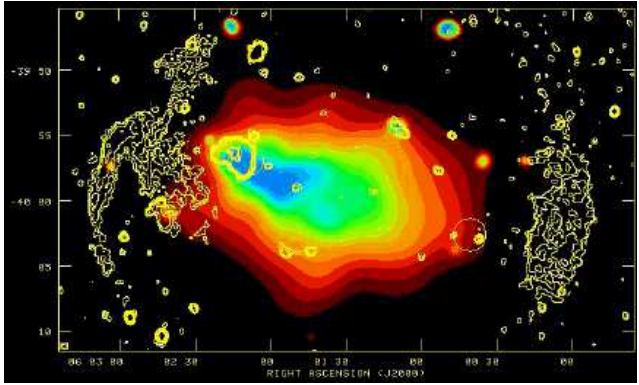}
\caption{Double radio relics in the cluster A3376:
The radio emission is represented by yellow contours (0.12, 0.24,
0.48, 1 mJy/beam) obtained from VLA observations at 1.4 GHz at the
resolution of 20''.  The color image depicts the X-ray emission
detected by ROSAT PSPC within 0.14 to 2.0~keV band (from \cite{Bagchi2006}).
}
\label{fig:a3376}
\end{figure*}

In the framework that relics are witnessing the presence of shock
waves originated by mergers between clusters with approximately equal
masses and low impact parameter, it is expected that relics should
often come in pairs and be located on opposite sides of the cluster
along the axis merger, with the extended radio structures elongated
perpendicularly to this axis (e.g. \cite{Roettiger1999b}).  The first
cluster where two almost symmetric relics have been detected to be
located on opposite sides with respect to the cluster center is A3667
\cite{Roettgering1997,Johnston-Hollitt2002,Johnston-Hollitt2003}.
Several other objects have
been found recently, so these cases are getting more and more common,
see e.g.  A\,3376 (Fig. \ref{fig:a3376}, \cite{Bagchi2006}),
RXCJ1314.4-2515 \cite{Feretti2005}, A\,1240 and A\,2345
\cite{Bonafede2009a}, CIZAJ2242.8+5301 \cite{Vanweeren2010}, ZwCl
0008.8+5215 \cite{Vanweeren2011c}, PLCK G287.0+32.9 \cite{Bagchi2011}
and CL0217+70 (see Fig.\ref{fig:cl0217} and \cite{Brown2011b}).

In the case of CIZAJ2242.8+5301, Van Weeren at
el. \cite{Vanweeren2011d} discuss, from hydrodynamical simulations and
comparison with the data, that the relic morphology arises naturally
from shocks produced by a head-on merger in the plane of the sky of
two roughly equal mass clusters.  In the clusters A1240 and A2345,
optical data support the outgoing merger shocks models
\cite{Barrena2009,Boschin2010}.  In the case of A3376
(Fig. \ref{fig:a3376}), Bagchi and collaborators \cite{Bagchi2006}
suggested the possibility that the two relics may be tracing shocks
induced by the accretion flows of the intergalactic medium during the
large-scale structure formation (accretion shocks
\cite{Miniati2003,Keshet2003}). This interpretation is under debate
for this cluster (e.g. \cite{Akamatsu2011}), however it may be invoked
to explain the origin of other sources as e.g. the outermost relics in A2255
\cite{Pizzo2008} and possibly the peripheral emission of the
Coma cluster \cite{Brown2011a}.

\subsubsection{Clusters with halo and relic(s)}
\label{sec:514}

\begin{figure*}
  \includegraphics[width=0.7\textwidth,angle=270]{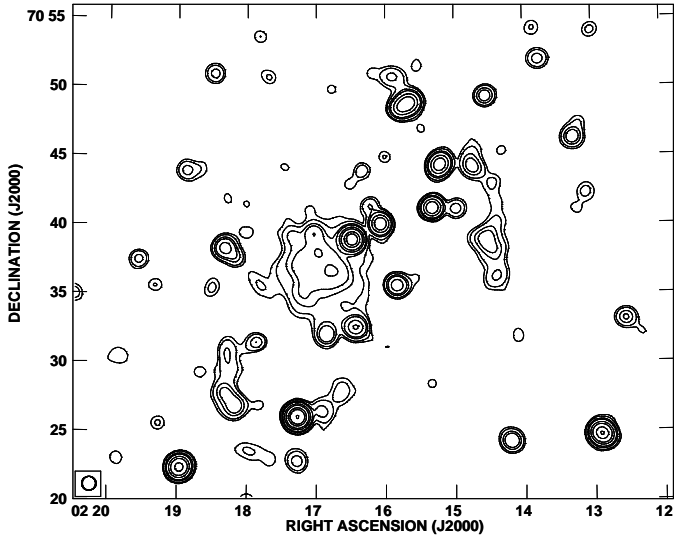}
\caption{Radio halo plus double radio relics in the cluster
CL0217+70. The HPBW is 60'', and levels are: 0.3, 0.5, 1, 1.5, 3,
 5, 10, 15, 30 mJy/beam.
For more details see \cite{Brown2011b}.
}
\label{fig:cl0217}
\end{figure*}

As the Coma cluster, several other clusters have been found to show a
peripheral relic radio emission in addition to a central halo. According
to the September2011-Relic collection, at present 11
clusters are known to show this feature.  Sometimes, the relic is
connected to the radio halo through a low brightness bridge of radio
emission (e.g. Coma cluster, A2255 and A2744), other cases are
characterized by complex structures, as in A754
(Fig. \ref{fig:clcollection}) and A2256.  In a few case, clusters may
host a central radio halo and double elongated relics. At present we
know 3 clusters showing these features: CL0217+70
(Fig. \ref{fig:cl0217}), RXCJ1314.4-2515
(Fig. \ref{fig:clcollection}), and CIZAJ2242.8+5301.

These structures are the most extreme examples of the connection
between halo sources and cluster mergers (see Sect. \ref{sec:432}).
Indeed the merger that gives rise to the radio halo, could also be
the origin of shock waves which supply the energy to the peripheral
relics. In this scenario, the origin of the bridge of radio emission
connecting the halo to the relic source in some clusters has still to
be clarified.

We note that in most clusters showing at least a relic source, no
central halo is present. This could suggest that relics
could be created also by minor mergers, which are not strong enough to
produce a radio halo at the cluster center.  This is in agreement with
the presence of relics in a few cool-core clusters, since a minor and
off-center merger could create a peripheral shock wave, without
destroying the central cool-core. As discussed by Van Weeren et al.
\cite{Vanweeren2011d}, double relics should be located in merging
systems with a small mass ratio, while single relics should be found
in mergers with larger mass ratios.

\subsection{Radio Power, Size and Projected Distance}
\label{sec:52}

To better understand the properties of relics, we derived correlations
between their observational radio properties and their location in the
cluster, using the data of the September2011-relic collection.
Double relics in the same clusters are included separately. In the
plots, we use different symbols for the elongated and roundish relics,
and find that in general elongated and roundish relics are rather
separated, with the exception of a few sources.

\begin{figure*}
  \includegraphics[width=0.5\textwidth]{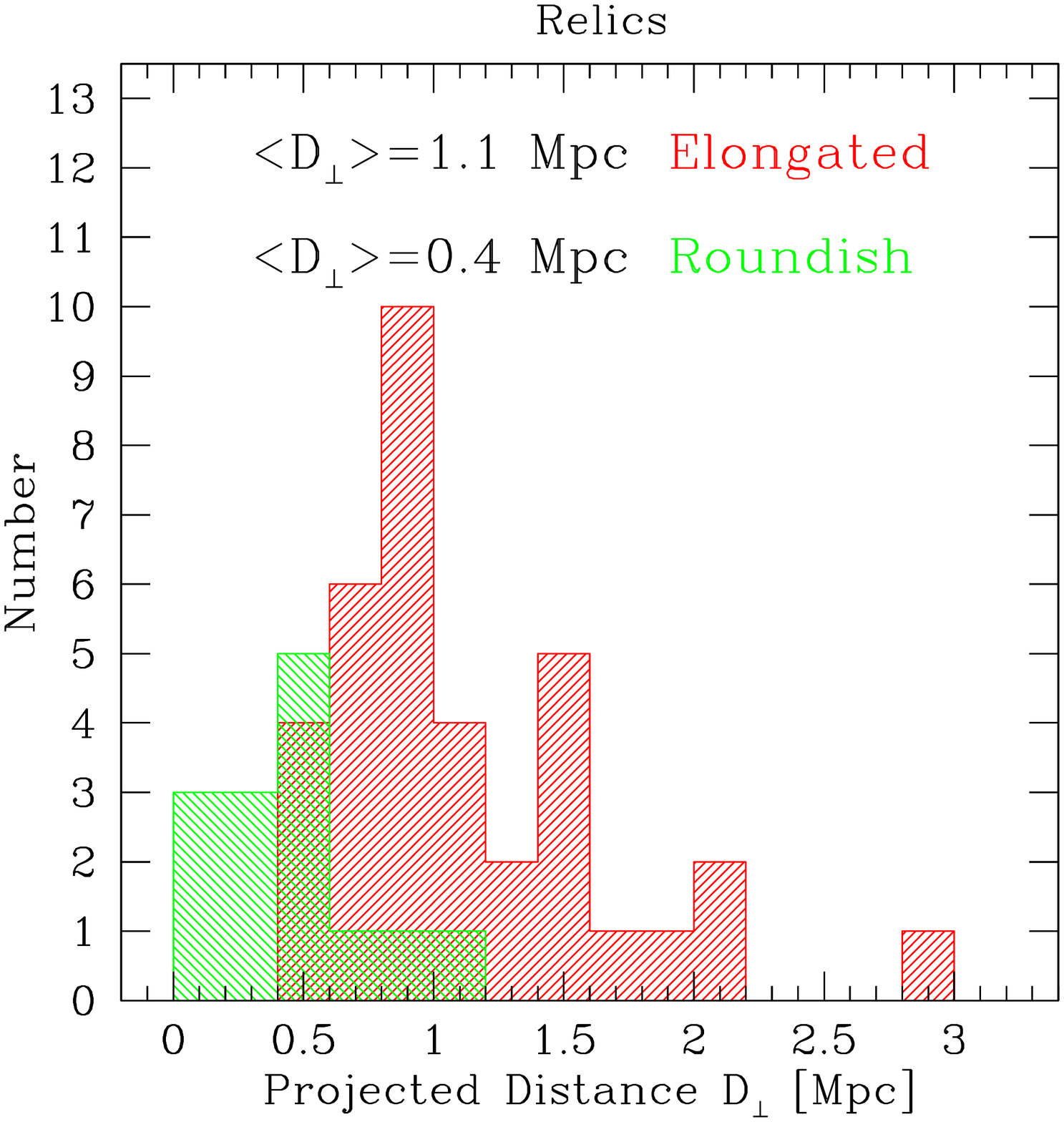}
  \includegraphics[width=0.5\textwidth]{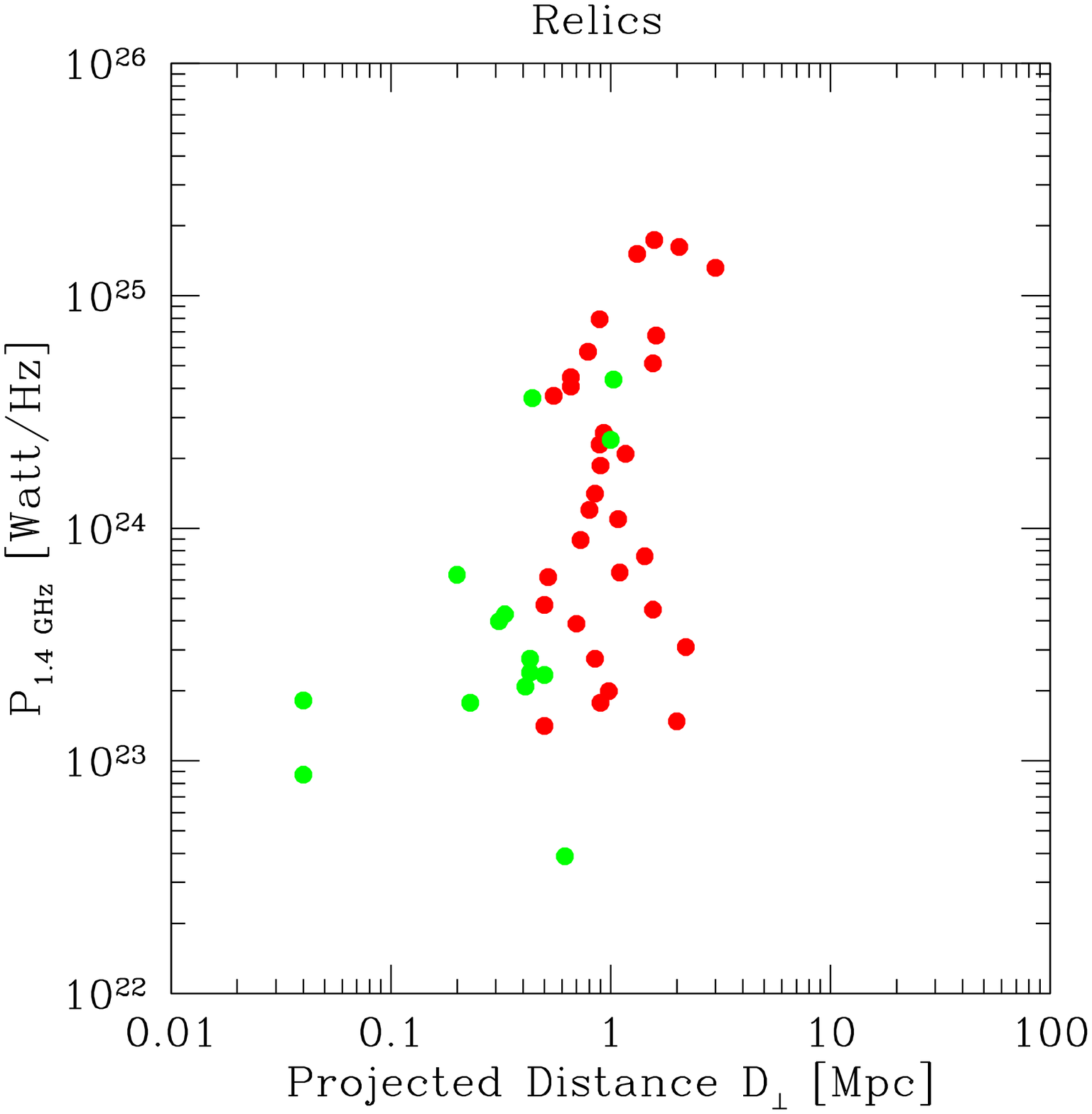}
\caption{
{\bf Left panel:} Distribution of relics according to their
projected distance from the
cluster center. {\bf Right panel:}
Monochromatic radio power of relics at 1.4 GHz versus their projected
distance from the cluster center. In both panels, red color refers
to the elongated relics, green color to the roundish relics.
}
\label{fig:reldist}
\end{figure*}

In the left panel of Fig. \ref{fig:reldist}, the distribution of
projected distance of relics from the cluster center is presented.  We
note that roundish sources are more concentrated toward the cluster
center, as expected since this class includes the relics located near
the FRG (see Sect. \ref{sec:512}). Their distribution is different
from that of elongated relics, with an average projected distance from
the cluster center R$_{cc}$ $\sim$ 0.4 Mpc suggesting that most of
them are really at a small distance from the center.  Elongated relics
are mostly located between 0.5 and 1.5 Mpc from the center, up to a
distance of 3
Mpc.  They avoid cluster centers and show a distribution peaked at the
cluster periphery.  From the figure it can be deduced that projection
effects are unlikely to play a crucial role, i.e.  elongated relics
are actually located in the cluster outskirts. This may be due to the
fact that high Mach number shocks needed to supply energy to the radio
emitting particles are only present in cluster peripheral regions
\cite{Gabici2003}. This point has been recently discussed in detail by
Vazza et al. \cite{Vazza2011}.

The monochromatic radio power of relics as a function of the projected
distance from the cluster center is given in the right panel of
Fig. \ref{fig:reldist}. No clear correlation is present.  Roundish
relics are less powerful than elongated relics, except for a few
cases. It is derived that relics, both elongated and roundish, located
at distances $\gtrsim$ 1 Mpc from the cluster center, show powers in
the range from 10$^{23}$ to 10$^{25}$ W/Hz, whereas there is a lack of
powerful ($>$ 10$^{24}$ W/Hz) relics near the cluster center ($<$ 0.4
Mpc), which could be related to the different relic morphology and to
the low efficiency of shocks in cluster central regions (see
e.g. \cite{Vazza2011}).

\begin{figure*}
  \includegraphics[width=0.5\textwidth]{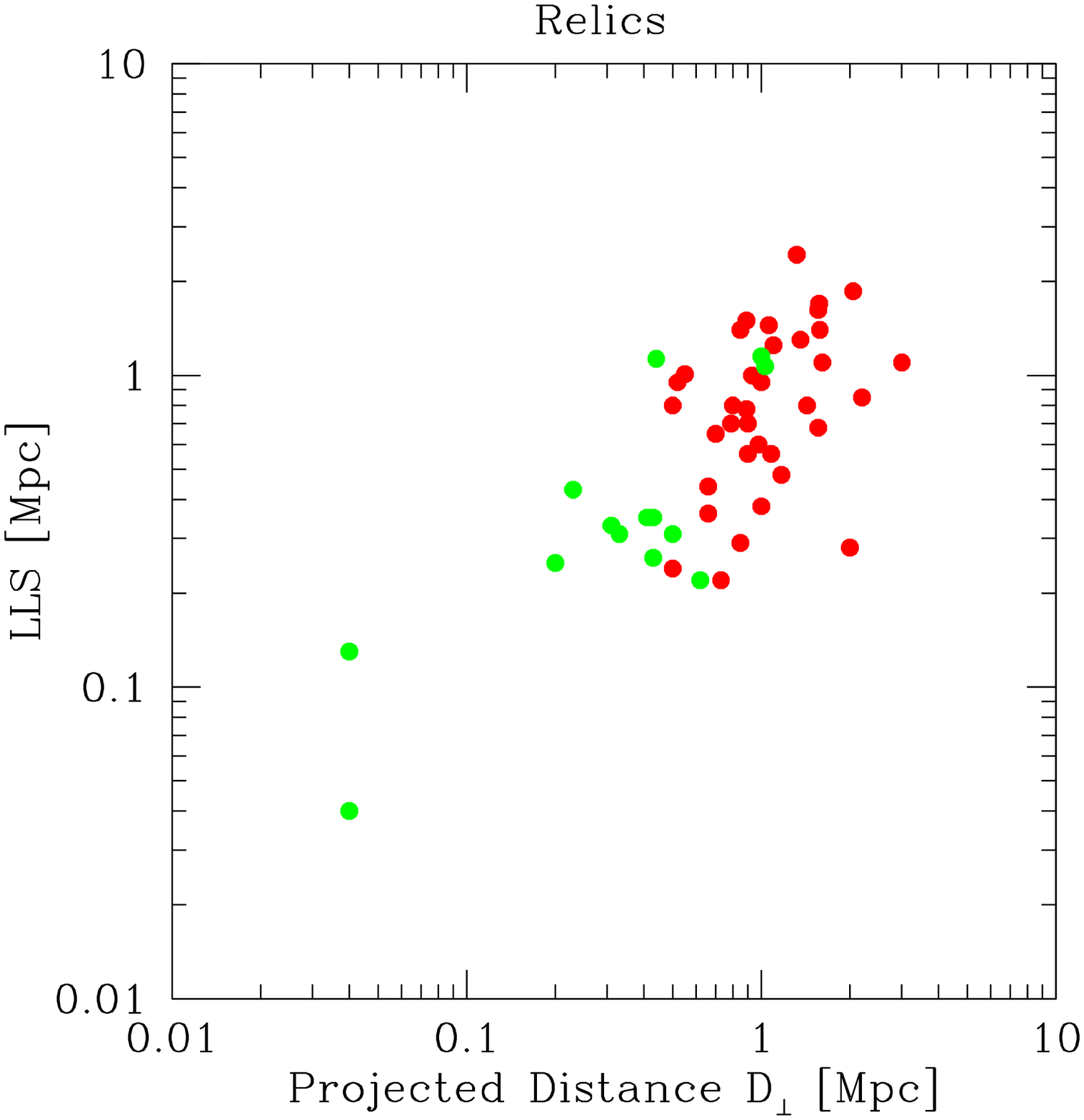}
  \includegraphics[width=0.5\textwidth]{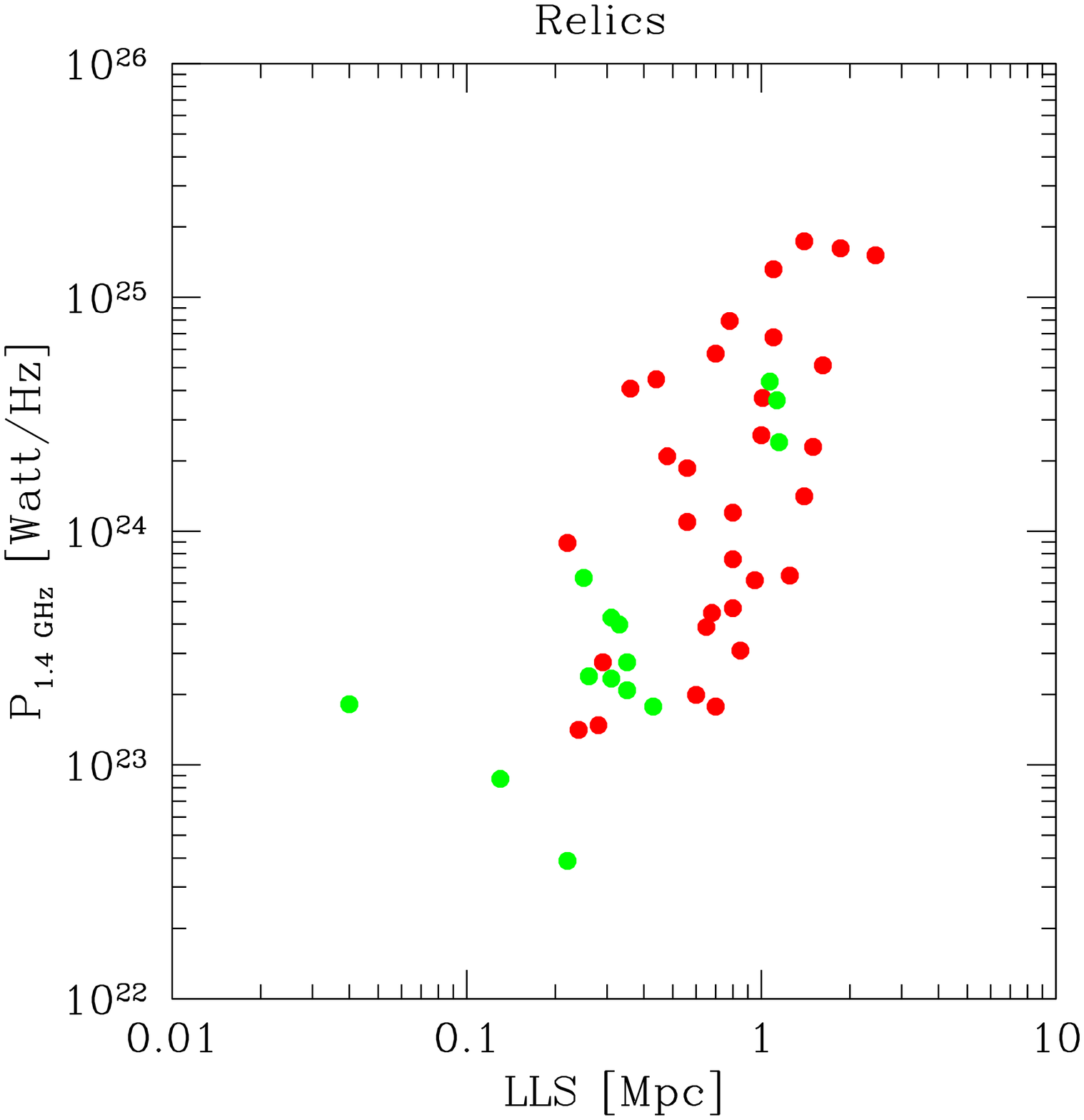}
\caption{
{\bf Left panel:} Relic largest linear size (LLS) versus the
projected distance from the cluster center;
{\bf Right panel}: Monochromatic radio power of relics at 1.4 GHz versus the
relic largest linear size. In both panels, red color refers
to the elongated relics, green color to the roundish relics.
}
\label{fig:relsize}
\end{figure*}

The left panel of Fig. \ref{fig:relsize} gives the largest linear size
of relics as a function of their projected distance from the cluster
center (see also \cite{Vanweeren2009c}).  As in the plot of
monochromatic radio power versus projected distance
(Fig. \ref{fig:reldist}, right panel), elongated and roundish relics
are well separated, except for a few roundish objects, the largest
ones, which are in the same region of the plot as the elongated
relics.

The monochromatic radio power of relics at 1.4 GHz versus the relic
largest linear size (LLS) is presented in the right panel of
Fig. \ref{fig:relsize}. Elongated relics show a similar behaviour to
the roundish relics, although they are somewhat separated on the plot,
being on average more powerful and more extended.  In particular,
relics with low radio powers ($<$ 10$^{24}$ W/Hz) can be either
roundish or elongated, but the latter have larger size. The most
powerful relics are mostly large elongated relics.
By considering elongated and roundish relics together, a
marginal correlation could be present between LLS and projected
distance from the cluster center, in the sense that larger relics are
at larger distance from the cluster center.

The summary of the above plots is as follows:

1) Roundish relics are more often located near the cluster center, while
elongated relics are more distributed in peripheral regions.

2) The properties of roundish and elongated relics are generally rather
different. While roundish relics are on average of low power and
small size, elongated relics cover a large interval of
radio powers and sizes, and are on average more powerful and
more extended. Only a few roundish relics show  sizes and
radio powers similar to those of elongated relics: these objects are
the relics in A1664, A2256 and A2345 (western).

3) In central cluster regions ($<$ 0.4 Mpc), high power ($>$ 10$^{24}$
W/Hz) relics are missing.  In peripheral regions, on the contrary,
there are both small and large size relics, as well as low and high
power objects. In general relic properties are in agreement with
current models, where shocks are more efficient in low density
peripheral regions, where they would give rise to larger and more
powerful relics.

\subsection{Radio spectra of relics}
\label{sec:53}

Spectral data for relics are available for several
objects. Indeed, the elongated structure of most relic favours
their detection with interferometric observations, moreover
their position at the cluster periphery, where the number of confusing
sources is lower, allows more easily observations at different frequencies.
In Tab. \ref{rel-alpha} we present the available spectral
index information for the objects of the September2011-relic collection.
For homogeneity with radio halos, we report an average
spectral index value in the range $\sim$ 327 - 1415 MHz, and
refer to single papers on the cluster of galaxies hosting a relic
(see Tab. \ref{table-rel}) for more detailed information.
The note in the last column of the table indicates the
spectral shape, in the case that flux density measurements are
available at more than 2 frequencies.

\subsubsection{Integrated spectrum}
\label{sec:531}

\begin{figure*}
  \includegraphics[width=0.5\textwidth]{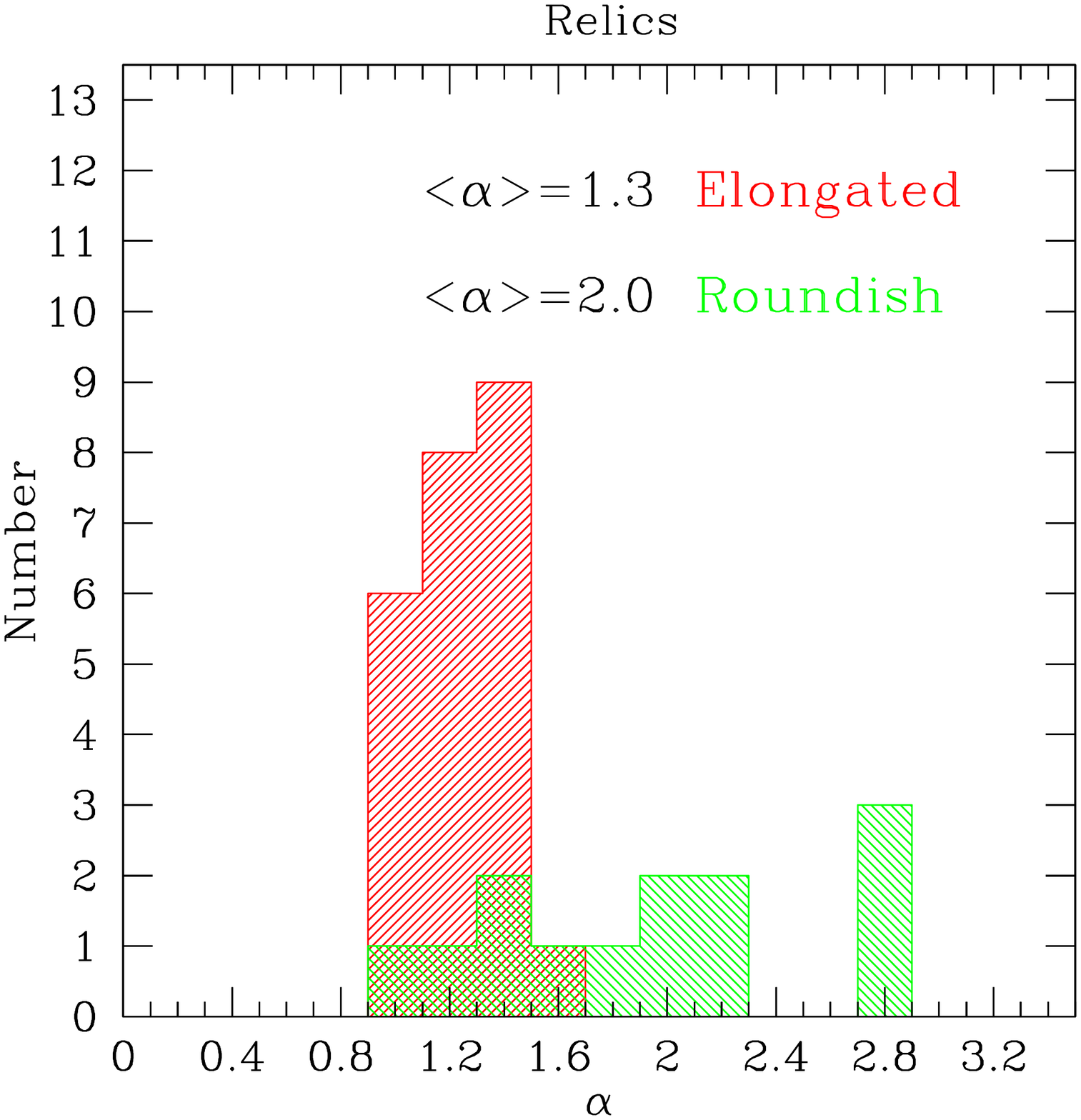}
  \includegraphics[width=0.5\textwidth]{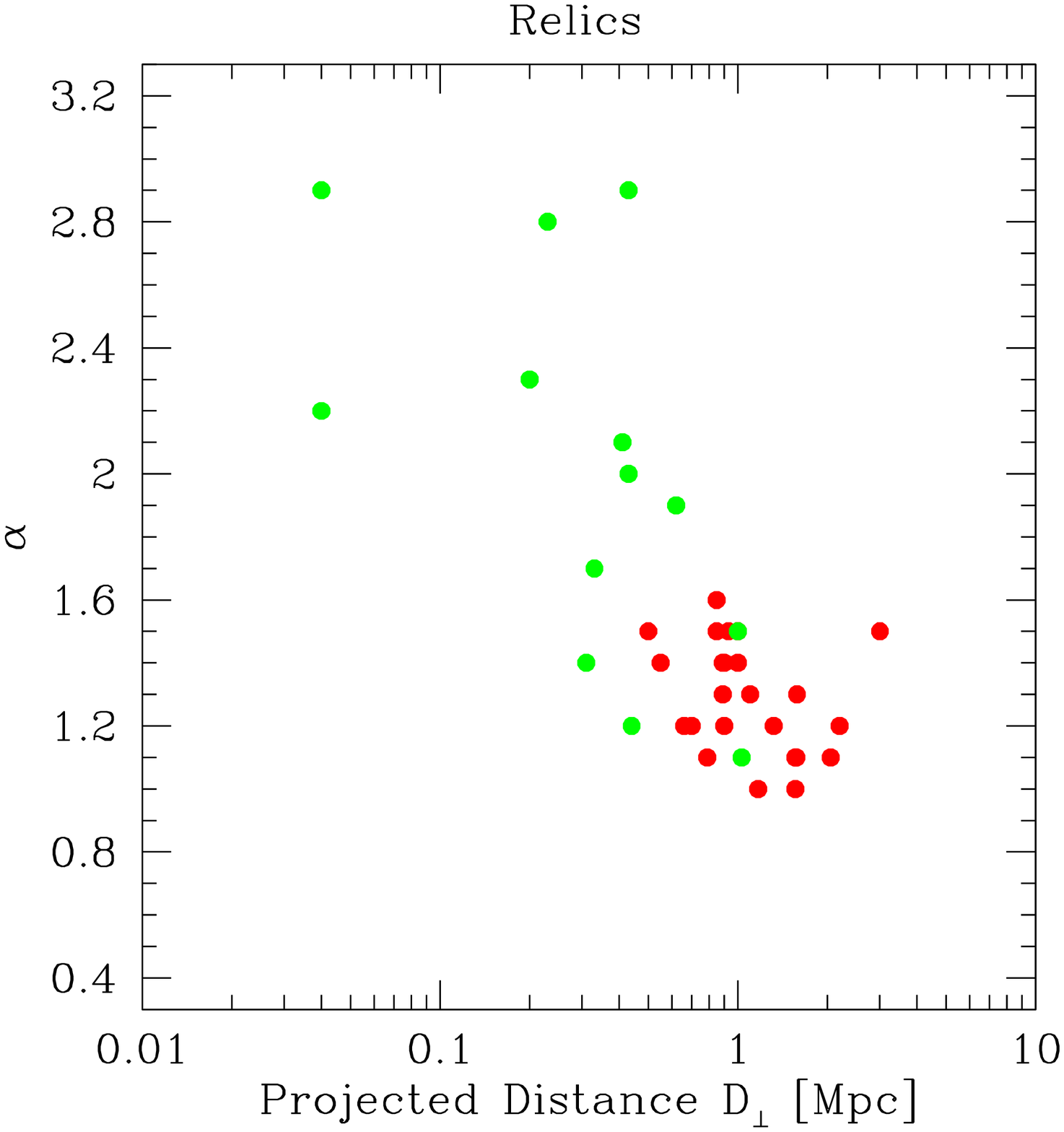}
\caption{{\bf Left panel:} Spectral index distribution of relics.
{\bf Right panel:} Plot of the relic spectral index as a function of the
relic projected distance from the cluster center. In both panels,
red color refers to the elongated relics, green color to the roundish relics.
In these plots, the lower limit $\alpha >$ 2 for A\,548b-B
is not shown.
}
\label{fig:alpha-1}
\end{figure*}

Similarly to radio halos, relics show quite steep spectra.  This is
evident from the spectral index distribution presented in the left
panel of Fig. \ref{fig:alpha-1}.  In this plot, as in the following
plots, the lower limit for
A548b-B ($\alpha >$ 2) is not shown.  Elongated relics show spectral
index $\alpha$ in the range 1 -- 1.6, with an average value of 1.3.
Spectral indices of roundish relics are in a much larger range of
values: $\alpha$ = 1.1 -- 2.9, with an average value of 2.0.
Therefore, there is a remarkable difference in spectral index
properties between elongated and roundish relics, indeed spectra of
roundish relics are steeper than those of elongated relics, with very
few exceptions. It is interesting to note that among the roundish
relics with $\alpha$ $<$ 1.6, there are the 3 objects, whose
properties have been outlined in Sect. \ref{sec:52}, characterized by
power, size and distance from the cluster center similar to those of
elongated relics, i.e.  A\,1664 ($\alpha$ = 1.1), A\,2256 ($\alpha$
= 1.2), and A\,2345W ($\alpha$ = 1.5).
For objects with measurements at more than two frequencies, the radio
spectrum is straight for elongated relics, while it shows
a high frequency steepening for roundish relics.

\begin{table*}
\caption{Spectral index of radio relics} 
\label{rel-alpha}
\centering
\begin{tabular}{lrcl}
\hline\hline
Name    & $\alpha$ & Type &  Notes \\
\hline
A13     & 2.3 & roundish   & steepening \\
A85     & 2.9 & roundish   & steepening \\
A115    & 1.2 & elongated &   \\
A521    & 1.5 & elongated & straight \\
A548b-A & 2.0 & roundish   & \\
A548b-B &$>$2 & roundish   & \\
A610    & 1.4 & roundish   & \\
AS753   & 2.1 & roundish   & steepening \\
A754    & 1.5 & elongated & complex \\
A781    & 1.2 & elongated &  \\
A1240N  & 1.2 & elongated & \\
A1240S  & 1.3 & elongated & \\
A1300   & 1.1 & elongated & \\
A1367   & 1.9 & roundish   & \\
A1612   & 1.4 & elongated & straight \\
A1664   & 1.1 & roundish& \\
A1656   & 1.2 & elongated & straight \\
A2048   & 1.7 & roundish   & steepening \\
A2061   & 1.0 & elongated & \\
A2063   & 2.9 & roundish   & steepening \\
A2163   & 1.0 & elongated & \\
A2255   & 1.4 & elongated & \\
A2256   & 1.2 & roundish   & complex \\
A2345W  & 1.5 & roundish   & \\
A2345E  & 1.3 & elongated & \\
A2443   & 2.8 & roundish   & steepening \\
A2744   & 1.1 & elongated & \\
A3667NW & 1.1 & elongated & straight \\
A4038   & 2.2 & roundish   & steepening \\
ZwCl0008-W& 1.5&elongated & straight \\
ZwCl0008-E& 1.6&elongated & straight \\
CL0217+70-NW& 1.4&elongated & \\
CL0217+70-SE& 1.5&elongated & \\
RXCJ1314-E& 1.2&elongated & \\
RXCJ1314-W& 1.4&elongated & \\
CIZAJ2242-N&1.1&elongated & straight \\
PLCK G287.0+32.9-N&1.3&elongated& \\
PLCK G287.0+32.9-S&1.5&elongated& \\
\hline
\multicolumn{4}{l}{\scriptsize Col. 1: Cluster name; Col.2:
Total spectral index value; }\\
\multicolumn{4}{l}{\scriptsize Col. 3: Relic type; Col.4:
Note on the spectral shape }\\
\end{tabular}
\end{table*}

\begin{figure*}
  \includegraphics[width=0.5\textwidth]{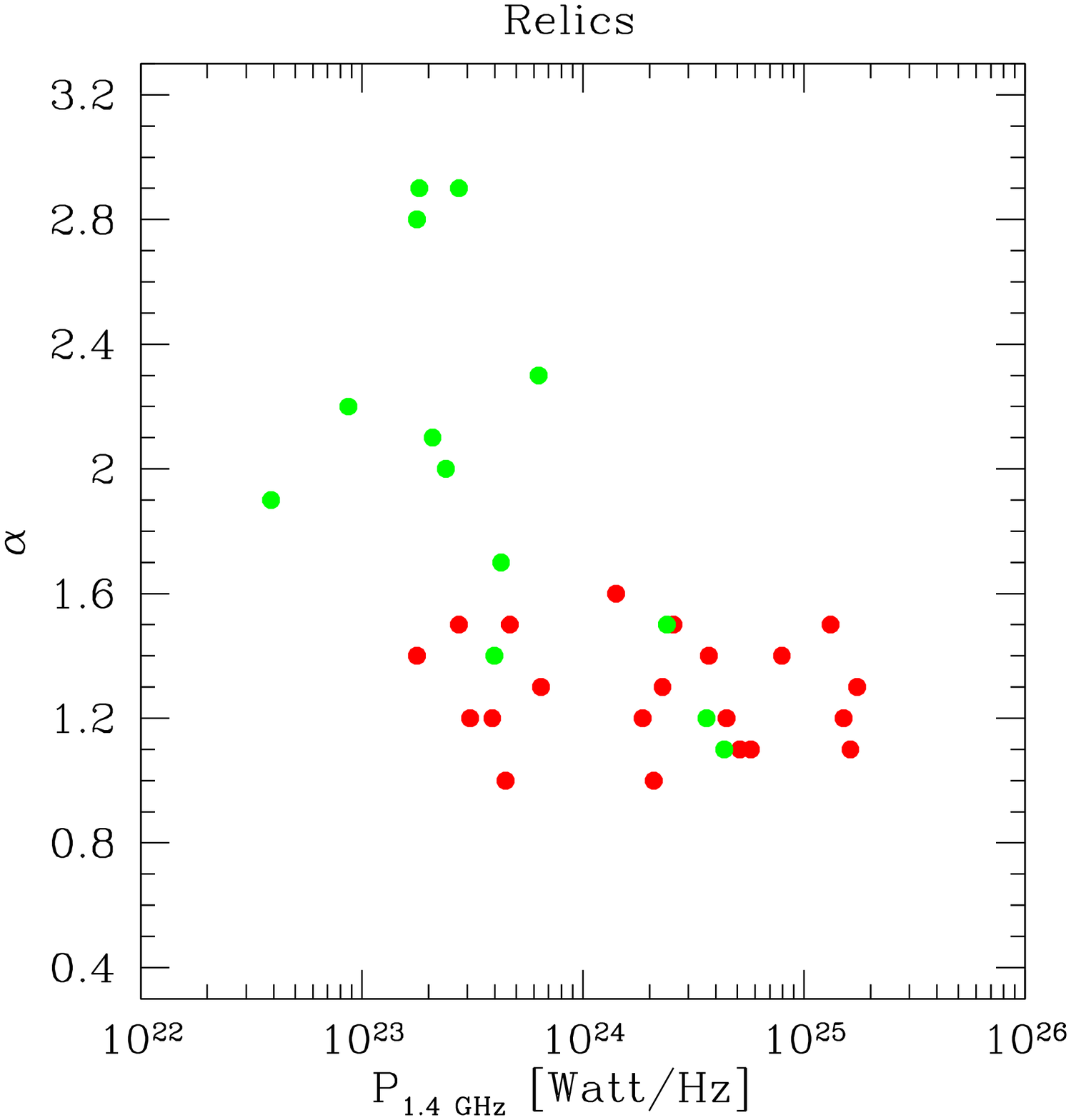}
  \includegraphics[width=0.5\textwidth]{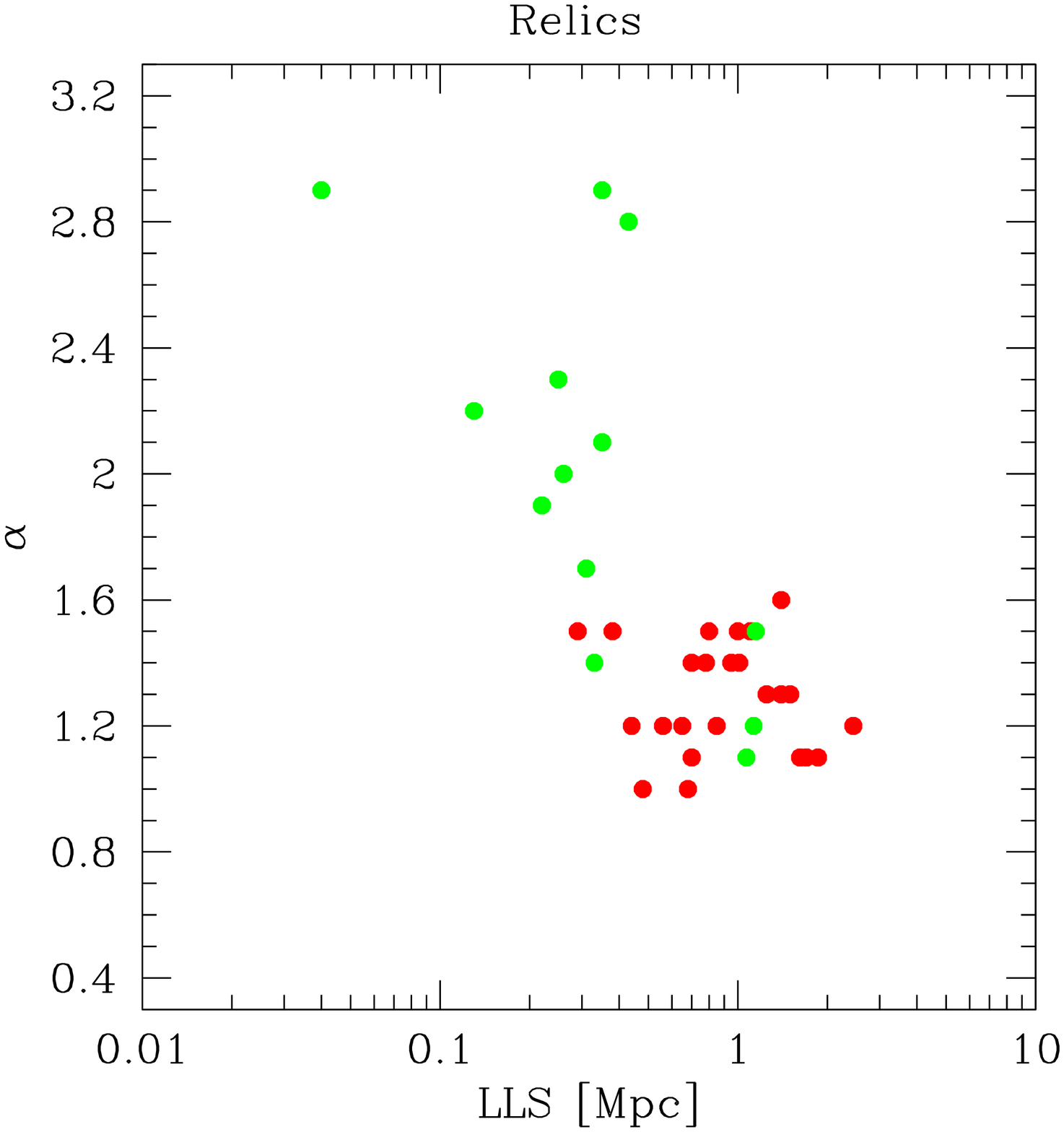}
\caption{{\bf Left panel:} Relic spectral index as a function of the
monochromatic radio power at 1.4 GHz;
{\bf Right panel:} Relic spectral index  versus the largest
linear size (LLS). In both panels, red color refers
to the elongated relics, green color to the roundish relics.
In these plots, the lower limit $\alpha >$ 2 for A\,548b-B
is not shown.
}
\label{fig:alpha-2}
\end{figure*}

In the right panel of Fig. \ref{fig:alpha-1}, we show the relic
spectral index versus the relic projected distance from the cluster
center. We note that relics with steepest spectra are at small
distances from the cluster center.  From the plots of the spectral
index versus the relic power, and of the spectral index versus the
LLS, presented in the 2 panels of
Fig. \ref{fig:alpha-2}, it is derived that most
powerful (P$_{1.4}$ $>$ 10$^{24}$ W/Hz) and most extended (LLS $>$ 0.5
Mpc) relics are characterized by less steep spectra ($\alpha$ $<$
1.6).  Smaller/low power relics have a spectral index in the range
$\sim$ 1.0 -- 2.8 (see also \cite{Vanweeren2009c}).

These results are in agreement with the expected origin of all relic
sources from strong shock waves: in peripheral regions they should
have a larger Mach number and therefore they should produce flatter
spectra (see e.g.  \cite{Vanweeren2009c}).

\subsubsection {Spectral index distribution}
\label{sec:532}

Spectral index maps have been obtained so far only for a few relics.
Elongated relics show a well defined spectral index
distribution, with a transverse steepening toward the cluster center.
This steepening is clearly detected e.g. in A\,2744 \cite{Orru2007}
(Fig.  \ref{fig:alfa2744}, left panel), and in CIZAJ2242.8+5301
\cite{Vanweeren2011d} (Fig. \ref{fig:sausage}, right panel).  This
trend is in agreement with the relic brightness profile and strongly
supports the shock model (see e.g.  \cite{Bruggen2011,Vanweeren2011d}
for a recent discussion).

Among roundish sources, a spectral index map has been obtained for
A\,2256 \cite{Clarke2006}, which shows spectral steepening from NW to
SE. In other clusters, available images of the spectral index do not
show any obvious distribution. The radio morphology at high resolution
is filamentary and the spectral index distribution is irregular.

\subsubsection {Ultra-steep spectrum relics}
\label{sec:533}

As outlined in Sect. \ref{sec:531}, roundish relics may exhibit
extremely steep spectra ($\alpha$ $\gtrsim$ 2).  Examples of these
sources are in A13, A85, A4038 (see \cite{Slee2001}), and A2443 (see
\cite{Cohen2011}). The extremely steep spectra also show a high
frequency cutoff.  These objects have different properties from the
giant elongated relics, being on average less powerful and less
extended.
The number of these sources is very small.  A
reason could be that they are the tip of the iceberg of a population
of ultra-steep relics that may be detected at low frequency, similarly
to the ultra-steep halos illustrated in Sect. \ref{sec:413}.
Alternatively, we could also consider possible selection effects:
indeed small size relics which are marginally resolved may be
often classified as background discrete radio galaxies.

We can argue that the nature of roundish, ultra-steep relics is different
from that of other relics, either elongated or roundish, and that the
spectral index is likely a key parameter for the classification
of relics.

\subsection{Radio - X-ray connection and comparison with halos}
\label{sec:54}

\begin{figure*}
  \includegraphics[width=0.7\textwidth]{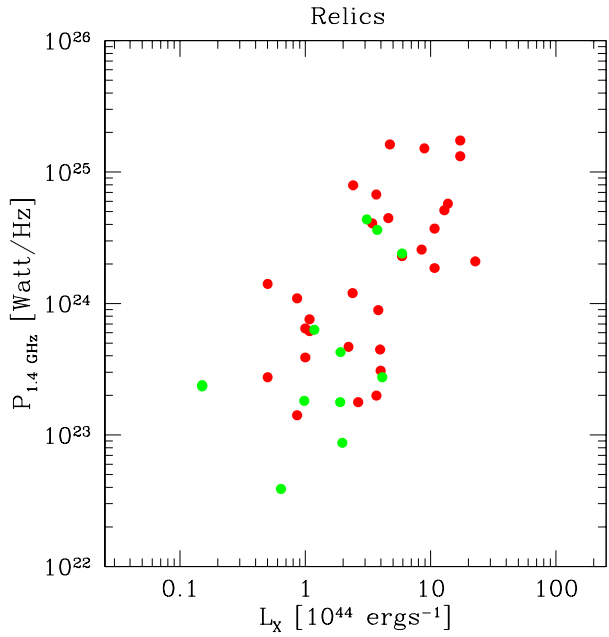}
\caption{Monochromatic radio power of relics at 1.4 GHz versus the
cluster X-ray luminosity between 0.1 -- 2.4 keV. Red color refers
to the elongated relics, green color to the roundish relics.
}
\label{fig:relradiox}
\end{figure*}

The properties of relics have been found to be related to
the properties of their parent clusters. Indeed, similarly to
radio halos, a correlation is found between the relic
radio power at 1.4 GHz and the X-ray luminosity of clusters
(Fig. \ref{fig:relradiox}). The correlation holds for both elongated
and roundish relics, although with somewhat large dispersion;
no systematic difference is visible between the two
classes of relics. We find P$_{1.4} \propto L_x^{1.2}$.

\begin{figure*}
  \includegraphics[width=0.5\textwidth]{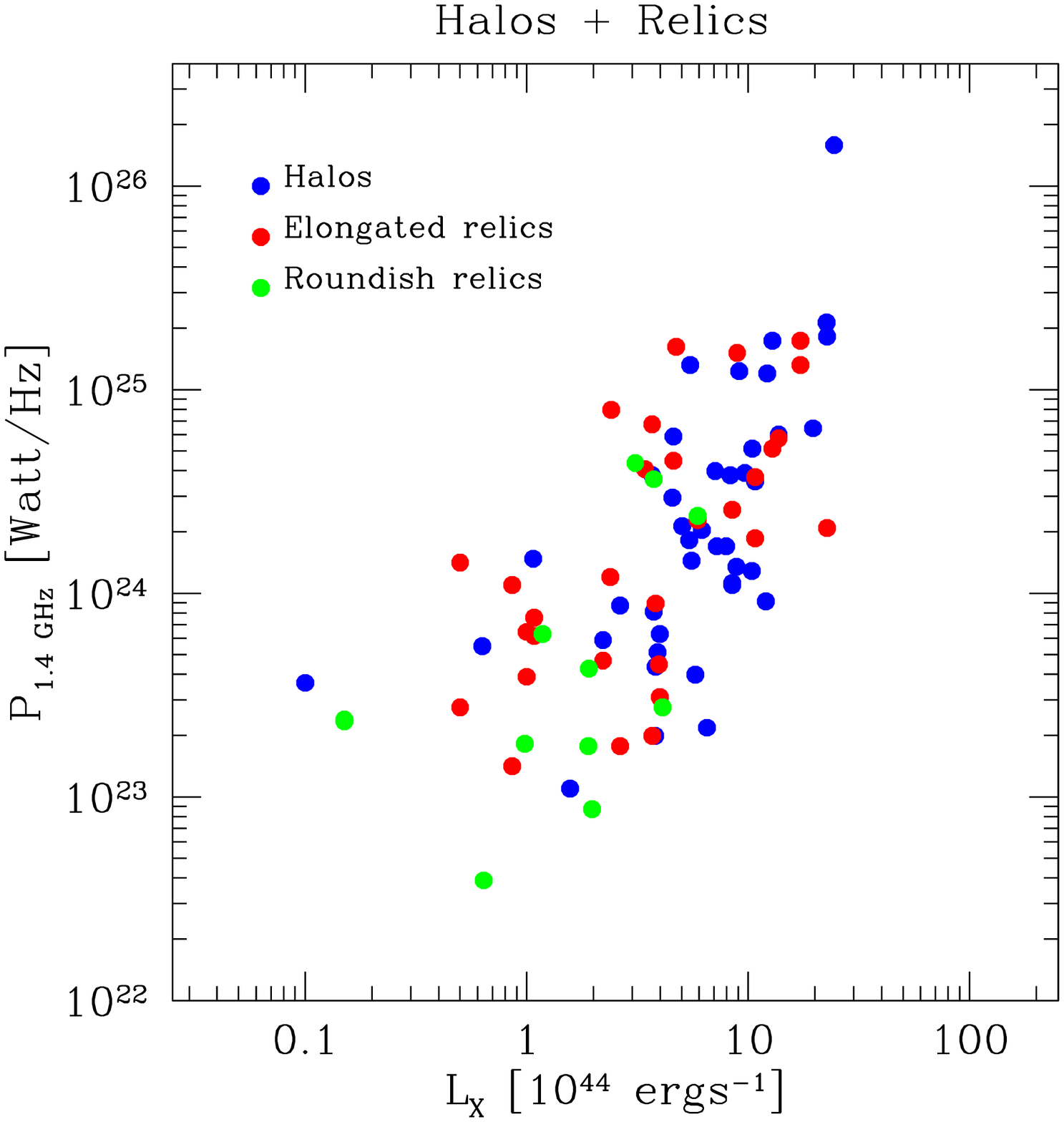}
  \includegraphics[width=0.5\textwidth]{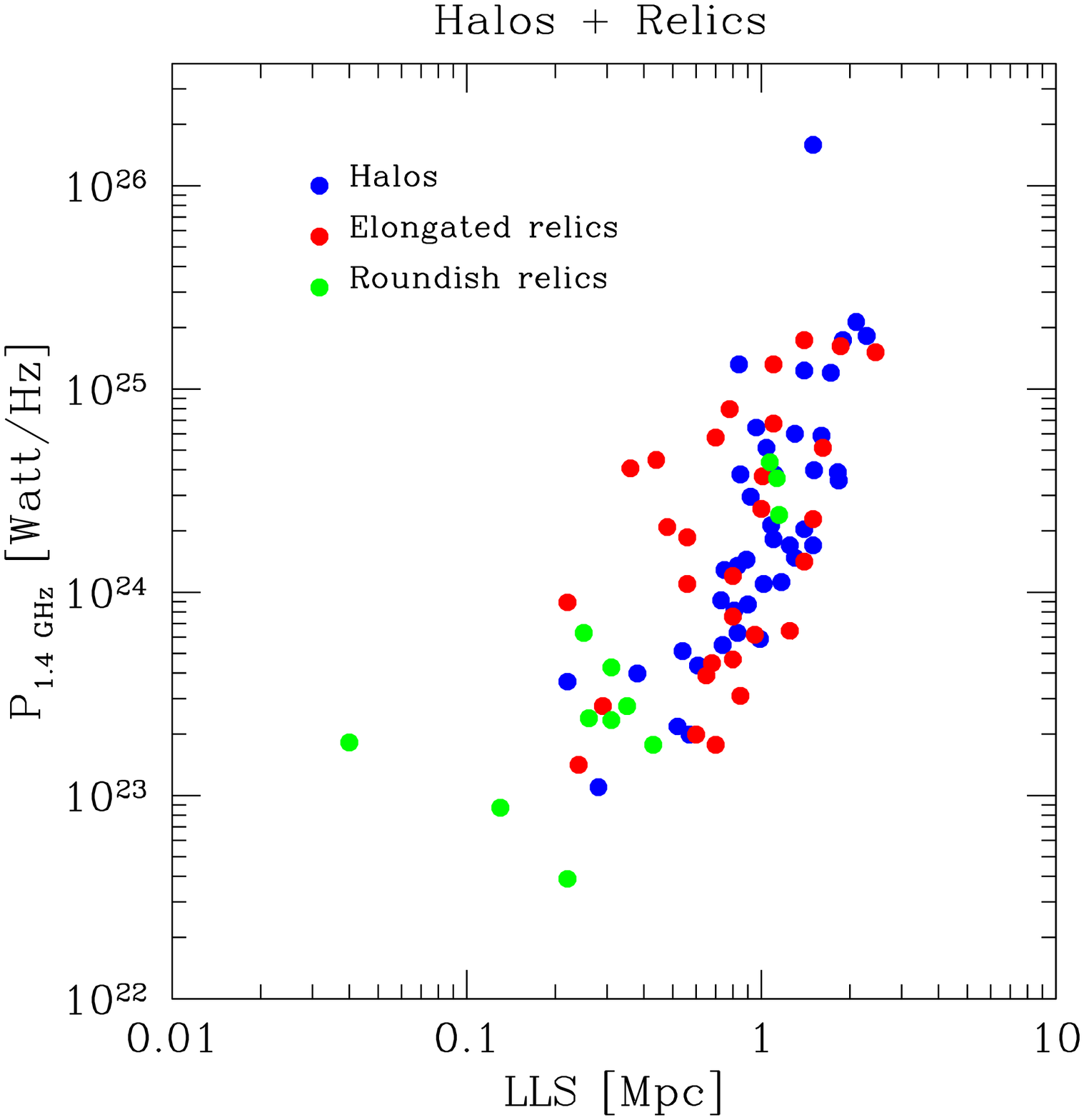}
\caption{{\bf Left panel:} Monochromatic radio power of halos and relics
at 1.4 GHz versus the cluster X-ray Luminosity.
{\bf Right panel: } Monochromatic radio power of halos and
relics at 1.4 GHz versus their
largest linear size measured at the same frequency.
In both panels, blue dots represent halos, while red and green dots represent
elongated and roundish relics, respectively.
}
\label{fig:harel}
\end{figure*}

The similarity between the properties of halos and relics has been
already noted in the past (e.g. \cite{Giovannini2002}).  In
Fig. \ref{fig:harel} we show the halo and relic radio power as a
function of the X-ray cluster luminosity (left panel), and as a
function of their linear size (right panel).  For clarity, we use
different symbols for halos, elongated relics and roundish relics. The
most powerful object in the plots is the radio halo in the cluster
MACS J0717.5+3745.

It turns out that halos and relics show a very similar behaviour in
both plots, supporting the connection between halos and relics, and
their link to cluster mergers.  From the left panel of
Fig. \ref{fig:harel}, we note that the most powerful halos and relics
show similar values in radio power and cluster X-ray luminosity (with
the exception of MACS J0717.5+3745), with similar dispersion.  At
lower radio powers, there is a large dispersion in both classes, and
in particular it is interesting to note that, while halos in poor
X-ray clusters are only a few (Sect. \ref{sec:431}), relics are easily
detected also in clusters of low X-ray luminosity.  Indeed,
elongated and roundish relics fill the region of radio power P$_{1.4}
< $ 10$^{24} W/Hz$ and cluster X-ray luminosity L$_x <$ 3 $\times$ 10$^{44}$.
The flatter correlation found for relics with respect to halos likely
reflects this behaviour.

Halos and relics show similar behaviour also in the plot of radio
power versus largest linear size (right panel of
Fig. \ref{fig:harel}).  As for the previous correlation, halos show a
lower dispersion in the data, and at a given radio power they tend to
be more extended than relics.  We finally note the peculiarity of MACS
J0717.5+3745 \cite{Bonafede2009b,Vanweeren2009b} (see
Sects. \ref{sec:42} and \ref{sec:43}), which, even with the addition
of relics, remains the most powerful diffuse radio source, with a
radio power about an order of magnitude higher than that of other
known diffuse sources.

\section{Mini-Halos}
\label{sec:6}

\begin{figure*}
\includegraphics[width=0.7\textwidth, bb=1 295 590
840,clip]{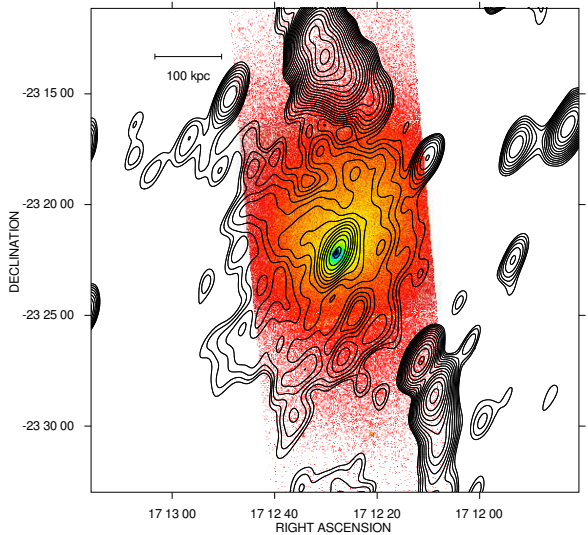}
\caption{VLA 1.4 GHz contours of the mini-halo in Ophiuchus overlaid on the
Chandra X-ray emission. The resolution of the radio image is
91.4$''\times$40.4$''$. The first contour level is drawn at
0.3 mJy/beam and the rest are spaced by a factor $\sqrt2$. Taken from
\cite{Govoni2009}.
}
\label{fig:ophiucus}
\end{figure*}

Some relaxed, cool-core, galaxy clusters exhibit signs of diffuse
synchrotron radio emission that extend
far from the dominant radio galaxy at the cluster center,
forming a so-called mini-halo. Mini-halos are extended on a moderate
scale ($\simeq$ 500 kpc) and in common with halos and relics they have a low
surface brightness and a steep spectrum.
The diffuse source in the Ophiucus cluster \cite{Govoni2009,Murgia2010b},
reported in Fig. \ref{fig:ophiucus}, is an example of a mini-halo.

Mini-halos are sometimes considered small versions of radio halos, but
their classification is more complex and fuzzy. Their emission
originates from relativistic particles and magnetic fields which are
believed to be deeply mixed with the thermal intracluster gas. As a
general guideline, this physical characteristic can be used to
distinguish mini-halos from other kind of steep spectrum radio sources
in clusters such as radio bubbles related to AGN activity.  We shall
not define as mini-halos those radio sources in which the ambient thermal
gas is clearly separated by the non-thermal plasma, as in the case of
AGN radio lobes whose expansion has created cavities or holes in the
intracluster X-ray emission.  An example of this situation is given by
the radio source 3C317 at the center of A2052
(e.g. \cite{Blanton2011}).  3C317 is composed of a compact core
surrounded by an amorphous halo. This kind of radio morphology is also
referred to as a {\it core-halo} (see e.g. \cite{Mazzotta2008}). Although
the radio emission at the center of A2052 resembles on a smaller scale
the typical mini-halo, almost all of the radio emission detected so
far is contained within cavities devoid of X-ray emitting gas.
Therefore, this type of radio sources will not be considered here.

The Perseus cluster is a different case.  The radio source 3C84 at the
center of Perseus shows X-ray cavities in the inner region where the
AGN radio activity interacts with the thermal gas
(e.g. \cite{Fabian2003}), but on larger scale the cluster exhibits a
diffuse radio emission mixed with the thermal intracluster gas.
Indeed Perseus cluster represents the prototype example of a mini-halo
(e.g. \cite{Noordam1982,Pedlar1990,Burns1992}).  When the cluster
X-ray information is lacking, as for the diffuse emission found in the
system MRC 0116+111 \cite{Bagchi2009}, the mini-halo classification
can be particularly troublesome.

In addition to the above mentioned difficulties in the mini-halo
classification, the relatively small angular size of mini-halos in
combination with a possibly strong emission of the central radio
galaxy, complicates their detection. Thus, our current observational
knowledge on mini-halos is limited to a few clusters. Continued
searches have so far revealed no more than 11 mini-halos. A list of
mini-halos is given in Tab. \ref{tab-minih}.  This collection, which
is updated to September, 2011, will be referred to henceforth as
September2011-Mini-halo collection.  Other mini-halos candidates have
been found in A1068, and A1413 (\cite{Govoni2009}), although to be
classified as mini-halos they would require further investigation.

The spectra of mini-halos are steep, as well as for halos, with some
evidence of radial spectral steepening (e.g. Perseus
\cite{Sijbring1993,Gitti2002}, A\,2626 \cite{Gitti2004}, Ophiucus
\cite{Murgia2010b}).

The origin of mini-halos is still poorly known.  Gitti et
al. \cite{Gitti2002} argued that the radio emitting particles in
mini-halos cannot be connected to the central radio galaxy in terms of
particle diffusion. They proposed that mini-halos result from a relic
population of relativistic electrons reaccelerated by MHD turbulence
via Fermi-like processes, the necessary energetics being supplied by
the cool-core region. This is supported by the correlation observed
between the mini-halo radio power and the cooling rate power \cite{Gitti2004}.

\begin{figure*}
  \includegraphics[width=0.7\textwidth]{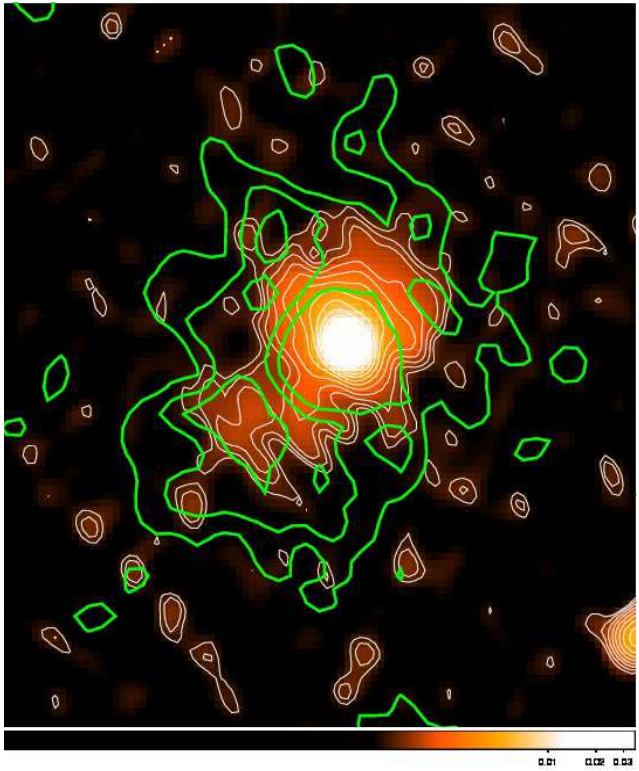}
\caption{GMRT 614 MHz total intensity map and contours (white)
of the mini-halo in RXJ1347.5-1145. The resolution is
4.8$''\times$3.5$''$. The first contour levels are drawn at
$-$0.3, 0.3 mJy/beam and the rest are spaced by a factor $\sqrt2$.
Contours of the MUSTANG SZE image \cite{Mason2010} of the cluster
are overlaid in green. Image taken from
\cite{Ferrari2011}.}
\label{fig:minihalo}
\end{figure*}

The analysis of the mini-halo present in the most X-ray luminous
cluster RXJ1347.5-1145 \cite{Gitti2007}, suggests that additional
energy for electron reacceleration in mini-halos might be provided by
sub-cluster mergers that have not been able to destroy the central
cluster cool core.  
In this cluster, the spatial correspondence
between a radio excess of the mini-halo emission and a hot region,
detected through both Sunyaev-Zel'dovich effect (see
Fig. \ref{fig:minihalo}) and X-ray observations, indicates that
electron reacceleration in this location is most likely related to a
shock front propagating into the intra-cluster medium
\cite{Ferrari2011}.  In addition to RXJ1347.5-1145, other clusters of
galaxies like Ophiuchus \cite{Govoni2009} and A2142
\cite{Giovannini2000} present a radio mini-halo, as well as a
cool core
that has likely survived a possible recent merging event. Indeed,
although cool-core clusters are generally considered relaxed
systems, when analyzed in detail they sometimes reveal peculiar X-ray
features in the cluster center which may be indicative of a link
between the mini-halo emission and some minor merger activity.

Finally, cold fronts associated to the core gas sloshing have been
observed in some clusters hosting a mini-halo. Thus,  the possibility
has been advanced that gas sloshing may generate turbulence in
the core, which in turn may reaccelerate the relativistic electrons
necessary to form a mini-halo \cite{ZuHone2011}.

\begin{table}
\caption{September2011-Mini-halos collection (published mini-halos, Sept.
2011)}
\label{tab-minih}
\begin{tabular}{lllrrcl}
\hline\noalign{\smallskip}
Cluster        & z      &  kpc/$''$     & S(1.4) & Log P(1.4)  &
Lx[0.1-2.4keV]   &  Ref  \\
               &        &               & mJy    &  W/Hz   & $10^{44}$
erg/sec &        \\
\noalign{\smallskip}\hline\noalign{\smallskip}
Perseus         & 0.0179 &  0.36         & 3020.0     &24.33 &  7.88 &
\cite{Sijbring1993}   \\
Ophiuchus       & 0.028  &  0.55         &  106.4     &23.27 &  6.11 &
\cite{Govoni2009} \\
A1835           & 0.2532 &  3.91         &    6.0     &24.06 & 24.15 &
\cite{Govoni2009} \\
A2029           & 0.0765 &  1.43         &   18.8     &23.42 &  9.24 &
\cite{Govoni2009} \\
A2142           & 0.0894 &  1.65         &   18.3     &23.55 & 11.47 &
\cite{Giovannini2000}  \\
A2390           & 0.228  &  3.62         &   63.0     &24.98 &  5.33 &
\cite{Bacchi2003} \\
A2626           & 0.0604 &  1.15         &   29.0     &23.39 &  1.18 &
\cite{Gitti2004} \\
MRC0116+111     & 0.131  &  2.31         &  140.0     &24.79 &  $-$ & 
\cite{Bagchi2009} \\
RBS 797         & 0.35   &  4.91         &   $-$     &  $-$  & 16.91 &
\cite{Gitti2006}  \\
RXJ1347.5$-$1145  & 0.451  &  5.74       &   25.2     &25.27 & 51.95 &
\cite{Gitti2007}  \\
RXCJ1504.1$-$0248 & 0.2153  &  3.46      &   20.0     &24.42 & 27.55 &
\cite{Giacintucci2011c} \\
\noalign{\smallskip}\hline
\multicolumn{7}{l}{\scriptsize Col. 1: Cluster name; Col.2:
Redshift; Col. 3: Angular to linear size conversion;}\\
\multicolumn{7}{l}{\scriptsize Col. 4: Radio flux density at 1.4 GHz; 
Col. 5: Logarithm of radio power at 1.4 GHz;  }\\
\multicolumn{7}{l}{\scriptsize Col. 6: Cluster X-ray luminosity
in the 0.1-2.4 keV band in 10$^{44}$ units; 
Col. 7: References to the radio flux density. }\\
\end{tabular}
\end{table}

\subsection{Comparison with halos}
\label{sec:61}

Cassano et al. \cite{Cassano2008} found that the synchrotron emissivity
of mini-halos is about a factor of 50 higher than that of radio halos.
In the framework of the particle reacceleration scenario,
they suggested that, an extra amount of relativistic electrons would be
necessary to explain the higher radio emissivity of mini-halos.
These electrons could be provided by the central radio galaxy
(e.g. \cite{Fujita2007}) or be of secondary origin as previously
suggested in the literature by some authors
(e.g. \cite{Pfrommer2004,Keshet2010}).
In this latter case relativistic electrons in mini-halos are expected to be
continuously produced by the interaction of cosmic ray protons with
the ambient thermal protons.

\begin{figure*}
  \includegraphics[width=0.7\textwidth]{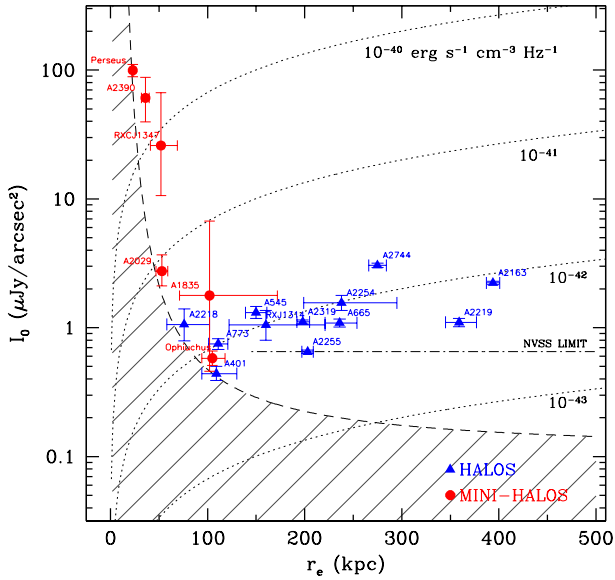}
\caption{
Best fit central brightness $I_0$ versus the length-scale $r_e$ both
for halos (blue triangles) and mini-halos (red dots) taken from Murgia
et al. \cite{Murgia2009}.  The dotted lines indicate regions of
constant emissivity, namely 0.1, 1, 10, 100 times the average
emissivity of radio halos which is $\langle J \rangle \simeq 10^{-42}$
erg\,s$^{-1}$cm$^{-3}$Hz$^{-1}$.  They have been traced by assuming a
putative redshift of $z=0.18$ and a spectral index $\alpha=1$. The
dashed line represents the detection limit expected for a mini-halo or
a halo at $z=0.18$ observed in a deep image with a beam of $25''$ and
a sensitivity level of 25 $\mu$Jy/beam.  The dot-dashed line
represents the $3\sigma$ sensitivity level of the NVSS.
}
\label{fig:comparison}
\end{figure*}

Murgia et al. \cite{Murgia2009} modeled the radio brightness profile
$I(r)$ of radio halos and mini-halos, with an exponential of the form
$I(r)=I_{0}e^{-r/r_e}$.  The proposed method to derive the central
radio brightness, the length-scale, and hence the radio emissivity of
diffuse sources is relatively independent of the sensitivity of the
radio observation.  Fig. \ref{fig:comparison} shows the best fit
central brightness $I_0$ versus the length-scale $r_e$ for a
representative set of radio halos and mini-halos.  These authors found
that radio halos can have quite different length-scales but their
emissivity is remarkably similar from one halo to another. In
contrast, mini-halos span a wide range of radio emissivity. Some of
them, like Perseus, RXJ1347.5-1145 and A2390 are characterized by a
radio emissivity which is more than two order of magnitude greater
than that of radio halos.  This finding seems to suggest that part of
the mini-halo radio emission is related to properties of the local
intergalactic medium and part is correlated with the AGN activity of
the central brightest galaxy, as suggested by the faint correlation
discussed by Govoni et al. \cite{Govoni2009}, between the mini-halo
and the cD radio power.  On the other hand there are also mini-halos
like A2029, Ophiuchus, and A1835 that have a radio emissivity which is
much more typical of halos in merging clusters rather than of the
other mini-halos.

\section{Magnetic fields and observational results}
\label{sec:7}

Thanks to new radio observations and to the improvement of
interpretative models, our understanding of magnetic fields associated
with the intracluster medium in clusters of galaxies has advanced
significantly in recent years and the presence of $\mu G$-level
magnetic fields in clusters of galaxies is now widely acknowledged.
Most of what we know about intracluster magnetic fields derives from
the study of cluster diffuse radio sources (halos, relics and
mini-halos) and Faraday Rotation Measures of polarized radio galaxies
located inside or behind galaxy clusters. One of the first
comprehensive reviews on extragalactic magnetic field was presented by
Kronberg \cite{Kronberg1994}, followed by more recent works
\cite{Carilli2002,Govoni2004b}.

Large-scale magnetic fields are studied through radio halos, which
reveal magnetic fields spread over extremely large volumes.
Sometimes, the
intracluster magnetic fields can be ordered on scales of hundreds of
kpc, as revealed in A2255 \cite{Govoni2005,Pizzo2011}, and MACS
J0717.5+3745 \cite{Bonafede2009b}, where a polarized signal has been
detected.  The brightness fluctuations and polarization level of radio
halos are strictly related to the intracluster magnetic field
structure.  For example, lack of polarization and a smooth and regular
surface brightness may indicate that the cluster magnetic field is
ordered on small scales, while a disturbed radio morphology and
presence of polarization could be related to a magnetic field ordered
on large scales \cite{Tribble1991a,Murgia2004}.  Currently, only in
the radio halos of A\,2255 and A\,665, brightness fluctuations and
polarization information have been used to put some constraints on the
magnetic field power spectrum of a galaxy cluster
\cite{Govoni2006,Vacca2010}.

The most promising technique to derive a detailed view of
the intracluster magnetic fields on small spatial scales
is the analysis of the Faraday rotation of radio galaxies located inside
and beyond clusters.
The Faraday effect is the rotation suffered by linearly polarized radiation
traveling through a magnetized plasma. The intrinsic polarization
angle $\psi_0$ is rotated by an amount:
\begin{equation}
\Delta \psi=\psi-\psi_0=\lambda^2 RM
\label{lambda2}
\end{equation}
where $\lambda$ is the radiation wavelength while
$RM$ is the so-called Rotation Measure.
The rotation measure is defined by:
\begin{equation}
RM=812 \int_{0}^{L}n_{e} B_{z} dl  ~~({\rm rad/m^2})
\label{rm}
\end{equation}

\noindent
where $n_e$ is the thermal electron density in cm$^{-3}$, $B_{z}$ is
the magnetic field component along the line-of-sight in $\mu$G, and
$L$ is the path length through the plasma in kpc \cite{Burn1966}.  If
we perform multi-frequency observations of a radio source embedded in
or behind a galaxy cluster, then by a linear fit of the
$\lambda^2$-law in Eq.\ref{lambda2} we can estimate the RM along that
direction.  By knowing the RM and given a model for the distribution
of the thermal electrons (e.g. from X-ray observations) we can infer
the intracluster magnetic fields.  However, because of the random and
turbulent nature of intracluster magnetic fields, inverting
Eq.\ref{rm} is not straightforward even in the cases of simple
electron density distributions.

A technique to analyze and interpret the RM data is the RM Synthesis
\cite{Brentjens2005}, which uses the RM transfer function to solve the
n$\pi$ ambiguity related to the RM computation, and allows
distinguishing the emission as a function of Faraday depth.

Detailed RM images of extended radio sources have been obtained
(e.g.
\cite{Perley1991,Taylor1993,Feretti1995,Feretti1999a,Feretti1999b,Govoni2001a,Taylor2001,Eilek2002,Govoni2006,Taylor2007,Guidetti2008,Laing2008,Guidetti2010,Bonafede2010,Govoni2010}).
The RM distributions seen across these sources present patchy
structures whose statistics is, in first approximation, Gaussian-like
with dispersions up to several hundred (and even thousand) rad/m$^2$.
The observed RM fluctuations indicate that the intracluster magnetic
field is not regularly ordered but turbulent on scales ranging from
tens of kpc to $\lesssim$ 100 pc.

The early interpretation of this finding was that the intracluster
magnetic field is composed by uniform cells of size $\Lambda_C$ with
random orientation in space. In this ``single-scale'' model
(e.g. \cite{Lawler1982,Tribble1991b,Feretti1995,Felten1996}), the RM
distribution is expected to be a Gaussian with zero mean and
dispersion given by:
\begin{equation}
\sigma_{RM}^2=812 \Lambda_C \int_{0}^{L}(n_{e} B_{z})^2 dl  ~~({\rm
rad^2/m^4}).
\label{sigmarm}
\end{equation}

Thus, by measuring the dispersion of the RM we can estimate the
cluster magnetic field provided that the cell size ${\Lambda_C}$ is
known.  The question is then what exactly is ${\Lambda_C}$ and how can
we find it.  This issue was investigated theoretically and in
numerical tests by different authors
\cite{Ensslin2003,Vogt2003,Murgia2004,Vogt2005}, who determined that
the correct value for $\Lambda_C$ is the magnetic field auto-correlation
length, $\Lambda_B$, which can be calculated if the power spectrum of
the magnetic field fluctuation is known.  Hence, if we want to
determine the strength of the magnetic field we ultimately need to
determine the power spectrum of its fluctuations.

\subsection{Faraday rotation modeling}
\label{sec:71}

Murgia et al. \cite{Murgia2004} and Laing et al. \cite{Laing2008}
developed dedicated software tools to attempt to constraint the
magnetic field power spectrum parameters.  These codes have been
designed to produce, starting from 3D magnetic field models, synthetic
polarization images that can be directly compared with the
observations.  To limit the degrees of freedom and the computational
burden, in many of these simulations the magnetic field power spectrum
is assumed to be a simple power-law of the form $\vert B_{\rm
  k}\vert^2\propto k^{-n}$.  However, in some cases high-order terms
in curvature have been also explored
\cite{Laing2008,Guidetti2010}.  Another more general method, relying
on a semi-analytical approach and Bayesian inference has been also
developed by En{\ss}lin \& Vogt \cite{Ensslin2003}.  The power
spectrum shape is decomposed with multiple spectral basis components
and it is free to assume, in principle, any form required by the data
\cite{Ensslin2003,Kuchar2011}.

In all the mentioned modeling, it is assumed that the power spectrum
normalization scales as a function of the gas density such that $B(r)=
B_0 \left( \frac{n_e(r)}{n_0} \right)^{\eta}$, where the index $\eta$
is in the range 0.5$-$1.0 \cite{Dolag2001}.  Indeed there are several
indications that the magnetic field intensity declines with radius
with a rough dependence on the thermal gas density. This is predicted
by both cosmological simulations e.g
\cite{Dolag1999,Bruggen2005,Xu2010} and by the comparison between
thermal and radio brightness profiles \cite{Govoni2001b} in clusters
hosting radio halos.

We note that in agreement with the absence of a preferred direction in
most of the RM images, the results of the modeling are based on the
assumption that the magnetic field is statistically isotropic and that
the Faraday rotation is occurring entirely in the intracluster medium
with a negligible local RM enhancement occurring at the interface
between the radio lobes and the surrounding medium (but
see \cite{Rudnick2003} for a contrary viewpoint).  However, recent RM
images \cite{Guidetti2011} of four radio galaxies (0206+35, 3C270,
3C353, M84) located in environments ranging from poor groups to rich
clusters show clearly anisotropic ``banded'' patterns.  In some cases,
these RM patterns coexist with regions of isotropic random variations.
These RM bands may be interpreted as a consequence of interactions
between the sources and their surroundings and are likely be caused by
magnetized plasma which has been compressed by the expanding radio
lobes.

\subsection{Cluster centers}
\label{sec:72}

The Faraday rotation modeling described above has been applied to
interpret the RM images of single extended radio galaxies located at
the center of galaxy clusters.  Some of them present strong RM
asymmetries between the two lobes. This is the Laing-Garrington
effect: the farthest lobe suffers a higher Faraday rotation due to the
largest amount of intervening material
\cite{Laing1988,Garrington1988,Garrington1991}.  Indeed, at least for
these sources, we know that the observed RM must originate mostly in
the foreground magnetized intracluster medium.

Radio sources at the center of dense cool-core clusters, like Hydra-A,
probe magnetic fields as high as 10$-$30$ \mu G$
\cite{Laing2008,Kuchar2011}.  The RM analysis of radio sources in the
less dense environments of galaxy groups, such us 3C31 or 3C449
\cite{Feretti1999c,Laing2008,Guidetti2010}, indicates lower central
magnetic field
strengths in the range 3$-$10 $\mu G$. The intracluster magnetic field
auto-correlation length is generally found to be of the order of a few
kpc.

Determining the detailed shape of the magnetic field power spectrum it
is not trivial because of the degeneracy existing between its slope
and the outer scale of the magnetic field fluctuations,
$\Lambda_{max}$ \cite{Murgia2004}.  In fact, the same RM images can be
explained either by a flat power spectrum with a large $\Lambda_{max}$
or by a steeper power spectrum with a lower $\Lambda_{max}$. Given the
limited spatial dynamic range of the current RM images, it's still not
easy to break the degeneracy.  According to \cite{Kuchar2011}, the
slope magnetic field power spectrum is close to the Kolmogorov index
in Hydra-A while other authors report a flatter slope
\cite{Laing2008}.

\subsection{Cluster peripheries}
\label{sec:73}

To get information on the radial trend of the cluster magnetic field,
one has to take into account that in the previously described
modeling, another important degeneracy exists between the central
magnetic field strength $B_0$ and the index $\eta$, which describes
the scaling of $B$ with the gas density.  The same RM image can be
explained either by a higher $B_0$ with a rapid radial decrease (high
$\eta$) or by a lower $B_0$ with a flatter radial profile.  The
degeneracy can be solved if RM images of different radio sources at
different impact parameters from the cluster center are available.

In A2382 the RM of two tailed sources have been investigated
\cite{Guidetti2008}.  In this cluster, the power spectrum of the RM
fluctuation is compatible with the Kolmogorov index. The observed
$\sigma_{RM}$ decreases in the cluster outskirts and can be explained
if $B_0\simeq 3.5$ $\mu G$ and $\eta=0.5$. But it is also marginally
compatible with a flat magnetic field profile with $B_0\simeq 0.9$
$\mu G$ or with a higher central magnetic field strength of $B_0\simeq
13$ $\mu G$ which decreases with radius following the thermal gas
density, $\eta=1$.

In a few other Abell clusters, the RM of radio sources at different
impact parameters have been obtained, In A119
\cite{Feretti1999b,Murgia2004}, the RM analysis indicates a central
magnetic field strength of $B_0\simeq 5.5$ $\mu G$ which decreases
with radius roughly following the gas density.  The case of A2255 is
particularly interesting since this cluster hosts a Mpc-scale radio
halo.  By analyzing the RM of three different radio sources and the
radio halo properties, a central magnetic field strength of about 2
$\mu G$ for $\eta=0.5$ has ben obtained \cite{Govoni2006}.  These
results would imply that the magnetic field strength in clusters with
or without a diffuse halo is indeed very similar.

\begin{figure*}
 \includegraphics[width=1\textwidth, bb= 1 214 542 580,
clip]{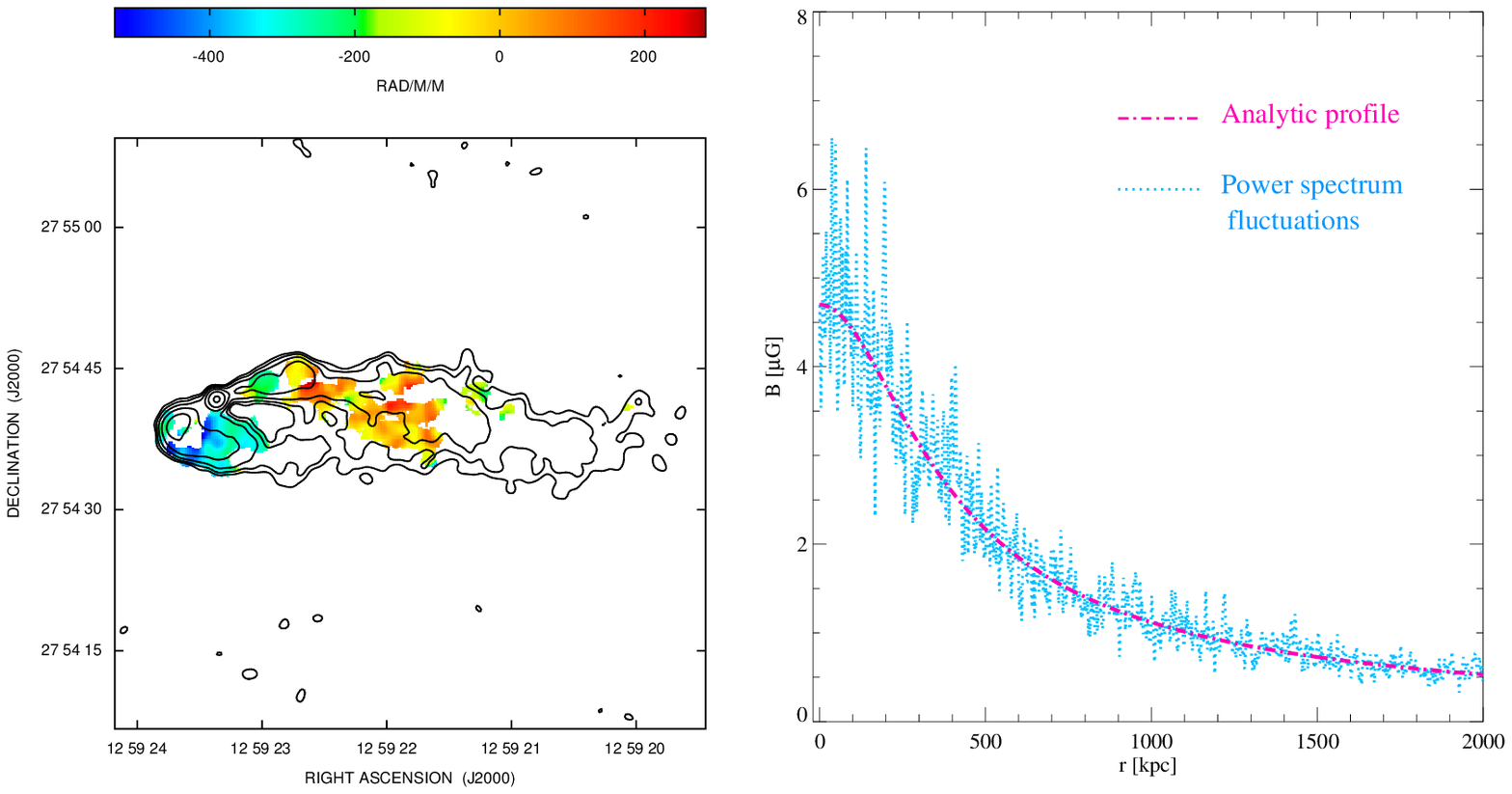}
\caption{{\bf Left panel:}
Rotation Measure image (along with the total intensity radio 
contours at 5 GHz) of the radiogalaxy NGC 4869. This is one of the
seven radio sources 
used for the analysis of the Coma cluster magnetic field \cite{Bonafede2010}.
{\bf Right panel: }  Profile of the best magnetic field model ofthe Coma
cluster \cite{Bonafede2010}. Magenta line refers to the analytic profile
$B(r)= B_0 \left( \frac{n_e(r)}{n_0} \right)^{\eta}$,
while the blue line refers to a slice extracted from the simulated
magnetic field numerical model. Power
spectrum fluctuations in the profile are shown.
}
\label{fig:Coma}
\end{figure*}

The RM of radio sources in the Coma cluster, are probably the best
studied so far \cite{Kim1990,Feretti1995,Bonafede2010}.  Bonafede et
al. \cite{Bonafede2010}, by comparing the RM observed in seven
radio sources (see Fig. \ref{fig:Coma}) with simulations, found a
cluster magnetic field in the range $B_0$=3.9 $\mu$G; $\eta=0.4$ and
$B_0$=5.4 $\mu$G; $\eta=0.7$.  These values correspond to models where
the magnetic field energy density scales as the gas energy density, or
the magnetic field is frozen into the gas, respectively.  This is
expected since the energy in the magnetic component of the
intracluster medium is a tiny fraction of the thermal energy.  It is
important to note that the average magnetic field intensity over a
volume of $\sim$1 Mpc$^3$ is $\sim$2 $\mu$G.  Indeed, the model
derived from RM analysis gives an average estimate that is compatible
with the minimum energy condition in the radio halo
\cite{Thierbach2003}.  A direct comparison with the magnetic field
estimate derived from the inverse Compton emission is more difficult
since the hard-X detection is debated.  The field strength derived
from RM analysis gives a magnetic field estimate that is consistent
with the present lower limits obtained from hard X-ray observations
\cite{Wik2009} while it is still a bit higher than the estimate
obtained on the basis of the non-thermal hard X-ray detection from
RXTE \cite{Rephaeli2002} and BeppoSAX \cite{Fusco2004}.

\subsection{Statistical studies}
\label{sec:74}

In addition to detailed RM studies focused on single clusters,
interesting analysis have been carried out also by obtaining RM
of radio galaxies through a sample of clusters.
Clarke et al. \cite{Clarke2001} analyzed the average RM values
as a function of source impact parameter for a sample of 16 Abell clusters.
They found a clear broadening of the RM distribution toward small
projected distances from the cluster center clearly indicating,
as suggested also by other statistical
investigations (e.g.
\cite{Lawler1982,Valle1986,Kim1991,Johnston-Hollitt2004}),
that most of the RM contribution comes from the intracluster medium
and proving that magnetic fields are present in all galaxy clusters.
A similar correlation has been recently found by Govoni et al.
\cite{Govoni2010}.
They analyzed a sample of 12 galaxy clusters in which they included
only those clusters hosting radio sources for which high-quality,
high-resolution, extended RM images are available.
Although the statistic is still limited, they found that the
RM of radio galaxies in clusters containing a radio halo seem to show a
similar behavior to that of clusters without diffuse synchrotron emission.

Increasing attention is given in the literature to the connection
between the magnetic field strength and the cluster temperature.
These studies are based on cosmological simulations
e.g. \cite{Dolag1999,Dolag2001,Bruggen2005,Xu2010} or plasma physical
considerations \cite{Kunz2011}.  The comparison between clusters with
high and low temperatures \cite{Govoni2010} does not yields
significant differences in the RM data, indicating that a possible
connection between the magnetic field strength and the gas
temperature, if present, is very weak.

Very recently Bonafede et al. \cite{Bonafede2011} selected a
sample of 39 massive galaxy clusters from the HIghest X-ray FLUx
Galaxy Cluster Sample \cite{Reiprich2002}, and used NVSS
data \cite{Condon1998} to analyze the trend of the fractional
polarization of radio sources lying at different projected distances
from the cluster center.  They detected a clear trend of the
fractional polarization, being smaller for sources close to the
cluster center and increasing with increasing distance form the
cluster central regions.  The higher RM suffered by sources in the
inner regions, where both magnetic field strength and gas density are
higher, causes in fact a higher beam depolarization.  Sources at
larger radii, instead, are subject to less depolarization, since their
emission is affected by lower RM. Such a trend can be reproduced by a
magnetic field model with central value of few $\mu$G.  In addition
there seems no differences in the depolarization trend observed in
cluster with and without radio halo, indicating that magnetic fields
in galaxy clusters are then likely to share the same properties
regardless of the presence of radio emission from the intracluster
medium.

\section{Beyond clusters}
\label{sec:8}

Cosmological theories and simulations predict that galaxy clusters
are connected by intergalactic filaments along which they accrete mass.
Shocks from infall into and along the filaments are expected
to accelerate particles. These accelerated particles can emit
synchrotron radiation if cosmic magnetic fields are present.
Attempts to detect diffuse radio emission beyond
clusters, i.e. in very rarefied regions of the intergalactic space,
have shown recent promise in imaging diffuse synchrotron radiation of
very low level.

\begin{figure*}
  \includegraphics[width=0.7\textwidth]{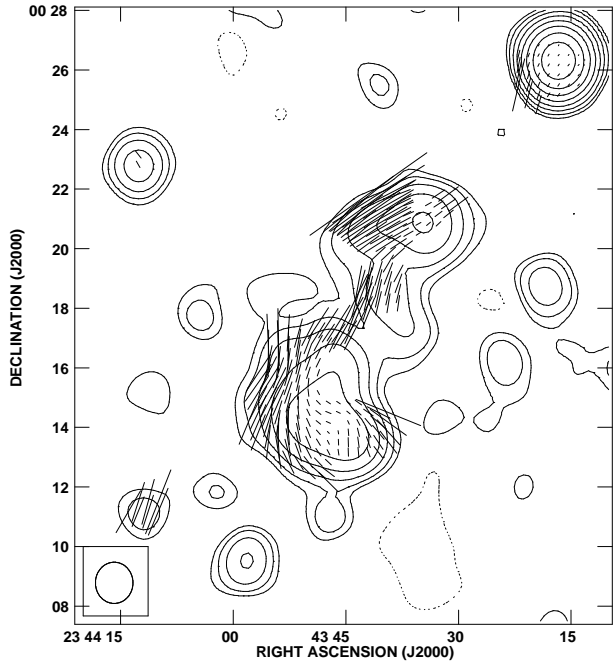}
\caption{
Large-scale radio emission of the filament ZwCl 2341.1+0000, obtained
with the VLA at 1.4 GHz (see \cite{Giovannini2010}), after subtraction
of unrelated point-sources, at the resolution of 83$'' \times$75$''$
(HPBW).  Contours show the total-intensity emission, at levels of 0.15
mJy/beam and following contours spaced by a factor of 2.  Lines refer
to the polarized emission: the orientation represents the projected
E-field (not corrected for galactic rotation), the length is
proportional to the fractional polarization (1$''$ corresponds to
0.2\%.)
}
\label{fig:filament}
\end{figure*}

Bagchi et al.\cite{Bagchi2002} suggested the presence of radio
emission from the filament surrounding the cluster
ZwCl2341.1+0000. This finding has been recently confirmed with deeper
observations by \cite{Giovannini2010} (Fig. \ref{fig:filament}),
thus proving that intergalactic filaments
can be magnetized.  Another filamentary structure extending over 3.3
Mpc is centered on the cD galaxy of the cluster A3444
\cite{Giovannini2009}. The large size of this radio feature, as well
as its morphology, rule out a direct connection with the central
galaxy and invite comparison with the filamentary structures found in
ZwCl2341.1+0000 (see also \cite{Vanweeren2009d}).
The presence of large-scale radio emission in regions in between
galaxy clusters, (e.g. 0917+75 \cite{Harris1993,Giovannini2000}), and
giant ring-like features (e.g. A3376 \cite{Bagchi2006}) seems to be
related to the large-scale structure formation process, rather than to
the cluster.

We note also that diffuse bridges of radio emission connecting the
halo to the relic have been observed in a few clusters, including Coma
\cite{Kim1989,Giovannini1990} and A2255 \cite{Govoni2005}.  In the Coma
cluster Kronberg et al. \cite{Kronberg2007}, confirmed by
Brown \& Rudnick \cite{Brown2011a}, detected a radio cloud
surrounding both the halo and the relic source. The radio cloud size
corresponds to about 4 Mpc if it is associated with the Coma cluster,
and is the most extensive complex of synchrotron emitting plasma yet
seen in cosmic large-scale structure.  In A2255, in addition to the
radio bridge, low brightness emission located far away from the
cluster center has been detected \cite{Pizzo2008}.

All the mentioned large-scale diffuse radio emissions might trace the
process of a large-scale structure formation, where cosmic shocks that
originated in complex merger events are able to amplify magnetic
fields and accelerate synchrotron electrons along cluster filaments.
In many cases the physical nature of these structures remains,
however, ambiguous; candidates include cosmological filaments,
unevolved protoclusters in the process of formation, and or
large-scale debris near rich, merging clusters.

In any case, all these data support the existence of an intergalactic
magnetic field more widespread and somewhat lower than that in the
intracluster medium within clusters.  This field may represent the
seed field for galaxies and clusters, and may play an important role
in the formation of large-scale structure.  These are important
results since they allow expanding our knowledge of magnetic fields
from clusters to filaments, and are a fundamental step to understanding
the origin and properties of cosmological scale magnetic fields.  We
are moving from magnetic field structures up to 1 Mpc scale to
structures 3 - 5 times more extended.  Attempts to probe magnetic
field strength on larger local-Universe scales have also been done via
Faraday rotation studies \cite{Xu2006}.

The measurement of extremely weak magnetic fields in the voids and
large scale structure is a challenging task and up to now only upper
bounds have been derived using various techniques. The tightest upper
bounds come from the search for the Faraday rotation of polarization
of the radio emission from distant quasars (e.g. \cite{Blasi1999a}),
and from the effect of magnetic fields on the anisotropy of Cosmic
Microwave Background radiation (e.g. \cite{Barrow1997,Durrer2000}).  A
recent approach, using the cascade emission from blazars has been
found to be promising (e.g.
\cite{Neronov2010,Dolag2011,Tavecchio2011,Taylor2011}).  The multi-TeV
gamma-ray flux from distant blazars is strongly attenuated by pair
production of the infrared/optical extragalactic background light,
initiating electromagnetic cascades in the intergalactic space, whose
charged component is deflected by the extragalactic magnetic fields.
Observable effects of these cascades include the delayed ``echoes'' of
multi-TeV gamma ray flares or gamma-ray bursts and the appearance of
extended emission around point-like gamma-ray sources.  From Fermi
observations, lower bounds are at levels of 10$^{-17}$ to 10$^{-15}$
G, depending on the assumptions.  Dolag et al.  \cite{Dolag2011}
derive that these magnetic fields are likely to fill at least 60\% of
space, thus imposing constraints on the magnetic field origin: either
the magnetic fields are produced by a volume filling process
(e.g. primordially) or very efficient transport mechanisms must be
present to distribute the locally generated magnetic fields into the
space.

\section{Models of the origin of diffuse sources}
\label{sec:9}

The connection between extended diffuse radio sources and cluster
evolution, in particular the link between merger processes and
halos/relics on one hand, and the link between mini-halos and relaxed
clusters on the other hand, is well established as reported in this review.
Major cluster mergers 
can supply energy to the radio emitting particles,
as well as  amplify magnetic fields
\cite{Roettiger1999a,Dolag1999,Dolag2002,Ryu2008}. It has been shown that the
magnetic field strength can grow during the process of cluster
formation, from initial fields of the order of 10$^{-9}$ G at z = 15
to final fields of the order of $\mu$G at z = 0.  This magnetic field
amplification is spread over the entire history of cluster formation,
and related to the various merger processes, therefore it is not
expected to directly affect the formation of diffuse radio sources.

Since magnetic fields have been found to be ubiquitous in galaxy
clusters (see Sect. \ref{sec:74}), the crucial ingredient for the
existence of diffuse synchrotron radio sources is the presence of
relativistic particles.  Radio halos are most difficult to explain,
because of their very large size. Radiating electrons cannot travel
such large distances within their lifetime, because of strong
radiative losses by synchrotron and inverse Compton emission.
Two main classes of models have been suggested for the origin of
relativistic electrons present in the cluster volume and responsible
for the diffuse radio emission: the {\it primary electron} model which
predict that relativistic particles are continuously accelerated in
the cluster volume, and the {\it hadronic (secondary electrons)}
model, in which relativistic electrons are produced through the
cluster volume by pp collisions.

We introduce here these models and summarize their expectations, to
draw a picture of the formation scenario of diffuse radio sources. For
a detailed analysis of the theoretical studies on this topic we refer
to recent papers (e.g. \cite{Ensslin2011,Brunetti2011a} and
references therein).

\subsection{Primary electrons}
\label{sec:91}

Primary relativistic electrons are present in the cluster volume,
because they were injected by AGN activity (quasars, radio galaxies,
etc.), or by star formation in normal galaxies (supernovae, galactic
winds, etc.) during the cluster dynamical history (see
\cite{Blasi2007} for a review).  This population of electrons suffers
strong radiation losses mainly because of synchrotron and inverse
Compton emission, thus reacceleration is needed to maintain their
energy to the level necessary to produce the observed synchrotron
radio emission in relatively weak magnetic fields
(e.g.
\cite{Jaffe1977,Schlickeiser1987,Brunetti2001,Petrosian2001,Brunetti2011b}).
The possibility that electrons are accelerated from the sub-relativistic
electrons has also been considered in the literature
(e.g. \cite{Dogiel2007} and references therein).

For the above reasons, primary electron models can also be referred to as
reacceleration models.
There are two main ways that entail the transfer of energy from the
cluster ICM to the radiating particles: these are cluster turbulence
and cluster shocks.

\subsubsection{Reacceleration by turbulence: radio halos and mini-halos}
\label{sec:911}

Simulations show that during cluster mergers, turbulence is generated
throughout the cluster over $\sim$ Mpc scales.  Energy can thus be
transferred from the ICM into the non-thermal component through
resonant or non-resonant interaction of electrons with MHD turbulence
(see e.g.
\cite{Brunetti2001,Brunetti2004b,Petrosian2001,Fujita2003,Cassano2005,Xu2009,Xu2010}).
The emerging scenario is that turbulence reacceleration is likely the
major mechanism responsible for the
supply of energy to the electrons radiating in radio halos.
Turbulent acceleration is like a second-order Fermi process,
i.e. related with a random processes, and consequently
 not quite efficient.  The time during which the process is
effective is relatively short (a few 10$^8$ years), so that the radio
emission is expected to correlate with ongoing or most recent
merger events (e.g. \cite{Brown2011b,Feretti2004a,Ensslin2011}).

These models predict that reaccelerated electrons will have a maximum
energy ($\gamma \sim $ 10$^5$) which produces a high frequency cut-off
in the resulting synchrotron spectrum \cite{Brunetti2004a}.  Thus a
high frequency steepening of the integrated spectrum is expected, as
well as a radial steepening and/or a complex spatial distribution of
the spectral index between two frequencies, the latter due to
different reacceleration processes in different cluster regions.
The evidence of the presence of faint diffuse magnetic fields in all
galaxy clusters (see Sect. \ref{sec:74}), the
behaviour of radio spectra (see Sect. \ref{sec:41}), the existence of
ultra-steep radio halos (Sect. \ref{sec:413}), the correlation between
radio spectra and temperature (Sect. \ref{sec:412}), the association
between radio halos and cluster mergers (Sect. \ref{sec:432}), and the
fact that halos are not common in galaxy clusters \cite{Kuo2004},
are supportive of primary electron models for radio halos.
The radio power - X-ray luminosity correlation (Sect. \ref{sec:43})
is easily reproduced by reacceleration models (see e.g. \cite{Brunetti2009}).

In mini-halos, the MHD turbulence associated with the cool-core region has
been also suggested to be responsible for reacceleration of radiating
particles \cite{Gitti2002}. This is consistent with the spectral
properties and the observed correlation between the mini-halo radio
power and the cooling rate power \cite{Gitti2004}.  Another
possibility, recently suggested by Mazzotta \& Giacintucci
\cite{Mazzotta2008}
and ZuHone et al. \cite{ZuHone2011}, is that turbulence in cool-core
clusters could arise from the sloshing motions of the cluster core
gas.  This hypothesis is supported by the detection of spiral-shaped
cold-fronts in the X-ray emission of cool-core clusters, suggested to
be the signature of gas sloshing (see e.g. \cite{Giacintucci2011c}).

\subsubsection{Reacceleration by shocks: relics }
\label{sec:912}

Shock acceleration is a first-order Fermi process of great importance
in radio astronomy, as it is recognized as the mechanism responsible
for particle acceleration in supernova remnants.  The acceleration
occurs diffusively, in that particles scatter back and forth across
the shock, gaining at each crossing and recrossing an amount of energy
proportional to the energy itself.  The acceleration efficiency is
mostly determined by the shock Mach number.

Current observations globally favour the scenario that cluster shocks
are strictly linked to relics \cite{Sarazin1999,Keshet2004}).
The production of outgoing shock waves at the cluster periphery with
Mach Number $\sim$ 2 - 3 is indeed observed in numerical simulations of
cluster
merger events and more generally in the large-scale structure
formation \cite{Miniati2000,Ryu2003,Ryu2008,Vazza2009}.
Accelerated electrons have been suggested either to be thermal ICM
electrons \cite{Ensslin1998}, or relativistic electrons
released by a former active radio galaxy
\cite{Ensslin2001,Ensslin2002,Hoeft2004}.

Because of the short electron radiative lifetimes, radio emission is
produced close to the location of the shock waves.  These models also
predict that the magnetic field within the relic is aligned with the
shock front, and that the radio spectrum is flatter at the shock edge,
where the radio brightness is expected to decline sharply. These
expectations are consistent with the classic elongated structure of
relics, almost perpendicular to the merger axis, and their
polarization properties.
Vazza et al. \cite{Vazza2011} show that the detection frequency of radio
relics
and the fact that they are also detected in cluster peripheries is
consistent with their link to shocks.

The detection of shocks in the cluster outskirts is difficult with
current instruments, because of the very low gas density and thus
X-ray brightness in these regions (for a recent review see
\cite{Markevitch2007}).  In the Coma cluster, Feretti \&
Neumann \cite{Feretti2006b}, analyzing Newton-XMM data, did not detect the
temperature jump implied by the presence of a shock at the relic
location, although the brightness trend may be consistent with the
shock. Solovyeva et al. \cite{Solovyeva2008} presented
evidence of a shock at the location of the relics in A548b.  A weak
shock has been detected by Macario et al. \cite{Macario2011b}
in A754. Recently, shocks have been detected with Suzaku in
A3376 \cite{Akamatsu2011} and A3667 \cite{Akamatsu2011b}.

It cannot be excluded that shock acceleration may also be efficient
in some particular regions of a halo (e.g. \cite{Markevitch2010}),
however, the radio emission of halos can be very extended up to large
scales, thus it is hardly associated with localized shocks.
Moreover, some clusters exhibit a spatial correlation between the radio halo
emission and the hot gas regions \cite{Govoni2004}, however, this is
not a general feature, and in some cases the hottest gas regions do
not exhibit radio emission.

\subsection{Secondary electrons}
\label{sec:92}

In the hadronic model, secondary electrons are injected as secondary
particles by inelastic nuclear collisions between the relativistic
protons and the nuclei of the thermal ambient intracluster medium
(e.g. \cite{Dennison1980,Blasi1999b,Dolag2000,Keshet2010}).  The
protons diffuse on large scale because their energy losses are
negligible.  They can continuously produce {\it in situ} electrons,
distributed through the cluster volume. The possibility that the
high-energy electrons, responsible for the synchrotron emission, arise
from the decay of secondary products of the neutralino annihilation in
the dark matter halos of galaxy clusters has also been suggested
\cite{Colafrancesco2001}.

Secondary electron models have been proposed for the emission of radio
halos (see e.g. \cite{Dennison1980,Dolag2000}) and of mini-halos
\cite{Pfrommer2004}.  They cannot work for relics, since
peripheral cluster regions do not host a sufficiently dense thermal
proton population required as targets for the efficient production of
secondary electrons.

Secondary electron models can reproduce the basic properties of the
radio halos provided that the strength of the magnetic field averaged
over the emitting volume is larger than a few $\mu$G.  In this case,
they predict synchrotron power-law spectra, which are independent of
cluster location, i.e. do not show any features and/or radial
steepening, and the spectral index values are flatter than $\alpha$
$\sim$ 1.5 (e.g. \cite{Brunetti2004a}).  The profiles of the radio
emission should be steeper than those of the X-ray gas
(e.g. \cite{Govoni2001b}). Since the radio emitting electrons
originate from protons accumulated during the cluster formation
history, no correlation to recent mergers is expected, but halos
should be present in virtually all clusters.  Moreover, emission of
gamma-rays and of neutrinos is predicted.

Recently, a universal linear correlation between radio and X-ray
surface brightness in halos and mini-halos was found
\cite{Keshet2010}, assuming secondary electron models and a relatively
strong magnetized IGM ($>$ 3 $\mu$G). This model reproduces the
observed radio spectral steepening, but requires a very high central
magnetic field ($>$ 10 $\mu$G).

\subsection{Hybrid models}
\label{sec:93}

The intracluster medium is a complex mixture of thermal and
non-thermal protons and electrons. It has been suggested in that halos
could arise from both primary and secondary electrons.  Relativistic
protons are an important component of relativistic ICM particles, and
they are an important origin of relativistic electrons. MHD turbulence
can reaccelerate both primary and secondary particles, producing
radio emission in agreement
with observational results (see e.g. \cite{Miniati2001}).  The
relative contribution of these two components is constrained by the
observational properties.  Constraints on the energy budget of the
various components have been derived by Brunetti \& Blasi
\cite{Brunetti2005} and more recently by Brunetti \& Lazarian
\cite{Brunetti2011c}.  These authors discuss also that secondary
electrons could give rise to Mpc scale halos characterized
by very low surface brightness and radio power, not yet observed
because of observational limits of present radio telescopes, but
which should be detected by the incoming new radio telescopes.

\subsection{Constraints from gamma-ray emission}
\label{sec:94}

The collisions producing secondary electrons also lead to gamma-ray
emission through the production and subsequent decays of neutral
pions.  The implications of gamma-ray detections from galaxy clusters
are largely discussed in literature (see e.g. Blasi et
al. \cite{Blasi2007} and references therein).  However, diffuse
gamma-ray emission has not yet been detected from galaxy clusters
\cite{Reimer2003,Ackermann2010}.  These non detections constrain
the possible energy density in cosmic-ray protons and, therefore,
the density of relativistic electrons that may be produced from these.

The Fermi-LAT Gamma-ray Space Telescope greatly improved the
sensitivity of observations at MeV/Gev energies and reported gamma-ray
upper limits for a large sample of clusters \cite{Ackermann2010}.
Jeltema \& Profumo \cite{Jeltema2011} examined the secondary electron
models for the origin of halos in the light of these new Fermi
results.  The non detection of gamma-ray emission from clusters
puts constraints to
the possible density of secondary relativistic electrons and
imply a minimum magnetic field strength to reproduce the observed
emission from radio halos.  For clusters such as A1914 and A2256,
average magnetic fields at least of the order of several $\mu$Gauss
are needed  to avoid an overproduction of gamma-rays,
not seen by Fermi \cite{Jeltema2011}.
The implied magnetic fields in hadronic models are in
excess of, or close to, the largest cluster magnetic field values
obtained with Faraday Rotation Measures (see Sect. \ref{sec:7}).
This seems problematic for the
hadronic origin of radio halos.  However, at present, due to the
uncertainties, secondary
electron models are not ruled out by Fermi gamma-ray data even
if required magnetic fields are in some clusters higher than values
obtained with other observations. This result show the importance of
high energy data in understanding non-thermal emission in galaxy
clusters.

\section{Future}
\label{sec:10}

Our knowledge of non-thermal phenomena in galaxy clusters has
significantly improved in the past years, owing to good and deep
observations and advanced modeling.
Nevertheless, new observations will be necessary to solve the
outstanding issues faced in determining what is the dominating
mechanism in the formation and maintenance of the diffuse radio
emission, in particular:
1) search for new sources of the different classes, in particular at low
powers and at large redshifts, to improve the information on
the statistical properties of cluster diffuse sources;
2) obtain polarization information on halos, and perform
Rotation Measure studies of embedded and background radio
sources, to get a detailed knowledge of the magnetic field
strength, structure, and degree of ordering;
3) derive accurate integrated spectra on a large frequency range and
detailed spectral index distributions at high resolution.

New generation instruments, which are becoming available, as the EVLA
and LOFAR, or are planned, as the SKA precursors (ASKAP, MeerKAT), and
SKA, will allow a breakthrough in these studies.

Owing to the new capabilities of the EVLA, in particular sensitivity,
frequency agility, and spectral capability, it will be possible to
detect faint radio emission and trace the magnetic fields in galaxy clusters.

The new radio telescope LOFAR will explore the Universe at low
frequencies, as shown e.g. by the first results on the cluster A2256
\cite{Vanweeren2012}.
 As diffuse cluster radio sources are bright at low
frequency, LOFAR will be able to detect thousands of such sources up
to the epoch of cluster formation, in particular halos and relics with
ultra-steep spectra, which are missed at cm frequencies.
This will allow the investigation of the occurrence of diffuse cluster
radio sources. Their properties will be derived as a function of
cluster properties up to the epoch at which the first massive clusters
assembled (z $\sim$ 1), testing predictions that cluster merging is
rampant at high redshift.
The detection of synchrotron radiation at the lowest possible levels
will allow the measurement of magnetic fields in clusters and in even
more rarefied regions of the intergalactic space, and the
investigation of the relation between the formation of magnetic fields
and the formation of the large-scale structure in the universe.  With
LOFAR, owing to the long wavelengths involved, it will also be
possible to obtain Faraday rotation down to 0.1-1 rad m$^{2}$, which
are very difficult to measure with the current instruments at cm
wavelengths. The detection of very low Rotation Measures implies in
turn the detection of very faint magnetic fields therefore LOFAR will
become the radio telescope to measure the weakest cosmic magnetic
fields so far.

Two projects planned with the  Australian Square Kilometre Array
Pathfinder (ASKAP), i.e. the deep radio sky survey project EMU
(Evolutionary Map of the Universe) and the polarization survey
POSSUM (Polarisation Sky Survey of the Universe's Magnetism)
will allow the detection of diffuse radio sources and the improvement
of Faraday rotation information in clusters and in the overall
intergalactic medium.
The first science with the MeerKAT array in South Africa
region, in particular the MIGHTEE (MeerKAT International GigaHertz
Tiered Extragalactic Exploration Survey) project will
cover a wide range of astrophysics, including non-thermal
phenomena in clusters.

APERTIF phased arrays that are being installed in WSRT will
dramatically increase  the survey speed for WSRT. This will
enable the WODAN (Westerbork Observations of the Deep APERTIF
Northern-Sky) project, a survey complementary to EMU.

The most innovative studies will be performed with SKA: they are summarized
in ''Science with the Square Kilometre Array'' \cite{Carilli2004}
(for cluster and intergalactic medium studies see e.g.
\cite{Feretti2004c,Feretti2004d,Keshet2004,Gaensler2004,Beck2011}).
Recently, Krause et al. \cite{Krause2009}
investigate the potential of the Square
Kilometer Array (SKA) for measuring the magnetic fields in clusters
of galaxies via Faraday rotation of background polarized sources.
They derive the distribution of RM due to clusters, by using
an average gas density distribution and adopting models for the
population of cluster radio sources, for the cluster mass distribution,
and for the source polarization.
They calculate the statistics for background RM measurements
in clusters at different redshift and different SKA exposures,
arrays and completion levels in the mid frequency range.
They conclude that the full SKA will be able to make very detailed
magnetic field structure measurements of clusters with more than
100 background sources in clusters up to z = 0.5, and more
than 1000 background radio sources in nearby clusters.
Bogdanovic et al. \cite{Bogdanovic2011} analyze different physical mechanisms
operating in the ICM, i.e. conduction-driven MHD instabilities and
turbulence-dominated instabilities, and derive that these deeply
affect the structure of the cluster magnetic field producing different
field topologies.  They evaluate the effects of the two mechanisms on
the Faraday RM of background radio sources and conclude that future
spectropolarimetrimetric measurements with EVLA and SKA will have
sufficient sensitivity to discriminate between them.

Observations in the radio domain will need to be complemented
by data at all other wavelenghts, in particular optical, X-ray and
gamma-ray, to establish
the cluster conditions, the merger evolutionary stage, the presence
and properties of shocks, the signatures of cluster turbulence,
and to derive firm correlations on large samples and over a large range of
parameters between radio properties and
cluster parameters.
Finally, the study of the non-thermal inverse Compton
emission in the hard X-ray and extreme ultraviolet domains needs to be
pursued on several clusters, with and without diffuse emission, to get
independent information on the existence of relativistic particles.

\begin{acknowledgements}                                                       
We are grateful to W.D. Cotton for a careful and 
critical reading of the manuscript. We thank R. Morganti and T. Courvoisier
for helpful suggestions. 
\end{acknowledgements}

\end{document}